\documentclass[%
reprint,
superscriptaddress,
bibnotes,
amsmath,amssymb,
aps,
floatfix
]{revtex4-1}

\usepackage[dvipdfmx]{graphicx}
\usepackage{graphicx}
\usepackage{color}
\usepackage{amsmath,amssymb,amsthm,mathrsfs,amsfonts,dsfont}
\usepackage{subfigure, epsfig}
\usepackage{braket}
\usepackage{bm}
\usepackage{bbm}
\usepackage{enumerate}
\usepackage{comment}
\usepackage{appendix} 
\usepackage[colorlinks,linkcolor=blue,citecolor=blue]{hyperref}
\newcommand{\tr}{\mathrm{Tr}}

\newtheorem{theorem}{Theorem}

\usepackage[utf8]{inputenc}
\usepackage{array}       
\usepackage{cellspace}   
\usepackage{graphicx}    

\newcolumntype{C}[1]{>{\centering\arraybackslash}m{#1}}
\addparagraphcolumntypes{C}   

\newcommand{\blue}[1]{\textcolor{blue}{#1}}

\begin{document}
\title{Measuring multipartite quantum correlations by thermodynamic work extraction}

\author{Toshihiro Yada}
\email{yada@noneq.t.u-tokyo.ac.jp}
\affiliation{Department of Applied Physics, The University of Tokyo, 7-3-1 Hongo, Bunkyo-ku, Tokyo 113-8656, Japan}

\author{Nobuyuki Yoshioka}
\affiliation{Department of Applied Physics, The University of Tokyo, 7-3-1 Hongo, Bunkyo-ku, Tokyo 113-8656, Japan}
\affiliation{Theoretical Quantum Physics Laboratory, RIKEN Cluster for Pioneering Research (CPR), Wako-shi, Saitama 351-0198, Japan}
\affiliation{JST, PRESTO, 4-1-8 Honcho, Kawaguchi, Saitama, 332-0012, Japan}

\author{Takahiro Sagawa}
\affiliation{Department of Applied Physics, The University of Tokyo, 7-3-1 Hongo, Bunkyo-ku, Tokyo 113-8656, Japan}
\affiliation{Quantum-Phase Electronics Center (QPEC), The University of Tokyo, 7-3-1 Hongo, Bunkyo-ku, Tokyo 113-8656, Japan}


\begin{abstract}
Quantum correlations are at the core of quantum mechanics and play a crucial role in various fields such as quantum information, condensed matter physics, high energy physics, and quantum thermodynamics. 
While bipartite quantum correlations have been extensively studied, multipartite quantum correlations in many-body systems remain elusive due to their complex structure. 
In particular, a primary challenge lies in the fact that the calculation of multipartite quantum correlation measure often requires exponential cost with respect to the system size.
In this work, we tackle this problem by adopting a thermodynamic approach; we introduce a measure of multipartite quantum correlations based on the difference in extractable thermodynamic work by global operations and LOCC (local operations and classical communication). This can be regarded as a multipartite generalization of the work deficit, which has attracted attention as a thermodynamic measure of bipartite quantum correlation.
A distinguishing feature of the thermodynamic approach to multipartite quantum correlation is that we can compare the degree of quantum correlations with clear operational meaning; for example, it is shown that W state has more quantum benefit in work extraction than GHZ state.
Importantly, we develop an efficient calculation method of the multipartite work deficit. 
This efficient method works for a special class of quantum many-body systems described by MPS (matrix product states), where the numerical cost is shown to be proportional to the system size, significantly reducing the exponential cost required for the direct calculations. 
We demonstrate this efficient method in representative MPS such as the AKLT state and the cluster state, and analytically obtain the exact values of this measure.
We further show that a quantum phase transition described by MPS is well captured by the multipartite work deficit, where a singular behavior appears and becomes sharper under coarse-graining, at the transition point.
This shows that the multipartite work deficit does not only highlight the fundamental connection between multipartite quantum correlations and quantum thermodynamics, but also serves as an efficiently-computable probe of the structures of quantum many-body systems including their quantum phase transitions.
\end{abstract}
\maketitle

\section{Introduction}
In light of recent developments in quantum technologies \cite{gross2017quantum,schafer2020tools,kjaergaard2020superconducting,krantz2019quantum,lukin2020integrated,altman2021quantum}, quantum correlations are becoming increasingly important not only in quantum computation and quantum communication \cite{gisin2007quantum,nielsen2010quantum,briegel2009measurement} but also in various fields of physics including condensed matter physics \cite{amico2008entanglement,de2018genuine} and high energy physics \cite{ryu2006aspects,ryu2006holographic,takayanagi2012entanglement}.
For example, quantum correlations have been utilized as useful probes for understanding critical phenomena \cite{vidal2003entanglement,calabrese2004entanglement,calabrese2009entanglement}, topological order \cite{kitaev2006topological,levin2006detecting}, and information scrambling \cite{swingle2016measuring}.
While the understanding on bipartite quantum correlations has been well-established through various measures with clear operational meanings \cite{plenio2005introduction,horodecki2009quantum,modi2012classical,bera2017quantum}, multipartite quantum correlations remain a very active research avenue \cite{plenio2001bounds,wei2004connections,barnum2001monotones,wei2003geometric,shimony1995degree,orus2008universal,shi2010finite,orus2010visualizing,huang2010multipartite,mousolou2013unifying,orus2014topological,shi2016geometric,orus2010geometric,rulli2011global,campbell2011global,campbell2013global,power2015nonclassicality,sun2015global,huang2018global,braga2012monogamy,xu2013analytical,liu2014general,coffman2000distributed,radhakrishnan2020multipartite,hu2018quantum,hyllus2012fisher,toth2012multipartite,frowis2012measures,hauke2016measuring,mermin1990extreme,ardehali1992bell,pezze2017multipartite}, mainly because of their rich and complicated structure \cite{guhne2009entanglement,bengtsson2016brief}. For example, even in tripartite quantum states, there are different types of quantum correlations represented by the W state and the GHZ state.

\begin{figure}[]
    \centering
    \includegraphics[width=0.5\textwidth]{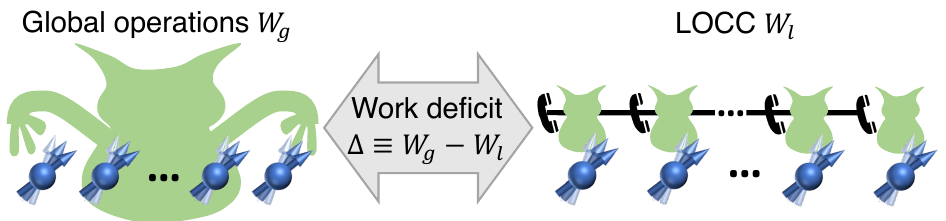}
    \caption{Schematic of the definition of the multipartite work deficit. The multipartite work deficit, denoted as $\Delta$, is defined as the difference between the extractable works through global operations, $W_g$, and that through LOCC, $W_l$, expressed as $\Delta \equiv W_g - W_l$.}
    \label{fig:intro}
\end{figure}

Another fundamental aspect of quantum correlations is its connection to thermodynamics \cite{binder2018thermodynamics}. 
A crucial issue is how quantum correlations affect energetic quantities such as extractable work and free energy. 
This connection between quantum correlations and thermodynamics is twofold: it not only reveals the extractable work from quantum correlations \cite{sagawa2008second,funo2013thermodynamic,kim2011quantum,park2013heat,hovhannisyan2013entanglement,perarnau2015extractable,huber2015thermodynamic,bruschi2015thermodynamics,giorgi2015correlation,andolina2019extractable,elouard2017role,manzano2018optimal}, but also enables the quantification of quantum correlations themselves based on thermodynamics. Specifically, a measure of quantum correlation, called the work deficit, is introduced as the difference in extractable works through global operations and LOCC (local operations and classical communication) \cite{oppenheim2002thermodynamical,horodecki2003local,horodecki2005local,devetak2005distillation,zurek2003quantum,brodutch2010quantum,lang2011entropic}.
While the work deficit is widely investigated in bipartite systems, the multipartite work deficit is almost unexplored except for simple demonstrations in the seminal works such as Refs.~\cite{oppenheim2002thermodynamical,horodecki2005local}.
Now, the critical questions are: $(\mathrm{i})$ how to formulate multipartite work extraction, $(\mathrm{ii})$ how the thermodynamic approach is useful for multipartite quantum correlations, $(\mathrm{iii})$ how one can efficiently compute the multipartite work deficit. 

In this work, we introduce the multipartite work deficit defined with work extractions as shown in Fig.~\ref{fig:intro}, for which the above points are clarified. By formulating the multipartite work extraction through non-adaptive LOCC, we obtain its expressions exactly calculable for general few-body states. For example, we show that the tripartite work deficit is given by $\ln 2$ in the GHZ state and $\ln 3$ in the W state. This quantitatively demonstrates that the W state has larger tripartite quantum correlation than the GHZ state, in terms of thermodynamically extractable work.

Furthermore, we develop an efficient calculation method of the multipartite quantum correlations for a special class of quantum many-body systems described by MPS (matrix product states) \cite{perez2006matrix,verstraete2008matrix,cirac2017matrix,cirac2021matrix,orus2014practical}.
While the cost of naive direct calculation of the multipartite work deficit grows exponentially with the system size as is the case for many other measures of multipartite quantum correlations \cite{plenio2001bounds,wei2004connections,barnum2001monotones,wei2003geometric,shimony1995degree,orus2008universal,shi2010finite,orus2010visualizing,mousolou2013unifying,huang2010multipartite,orus2014topological,shi2016geometric,orus2010geometric,rulli2011global,campbell2011global,campbell2013global,power2015nonclassicality,sun2015global,huang2018global}, the cost of our efficient method is proportional to the system size enabling the significant cost reduction (Table~\ref{tab:num_cost}).
It is noteworthy that this efficient method is not restricted to the work deficit but also is applicable to another measure of multipartite quantum correlation (i.e., global quantum discord \cite{rulli2011global}), thereby enabling the same reduction in numerical cost. This suggests the potential applicability of our method to various measures of multipartite quantum correlation.

\begin{table}[]
    \centering
    \begin{tabular}{C{2.8cm}C{2.6cm}C{3.0cm}} 
    \hline
     & Direct calculation & \begin{tabular}{@{}c@{}}  \fontsize{7}{7}\selectfont Efficient calculation \\[-0.7ex]  \fontsize{7}{7}\selectfont with error bound $\varepsilon$ \end{tabular}   \\ \hline\hline
    \ \ $N$-site MPS         & $e^{\mathcal{O}(N)}$    &   $N e^{\mathcal{O}\left(\frac{1}{\varepsilon}\right)}$  \\ \hline
    \begin{tabular}{@{}c@{}} \fontsize{7}{7}\selectfont Normal MPS in \\[-0.7ex] \fontsize{7}{7}\selectfont thermodynamic limit \end{tabular} & not applicable & $e^{\mathcal{O}\left(\frac{1}{\varepsilon}\right)}$ \\ \hline
    \end{tabular}
    \caption{Numerical costs for calculating the multipartite work deficit in $N$-site MPS and MPS in the thermodynamic limit, up to constant multiples. Direct calculation scales exponentially with the system size $N$, making it infeasible in the thermodynamic limit. On the other hand, by employing the efficient calculation method introduced in this work, the costs are reduced to scale linearly with respect to $N$. Furthermore, this method is applicable even in the thermodynamic limit, for translationally invariant normal MPS \cite{cirac2017matrix,cirac2021matrix}, achieving a constant cost dependent only on the desired error bound $\varepsilon$, which scales as $ e^{\mathcal{O}\left(\frac{1}{\varepsilon}\right)}$.}
    \label{tab:num_cost}
\end{table}

By utilizing our efficient method, we demonstrate the multipartite work deficit in several examples of quantum many-body systems. 
In the AKLT state \cite{AKLT_PRL,affleck1988valence} and the cluster state \cite{briegel2001persistent}, representative examples of MPS, we analytically obtain the exact values for the multipartite work deficit. 
We also conduct numerical calculation in an MPS family \cite{wolf2006quantum}, where we observe that the multipartite work deficit shows a sharp kink at the quantum phase transition point.
This showcases that the multipartite work deficit serves as an efficiently computable probe of quantum phase transitions.
We further show that the kink at the transition point becomes sharper under coarse-graining, which confirms that our probe indeed captures the singularity that is physically relevant (i.e. not from the artifact of optimization).

While the primary focus of this paper is on the multipartite work deficit defined with non-adaptive protocol, we also discuss the measure employing adaptive LOCC. Specifically, we introduce the measure called multipartite one-way work deficit, defined with one-way LOCC. We show that most of the results for the non-adaptive case can be extended to the one-way case, including the efficient calculation method, while there reamins an open question regarding the accuracy.

The present work would establish a fundamental relationship between multipartite quantum correlations and thermodynamics, introducing measures of quantum correlations with clear operational meaning based on thermodynamic work extraction. 
Since the introduced measures are efficiently computable for quantum many-body systems described by MPS, they would be useful for probing the structure of quantum many-body systems including quantum phase transitions.
It is also noteworthy that because of the operational definitions of the multipartite work deficits, our framework can also be applied to real experimental setups such as ultracold atoms \cite{gross2017quantum,schafer2020tools} and superconducting qubits \cite{kjaergaard2020superconducting,krantz2019quantum}.

This paper is organized as follows. 
In Sec. II, we introduce the multipartite work deficit in terms of thermodynamic work extraction, and calculate it in some simple examples. 
In Sec. III, we propose an efficient calculation method of this measure in many-body systems, and demonstrate it in the AKLT state, the cluster state, and an MPS family. 
In Sec. IV, we introduce the framework of coarse-graining and apply it to the MPS family. 
In Sec. V, we introduce the multipartite one-way work deficit and discuss the applicability of our efficient calculation method to this measure. 
In Sec. VI, we summarize our work and discuss the future perspectives.

\section{Multipartite work deficit} \label{s:multi_work_deficit}
In this section, we introduce the multipartite work deficit, which is defined as the difference in extractable works through global operations and LOCC.
We here focus on a class of LOCC where only the non-adaptive local measurements are allowed, and derive a formula for the multipartite work deficit, which is exactly calculable for general few-body states. We demonstrate it in some simple examples such as the GHZ state, the W state and the multipartite states composed of Bell pairs.

\subsection{Definition of multipartite work deficit}\label{ss:def_multi_wd}
We introduce the measure of multipartite quantum correlations based on the work extraction operation. This measure is defined as the gap between extractable thermodynamic work through global operations $W_g$ and LOCC $W_l$, as 
\begin{equation}
    \label{eq:opdef_WD}
    \Delta \equiv W_g - W_l.
\end{equation}
We consider the situation where the work is extracted under the interaction with a heat bath at inverse temperature $\beta$ (but we set $\beta=1$ for simplicity of notation), and the system's internal energy is negligible (in other word, system's Hamiltonian is trivial).
This framework of quantifying quantum correlations based on work extraction has been extensively studied in bipartite systems \cite{oppenheim2002thermodynamical,horodecki2003local,horodecki2005local,devetak2005distillation,zurek2003quantum,brodutch2010quantum,lang2011entropic}.
We note that our framework of considering work extraction from the heat bath is distinct from the previous works \cite{hovhannisyan2013entanglement,perarnau2015extractable,huber2015thermodynamic,bruschi2015thermodynamics,giorgi2015correlation,andolina2019extractable,allahverdyan2004maximal,skrzypczyk2014work,binder2015quantacell,campaioli2017enhancing}, where the work extraction from system's internal energy with unitary operation is discussed.

We first quantify the extractable work through the global operations $W_g$.
When we consider the state $\rho$ of the $N$-partite system composed of $d$-dimensional parties named as $\{a_i\}_{i=1}^N$, the extractable work by the global operations is calculated as
\begin{equation}
    \label{eq:m_global_ext_work}
     W_g = N\ln d - S(\rho),
\end{equation}
where $S(\rho)\equiv - \tr[\rho \ln \rho]$ denotes the von Neumann entropy.
This equation shows that the extractable work $W_g$ is characterized with the gap between the maximum entropy of the system $N\ln d$ and the entropy $S(\rho)$ of the total system. 
This can be understood as the consequence of the second law of thermodynamics, which states that a larger amount of work can be extracted from a state with high purity. The concrete protocols to achieve the work extraction in Eq.~(\ref{eq:m_global_ext_work}) have been studied in, e.g., Refs.~\cite{alicki2004thermodynamics,jacobs2009second,parrondo2015thermodynamics}.

\begin{figure}[]
    \centering
    \includegraphics[width=0.5\textwidth]{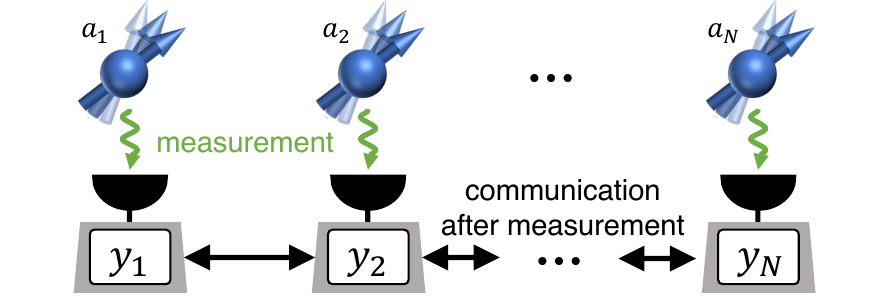}
    \caption{Schematic of non-adaptive local measurements and classical communication. The measurement outcomes can be shared among parties only after the local operations.}
    \label{fig:definition}
\end{figure}

We next consider the extractable work through LOCC $W_l$. 
When we denote the work extraction with a certain protocol $\Lambda$ as $W_\Lambda$, $W_l$ can be defined as the optimization of $W_\Lambda$ over the set of allowed LOCC protocols. In this section, we consider the LOCC class called zero-way closed LOCC, where the non-adaptive local projective measurement and classical communication are allowed, as shown in Fig.~\ref{fig:definition}. 
Correspondingly, we can refer to the work deficit (\ref{eq:opdef_WD}) mainly focused in this paper as the zero-way work deficit, while we often call it just the work deficit for simplicity.
The extractable work with such a protocol $\Lambda$ can be quantified as
\begin{equation}
\label{eq:work_lambda}
    W_\Lambda \equiv \sum_{n=1}^N \left\{\ln d - \sum_{y_n}p_{\Lambda}(y_n)S(\Tilde{\rho}_n^{\Lambda,y_n})\right\} - H_\Lambda (Y_N),
\end{equation}
where $Y_N \equiv (y_1, y_2, \dots, y_N)$ represents the measurement outcomes for all parties, with $y_n$ denoting the outcome of the $n$-th party.
Here, $H_{\Lambda}(Y_N)\equiv -\sum_{Y_N}P_{\Lambda}[Y_N]\ln P_{\Lambda}[Y_N]$ is the Shannon entropy for $Y_N$, where $P_{\Lambda}[Y_N]$ denotes the probability for the measurement outcomes $Y_N$ with measurement protocol $\Lambda$. The conditional density operator $\Tilde{\rho}_n^{\Lambda,y_n}$ represents the conditional state of $n$-th party when the measurement outcome $y_n$ is obtained, and $p_{\Lambda}(y_n)$ denotes the corresponding probability.

\begin{figure}[]
    \centering
    \includegraphics[width=0.5\textwidth]{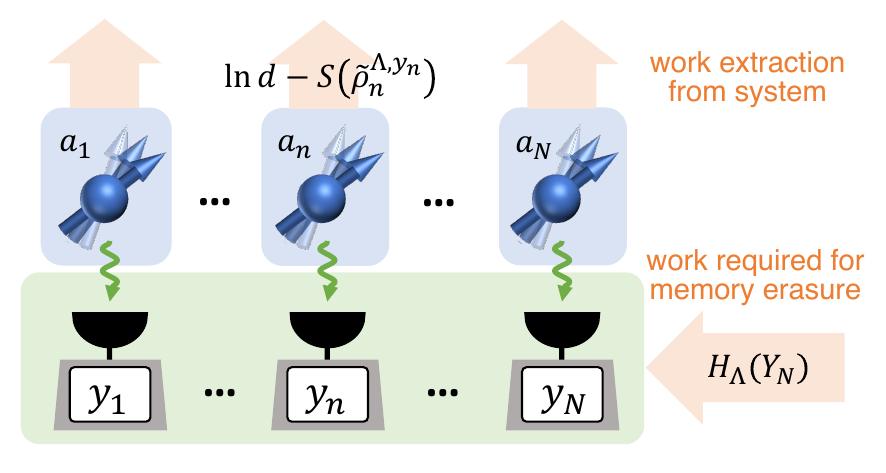}
    \caption{Schematic for extractable work through a measurement protocol $\Lambda$. Extractable work from $n$-th party $a_n$ is $\ln d - S(\Tilde{\rho}_n^{\Lambda,y_n})$ when the measurement outcome $y_n$ is obtained, and the averaged work extraction becomes $\ln d - \sum_{y_n} p_\Lambda (y_n) S(\Tilde{\rho}_n^{\Lambda,y_n})$. On the other hand, the work required to erase the measurement outcomes $Y_N$ is $H_\Lambda (Y_N)$. Therefore, the net work extraction can be quantified as the gap between these two contributions as shown in Eq.~(\ref{eq:work_lambda}).}
    \label{fig:WD_derivation}
\end{figure}

As shown in Fig.~\ref{fig:WD_derivation}, the net work extraction $W_\Lambda$ is decomposed into two contributions: work extraction from the system after the measurement, and the work needed to erase the memory. 
The first contribution corresponds to the terms in the curly bracket in Eq.~(\ref{eq:work_lambda}). Since we cannot feed back the measurement outcomes of the different party in the current setup, the extractable work from each party is the gap between the maximum entropy $\ln d$ and the entropy of conditional state $S(\Tilde{\rho}_n^{\Lambda,y_n})$. By taking the average over the measurement outcome $y_n$, we can derive the expression in Eq.~(\ref{eq:work_lambda}). The second contribution of the memory erasure corresponds to the Shannon entropy term $H_\Lambda (Y_N)$. 
Since the measurement outcomes can be communicated among parties, we can collectively erase all the outcomes $Y_N\equiv(y_1,y_2,\dots,y_N)$.
This enables the memory erasure with work cost $H_\Lambda (Y_N)$, which is less than that in the case without classical communication $\sum_{n=1}^N H_\Lambda(y_n)$.

The extractable work $W_l$ can be calculated by further optimizing the measurement protocol $\Lambda$.
Since we can show that the optimal local measurement becomes always rank-one as detailed in Appendix~\ref{apps:expressions}, $W_l$ can be represented as 
\begin{equation}
    \label{eq:Wl_zero-way}
    W_l = N\ln d - \min_{\Lambda\in \mathcal{M}_N} H_{\Lambda}(Y_N).
\end{equation}
Here, $\mathcal{M}_N$ is the set of measurement protocols whose measurement operators are denoted as $\{\Pi_{y_1}\otimes\Pi_{y_2}\dots \otimes \Pi_{y_N}\}$, where $\{\Pi_{y_n}\}$ is a set of rank-one projectors on $n$-th party.
By combining the expressions of Eqs.~(\ref{eq:m_global_ext_work}) and (\ref{eq:Wl_zero-way}), we can derive a formula for the multipartite work deficit as
\begin{equation}
\label{eq:zWD_Sh_def}
    \Delta = \min_{\Lambda\in \mathcal{M}_N} H_{\Lambda}(Y_N) - S(\rho).
\end{equation}
This equation shows that $\Delta$ is quantified with the optimization of the local measurements on $N$ parties $\Lambda\in \mathcal{M}_N$, which can be exactly performed with reasonable numerical cost for small systems.

By simple transformation, Eq.~(\ref{eq:zWD_Sh_def}) can be reformulated as the following expression:
\begin{equation}
\label{eq:zWD_zero_way_c}
    \Delta = \min_{\sigma \in \mathcal{C}_N } S(\rho\|\sigma),
\end{equation}
where $S(\rho\|\sigma) \equiv \tr[\rho (\ln \rho -\ln \sigma)]$ denotes the quantum relative entropy.
Here, $\mathcal{C}_N$ represents the set of classically correlated states, which can be represented as
\begin{equation}
\label{eq:cc_state}
    \sigma \equiv \sum_{Y_N} q(Y_N) \Pi_{y_1} \otimes \Pi_{y_2} \dots\otimes \Pi_{y_N},
\end{equation}
with a set of orthogonal rank-one projectors $\{\Pi_{y_n}\}$, and a certain probability distribution $q(Y_N)$.
From Eq.~(\ref{eq:zWD_zero_way_c}), we can see that the multipartite work deficit $\Delta$ is the optimization over the local bases of the relative entropy of coherence. 

We remark that multipartite work deficit is reduced to Eq.~(\ref{eq:zWD_zero_way_c}), which is equivalent to other measures of multipartite quantum correlations introduced in the previous works \cite{piani2008no,saitoh2008nonclassical,modi2010unified,yao2015quantum,braga2014maxwell}, in the contexts totally different from thermodynamics. Here, we have formulated the multipartite work deficit based on thermodynamics.
More importantly, we emphasize that the efficient calculation method discussed in Sec.~\ref{s:calc_many_body} has not proposed in any of the previous works.

We also remark the relationship between the multipartite work deficit and entanglement.
For the bipartite pure state, it has been shown in the previous work \cite{oppenheim2002thermodynamical} that the work deficit is reduced to the entanglement entropy $\Delta = S(\rho_1)$, where $\rho_1$ is the reduced density operator for a single party.
On the other hand, for the mixed states in bipartite and multipartite systems, the work deficit can take positive values even for separable states, as is the case for the quantum discord \cite{ollivier2001quantum,henderson2001classical}.
This indicates that the notion of multipartite quantum correlation for the work deficit is broader than that of entanglement theory.
This can be understood from the fact that the set of classically correlated states (\ref{eq:cc_state}) is included by the set of separable states, represented as
\begin{equation*}
    \sigma \equiv \sum_{x} q(x) \sigma_1^x \otimes \sigma_2^x \dots \otimes \sigma_N^x.
\end{equation*}

\subsection{Demonstration in three-qubit systems}\label{ss:cal_3_qubit}
We demonstrate the behavior of multipartite work deficit in some examples of three-qubit system. 
We first address the GHZ and W states, which are the distinct types of tripartite quantum correlation.
For the GHZ state, defined as
\begin{equation*}
\ket{\rm GHZ} \equiv \frac{1}{\sqrt{2}} (\ket{000} + \ket{111}),
\end{equation*}
the multipartite work deficit can be calculated as $\Delta = \ln 2$. The optimal measurement protocol is to perform the projective measurement on the basis $\ket{0}$ and $\ket{1}$ for all the parties.
For the W state, defined as
\begin{equation*}
\ket{\rm W} \equiv \frac{1}{\sqrt{3}} (\ket{001} + \ket{010} + \ket{100}),
\end{equation*}
the multipartite work deficit becomes $\Delta = \ln 3$, where the optimal measurement protocol is the same as that of the GHZ state.
Therefore, we can say that the W state has larger multipartite quantum correlation than the GHZ state in terms of the work deficit.

\begin{figure}[]
    \centering
    \includegraphics[width=0.5\textwidth]{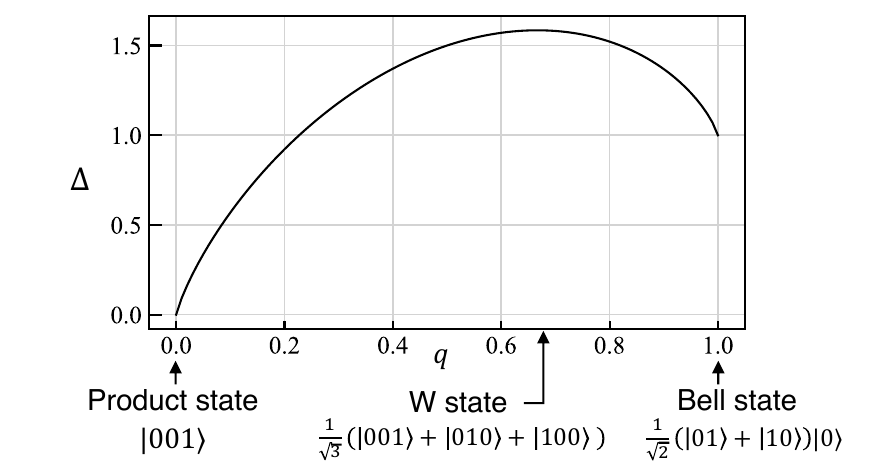}
    \caption{Multipartite work deficit $\Delta$ for the three-qubit state $\ket{\psi_q}$ as a function of the state parameter $q$.}
    \label{fig:3qubit}
\end{figure}

We next perform numerical calculations of the multipartite work deficit for a one-parameter family of three-qubit states,
\begin{equation}
\label{eq:3qubit_state_fam}
    \ket{\psi_q}\equiv \sqrt{1-q} \ket{001} + \sqrt{\frac{q}{2}}\ket{010} + \sqrt{\frac{q}{2}}\ket{100},
\end{equation}
where the parameter $q$ ranges from $0$ to $1$. This family includes the product state $\ket{001}$ at $q=0$, the W state at $q=\frac{2}{3}$, and the Bell state $\frac{1}{\sqrt{2}}(\ket{01}+\ket{10})\ket{0}$ at $q=1$.
As the result of the numerical calculation, it is observed that the optimal protocol for the multipartite work deficit remains the same across all values of $q$, which is the projective measurements on the basis $\{\ket{0},\ket{1}\}$ for all qubits.
Therefore, the multipartite work deficit for $\ket{\psi_q}$ becomes $-(1-q) \ln (1-q) - q \ln \frac{q}{2}$, as depicted in Fig.~\ref{fig:3qubit}.

\subsection{Calculation in multipartite states composed of Bell pairs} \label{ss:Decomp_bi_multi}
\begin{figure*}[]
    \centering
    \includegraphics[width=\textwidth]{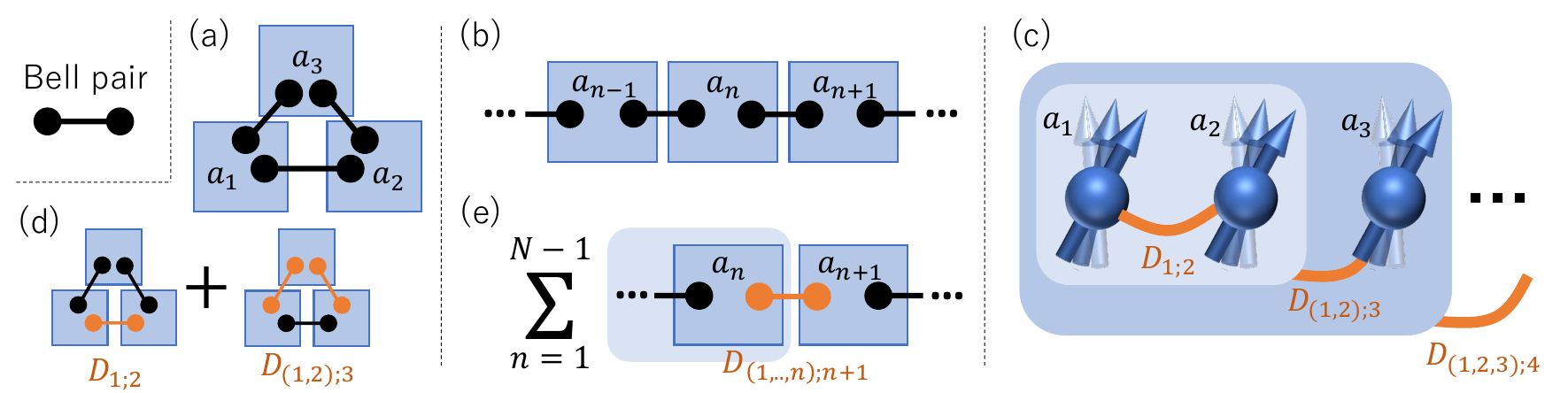}
    \caption{Role of bipartite quantum correlation in the multipartite work deficit. (a) and (b) show schematics for the tripartite state (\ref{eq:tri_bell_share}) and the $N$-partite state (\ref{eq:N_bell_share}), respectively, both composed of Bell pairs. (c) illustrates the accumulation of bipartite quantum correlations in general multipartite states, corresponding to the right-hand side of the inequality (\ref{eq:zWD_biD}). (d) and (e) illustrate that the right-hand side of the inequality (\ref{eq:zWD_biD}) is reduced to the number of Bell pairs in the states depicted in (a) and (b).}
    \label{fig:bipartite_div}
\end{figure*}

In this subsection, we evaluate the multipartite work deficit in the states composed of Bell pairs, and discuss its behavior based on the general relationship between the multipartite work deficit and bipartite quantum correlation. We first consider a tripartite state, illustrated in Fig.~\ref{fig:bipartite_div}(a). It can be described as
\begin{equation}
\label{eq:tri_bell_share}
    \ket{\psi_{1,2,3}} \equiv \frac{1}{\sqrt{2^3}} \sum_{i,j,k = 0,1} \ket{ij}_1\ket{jk}_2\ket{ki}_3,
\end{equation}
where $\{\ket{ij}_n\}_{i,j=0,1}$ denotes the orthonormal basis set of the $n$-th party. The multipartite work deficit of this state can be calculated as $\Delta = 3 \ln2$, where the optimal measurement protocol is to perform local projective measurements with the basis $\{\ket{ij}_n\}_{i,j=0,1}$ for all parties. 

We next address the $N$-partite state, depicted in Fig.~\ref{fig:bipartite_div}(b) and represented as 
\begin{equation}
\label{eq:N_bell_share}
    \begin{split}
        \ket{\psi_{1,\dots,N}} \equiv \frac{1}{\sqrt{2^{N-1}}} &\sum_{i_1,\dots, i_{N-1}= 0,1} \ket{i_1}_1\ket{i_1 i_2}_2\ket{i_2 i_3}_3 \\
     &\dots \ket{i_{N-2} i_{N-1}}_{N-1}\ket{i_{N-1}}_N.
    \end{split}
\end{equation}
For this state, the multipartite work deficit can be calculated as $\Delta = (N-1) \ln2$, where the optimal protocol is to perform a projective measurement with the basis $\{\ket{ij}_n\}_{i,j=0,1}$ for all parties. 

The calculations in states (\ref{eq:tri_bell_share}) and (\ref{eq:N_bell_share}) suggest that the multipartite work deficit for the states composed of Bell pairs is reduced to the number of Bell pairs.
This reduction can be well understood with the following inequality, which holds true for general multipartite states:
\begin{equation}
\label{eq:zWD_biD}
    \Delta \geq  \sum_{n=1}^{N-1} D_{(1,\dots,n);n+1}.
\end{equation}
where $D_{(1,\dots,n);n+1}$ is the quantum discord \cite{ollivier2001quantum,henderson2001classical}, a well-known measure of bipartite quantum correlation.  
Inequality (\ref{eq:zWD_biD}) indicates that the multipartite work deficit is lower bounded by the accumulation of bipartite quantum correlation, as shown in Fig.~\ref{fig:bipartite_div}(c).

The states composed of Bell pairs (\ref{eq:tri_bell_share}) and (\ref{eq:N_bell_share}) are the cases that the equality of Eq.~(\ref{eq:zWD_biD}) is achieved.
This can be shown from the fact that the right-hand side of the inequality (\ref{eq:zWD_biD}) is reduced to the number of Bell pairs for these states, as shown in Figs.~\ref{fig:bipartite_div}(d) and (e), and at the same time, these lower bounds can indeed be achieved with the aforementioned measurement protocols.
We remark that in general multipartite states, the equality of Eq.~(\ref{eq:zWD_biD}) is not satisfied due to the existence of genuinely multipartite quantum correlation. Therefore, in order to calculate work deficits in general many-body states, we need to go beyond these inequalities, which will be explained in the next section.

\section{Efficient calculation in many-body systems} \label{s:calc_many_body}
While the multipartite work deficit can be calculated exactly with a reasonable numerical cost in few-body systems, the calculations in many-body systems are challenging due to the exponential growth of the numerical cost. The reason for such a cost explosion is the necessity of optimizing all the local bases: we need to optimize $N$ sets of measurement bases in order to calculate the work deficit (\ref{eq:zWD_Sh_def}), which requires an exponential cost in the brute-force optimization. 
Such a difficulty in optimizing local bases is not unique to the multipartite work deficit but is a common issue across various measures of multipartite quantum correlations \cite{plenio2001bounds,wei2004connections,barnum2001monotones,wei2003geometric,shimony1995degree,orus2008universal,shi2010finite,orus2010visualizing,mousolou2013unifying,huang2010multipartite,orus2014topological,shi2016geometric,orus2010geometric,rulli2011global,campbell2011global,campbell2013global,power2015nonclassicality,sun2015global,huang2018global}.
While there have been some attempts to overcome this problem \footnote{In some previous works \cite{orus2008universal,shi2010finite,orus2010visualizing,mousolou2013unifying,huang2010multipartite,orus2010geometric,sun2015global,huang2018global}, the numerical costs are reduced by posing assumptions on the optimal local bases. For example, in the translationally invariant systems, the optimal bases are assumed to be also translationally invariant. While such an assumption is powerful in reducing the numerical cost and also seems to be natural, there is no solid theoretical guarantee on it, to the best of our knowledge.}, systematic methods to calculate these measures with polynomial costs are yet to be established. 
In this section, we propose an efficient calculation method for the multipartite work deficit, which achieves significant cost reduction in the systems described with MPS (Table~\ref{tab:num_cost}).
We note that this efficient calculation method is applicable to another measure of multipartite quantum correlation called global quantum discord \cite{rulli2011global}, as detailed in Supplemental Material \cite{SM}.

\subsection{Overview of efficient calculation method} \label{ss:overview_calc}
We first provide an overview of our efficient calculation method of the multipartite work deficit in many-body systems. 
Since the work deficit $\Delta$ increases proportionally with the number of party $N$ for typical $N$-partite states, we focus on the work deficit density, which is defined as
\begin{equation}
    \delta \equiv \frac{\Delta}{N}. \label{eq:zWD_dens_def}
\end{equation}
This quantifies the leading term of the work deficit in many-body systems, and does not diverge even in the thermodynamic limit.
Because the work deficit density $\delta$ is a leading term of $\Delta$ defined with the optimization over the entire system (Eq.~(\ref{eq:zWD_Sh_def})), the numerical cost of directly calculating $\delta$ increases exponentially with the system size.

In our efficient calculation method, instead of directly calculating $\delta$, we evaluate it through the lower and upper bounds as 
\begin{equation}
    L_l \leq \delta \leq U_l.\label{eq:zWD_LUBl}
\end{equation}
Here, these bounds $U_l$ and $L_l$ are derived by dividing the entire system into $l$-site segments, and by performing optimization in these segments.
The tightness of these bounds can be improved by increasing the locality parameter $l$, while the numerical costs also increase with $l$.
Especially for the systems described with MPS, we can derive the following inequality, which is also depicted in Fig.~\ref{fig:calc}(a):
\begin{equation}
\label{eq:WD_calc_presecion}
U_l - L_l \leq \mathcal{O}\left(\frac{1}{l}\right).
\end{equation}
This is one of the main results of this paper, along with Eq.(\ref{eq:zWD_LUBl}).
Since the computational costs for $U_l$ and $L_l$ are $N e^{\mathcal{O}(l)}$ up to constant multiples as discussed later in Sec.~\ref{ss:upp_low_bounds}, Eq.~(\ref{eq:WD_calc_presecion}) implies that the evaluation of $\delta$ with an arbitrary accuracy $\varepsilon$ can be achieved at the cost that scales as $N e^{\mathcal{O}(\frac{1}{\varepsilon})}$, as shown in Table~\ref{tab:num_cost}.

The crucial condition for the efficient calculation of these bounds is that the system is described by an MPS with fixed bond dimension $D_B$.
MPS is a class of states that can represent the ground state of gapped one-dimensional local Hamiltonians \cite{verstraete2006matrix,hastings2007area} and has played a significant role in quantum information and condensed matter physics.
An $N$-site MPS with bond dimension $D_B$ is represented as \cite{perez2006matrix,verstraete2008matrix,cirac2017matrix,cirac2021matrix,orus2014practical}
\begin{equation}
\label{eq:MPS_rep}
    \ket{\psi} \propto \sum_{\sigma_1,\dots,\sigma_N} \tr[A^{\sigma_1}A^{\sigma_2}\dots A^{\sigma_N} X]\ket{\sigma_1,\sigma_2,\dots,\sigma_N},
\end{equation}
where $A^{\sigma_n}$ is $D_B$-dimensional square matrix corresponding to the orthogonal basis $\ket{\sigma_n}$ for $n$-th site, and $X$ is the matrix of the same size which represents the boundary condition (e.g., $X=\mathbbm{1}_{D_B}$ for periodic boundary condition and $X=\bm{v}_R\bm{v}_L^T$ for open boundary condition, with certain $D_B$-dimensional vectors $\bm{v}_R$ and $\bm{v}_L$).
For such states, the numerical costs for calculating the lower and upper bounds are proportional to the system size $N$, which is much smaller than the exponential cost $e^{\mathcal{O}(N)}$ in the direct calculation.
Furthermore, these bounds are calculable even in the thermodynamic limit for translationally invariant normal MPS \cite{cirac2021matrix,cirac2017matrix}, where the direct calculation of $\delta$ is infeasible.
Normal MPS is a class of MPS with finite correlation length and represents the unique ground state of gapped one-dimensional local Hamiltonians \cite{perez2006matrix}.

\subsection{Upper and lower bounds for work deficit density} \label{ss:upp_low_bounds}
\begin{figure*}[]
    \centering
    \includegraphics[width=1\textwidth]{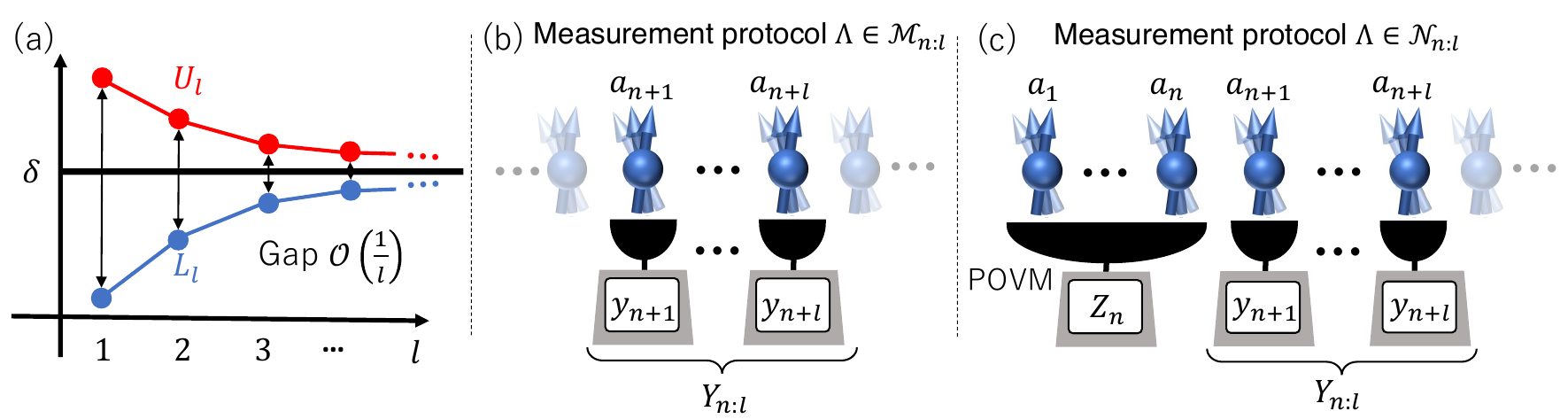}
    \caption{Efficient calculation method for the work deficit density $\delta$ with systematically improvable upper and lower bounds. (a) Schematic of the systematic improvement of the lower and upper bounds, $L_l$ and $U_l$, by increasing the locality parameter $l$. (b) The measurement protocol $\Lambda \in \mathcal{M}_{n:l}$, optimized to define the upper bound $U_l$. (c) The measurement protocol $\Lambda \in \mathcal{N}_{n:l}$, optimized to define the lower bound $L_l$.}
    \label{fig:calc}
\end{figure*}

We now explicitly express the lower and upper bounds for the multipartite work deficit density, which are utilized in its efficient calculation.
The upper bound $U_l$ is defined as
\begin{equation}
\label{eq:U^phi_l}
    U_l \equiv \frac{1}{N}\left\{\sum_{n \in I_l} \min_{\Lambda\in \mathcal{M}_{n:l}} H_{\Lambda}(Y_{n:l}) -S(\rho) \right\},
\end{equation}
where $I_l\equiv \{ml| m\in\mathbbm{Z}, 0 \leq m < \frac{N}{l}\}$ is the set of the party labels. Here, $\mathcal{M}_{n:l}$ represents the set of measurement protocols composed of the local projective measurements on $\{a_i\}_{i=n+1}^{n+l}$, and $Y_{n:l}\equiv (y_{n+1},\dots,y_{n+l})$ represents their outcomes, as shown in Fig.~\ref{fig:calc}(b) \footnote{We note that for $n$ such that $N-l+1 \leq n < N$, the definitions of $\mathcal{M}_{n:l}$ and $Y_{n:l}$ are changed reflecting the existence of the edge in the system: $\mathcal{M}_{n:l}$ is the set of protocols where local projective measurements on $\{a_i\}_{i=n+1}^{N}$ is performed, and $Y_{n:l}\equiv(y_{n+1},\dots,y_N)$ is the corresponding outcomes.}.
By denoting the measurement outcomes from parties $\{a_i\}_{i=1}^{n}$ as $Y_n \equiv (y_1,\dots,y_n)$, this upper bound can be derived as follows:
\begin{equation}
    \begin{split}
        \delta &= \frac{1}{N} \left\{\min_{\Lambda \in \mathcal{M}_{N}} \sum_{n \in I_l} H_{\Lambda}(Y_{n:l}|Y_{n}) -S(\rho)\right\}\\
    &\leq \frac{1}{N} \left\{\min_{\Lambda \in \mathcal{M}_{N}}\sum_{n \in I_l} H_{\Lambda}(Y_{n:l})  -S(\rho)\right\}\\
    &=  \frac{1}{N} \left\{\sum_{n \in I_l} \min_{\Lambda \in \mathcal{M}_{n:l}}  H_{\Lambda}(Y_{n:l}) -S(\rho)\right\}\\
    &= U_l.
    \end{split}
\end{equation}

In the upper bound $U_l$ defined as Eq.~(\ref{eq:U^phi_l}), the global optimization of $\delta$ is divided into the optimization of $l$-site segments. 
The cost for optimizing each term $\min_{\Lambda} H_{\Lambda}(Y_{n:l})$ is $e^{\mathcal{O}(l)}$, due to the necessity to optimize the local measurement bases on $l$ sites. Since $U_l$ is calculated as the sum of $\frac{N}{l}$ such terms, the total computational cost is represented as $N e^{\mathcal{O}(l)}$, capturing the dominant factors of $N$ and $l$.
Furthermore, for translationally invariant normal MPS in the thermodynamic limit, $U_l$ can be calculated with a cost of $e^{\mathcal{O}(l)}$. 
This is because almost all the terms in the summation $\{\min_{\Lambda} H_{\Lambda}(Y_{n:l})\}_{n \in I_l}$ are reduced to the same value due to the translation invariance and finite correlation length, eliminating the need for summation. 
More detailed discussion is provided as Theorem~\blue{3} in Supplemental Material.

We remark that despite the universal applicability of this type of upper bound $U_l$, different types of upper bounds are more effective for some multipartite states. 
In fact, since $\delta$ is defined as the minimization over the measurement protocol, its upper bounds can be derived in various ways, e.g., by utilizing the ansatz protocol or restricting optimization region reasonably. 
For example, by employing arbitrary ansatz measurement protocol $\Lambda^* \in \mathcal{M}_{N}$, we can define the upper bound of $\delta$ as
\begin{equation}
\label{eq:U_ansatz}
    U_{\Lambda^*} \equiv \frac{1}{N} \left\{ H_{\Lambda^*}(Y_N) -S(\rho)\right\},
\end{equation}
where the inequality $\delta \leq U_{\Lambda^*}$ directly follows from the definition of $\delta$.
In the calculation for the AKLT state and the cluster state in Secs.~\ref{sss:AKLT_state} and \ref{sss:cluster_state}, we employ this type of upper bound by choosing appropriate ansatz protocols $\Lambda^*$.
As another type of upper bound, we introduce $U_{l,k}$ with natural numbers $l$ and $k$, which is derived by restricting the optimization region of the measurement protocols.
This bound $U_{l,k}$ is the generalization of the upper bound $U_l$, and is utilized in the numerical calculation in Sec.~\ref{sss:calc_QPT}. The definition and derivation of this type of bound are provided in Appendix~\ref{apps:calc_method}.

The lower bound $L_l$ is defined as
\begin{equation}
\label{eq:L^phi_l}
    L_l \equiv \frac{1}{N} \left\{\sum_{n \in I_l} \min_{\Lambda \in \mathcal{N}_{n:l}} H_\Lambda(Y_{n:l}|Z_{n}) - S(\rho)\right\},
\end{equation}
where $\mathcal{N}_{n:l}$ denotes the set of measurement protocols composed of an arbitrary POVM (positive operator valued measurement) on the parties $\{a_i\}_{i=1}^{n}$ and local projective measurements on $l$ parties $\{a_i\}_{i=n+1}^{n+l}$ \footnote{For $n$ such that $N-l+1\leq n < N$, $\mathcal{N}_{n:l}$ becomes the set of protocols where a POVM on $\{a_i\}_{i=1}^{n}$ and local projective measurements on $\{a_i\}_{i=n+1}^{N}$ are performed.}, and $Z_{n}$ is the outcome of the POVM, as shown in Fig.~\ref{fig:calc}(c). 
By denoting the set of the local projective measurements on $\{a_i\}_{i=1}^{n+l}$ as $\mathcal{M}_{n+l}$, this bound is derived as follows:
\begin{equation}
    \begin{split}
       \delta &= \frac{1}{N} \left\{\min_{\Lambda \in \mathcal{M}_{N}} \sum_{n \in I_l} H_{\Lambda}(Y_{n:l}|Y_{n}) - S(\rho) \right\} \\
    &\geq \frac{1}{N} \left\{ \sum_{n \in I_l} \min_{\Lambda \in \mathcal{M}_{n+l}}  H_{\Lambda}(Y_{n:l}|Y_{n})  - S(\rho) \right\}\\
    &\geq  \frac{1}{N} \left\{\sum_{n \in I_l} \min_{\Lambda \in \mathcal{N}_{n:l}}  H_{\Lambda}(Y_{n:l}|Z_{n}) - S(\rho) \right\}\\
    &= L_l, 
    \end{split}
\end{equation}
where the inequality in the third line follows from the inclusion relationship of the set of measurement protocols $\mathcal{M}_{n+l} \subset \mathcal{N}_{n:l}$.

From Eq.~(\ref{eq:L^phi_l}), we can see that this lower bound $L_l$ is defined as the sum of the local terms $\min_{\Lambda} H_\Lambda(Y_{n:l}|Z_{n})$, which is similar to the upper bound $U_l$. The difference is that in the lower bound $L_l$, we need to optimize the POVM on the $d^{n}$-dimensional subsystem $\{a_i\}_{i=1}^{n}$, whose numerical cost seems to grow exponentially with the subsystem size $n$. However, in MPS with fixed bond dimension $D_B$, we can perform this optimization at a constant numerical cost, since the subsystem $\{a_i\}_{i=1}^{n}$ can be effectively regarded as a fixed-dimensional system ($D_B$-dimensional for open boundary conditions, $D_B^2$-dimensional for periodic boundary conditions). Because of this reduction of subsystem dimension, the numerical cost for $L_l$ becomes $N e^{\mathcal{O}(l)}$ up to constant multiple in $N$-site MPS, and $e^{\mathcal{O}(l)}$ for the translationally invariant normal MPS in the thermodynamic limit.

Finally, we discuss the tightness of these bounds $L_l$ and $U_l$, which is directly related to the accuracy of our calculation method.
In general states, we can show that the tightness of these bounds is improved with increasing the locality parameter $l$, as
\begin{equation}
\label{eq:LUb_improve}
    L_l \leq L_{kl} , \quad U_{kl}\leq U_{l},
\end{equation}
where $k$ is arbitrary natural number.
Furthermore, for systems described with MPS, we can derive Eq.~(\ref{eq:WD_calc_presecion}), which indicates that an arbitrary error bound is achieved with our method by increasing $l$.
Since the numerical costs for calculating the bounds $L_l$ and $U_l$ scale as $e^{\mathcal{O}(l)}$ with the locality parameter $l$, Eq.~(\ref{eq:WD_calc_presecion}) demonstrates that the cost scales with the inverse of the error bound, $\varepsilon$, as $e^{\mathcal{O}({\frac{1}{\varepsilon}})}$ (Table \ref{tab:num_cost}). 

\subsection{Demonstration in examples} \label{ss:demo_in_MPS}
We demonstrate our efficient calculation method of work deficit density introduced in Secs.~\ref{ss:overview_calc} and \ref{ss:upp_low_bounds} in some examples.
Especially, we address translationally invariant MPS in the thermodynamic limit, where the direct calculation is impossible, while our calculation method can be performed with reasonable numerical cost.

\subsubsection{AKLT state} \label{sss:AKLT_state}
We first calculate the work deficit density for the AKLT state \cite{AKLT_PRL,affleck1988valence} under open boundary condition.
The AKLT state is the ground state of the AKLT Hamiltonian
\begin{equation*}
    H_{\rm AKLT} \equiv \sum_{n=1}^N \hat{S}_n \cdot \hat{S}_{n+1} + \frac{1}{3} \left( \hat{S}_n \cdot \hat{S}_{n+1}\right)^2,
\end{equation*}
where $\hat{S}_n$ is the spin-1 operator on $n$-th site. It is represented in an MPS form with bond dimension $D_B =2$ with the following set of matrices:
\begin{equation}
\label{eq:MPS_AKLT}
    A^\uparrow = \begin{pmatrix}
        0 & 0 \\
        -1 & 0 
    \end{pmatrix}, 
    A^0 =\sqrt{\frac{1}{2}}\begin{pmatrix}
        1 & 0 \\
        0 & -1 
    \end{pmatrix},
    A^\downarrow = \begin{pmatrix}
        0 & 1 \\
        0 & 0
    \end{pmatrix}.
\end{equation}

While the direct calculations of $\delta$ is infeasible since we need to optimize infinite sets of bases in the thermodynamic limit, the calculation of the lower bound $L_1$ can be performed with the optimization in $2 \times 3$ dimensional system, due to the fixed bond dimension $D_B=2$. 
In this case, we can evaluate the lower bound through analytical calculation, as
\begin{equation}
\label{eq:AKLT_L1}
    L_1 \geq h\left(\frac{1}{3}\right),
\end{equation}
where $h(p)\equiv -p\ln p -(1-p)\ln(1-p)$ represents binary entropy function.

Furthermore, we can derive the tight upper bound $U_{\Lambda^*}$ by finding the effective ansatz measurement protocol $\Lambda^*$ in this case.
The employed ansatz protocol $\Lambda^*$ is to perform projective measurement with the basis $\{\ket{\uparrow},\ket{0},\ket{\downarrow}\}$ on all sites, where the corresponding upper bound becomes
\begin{equation}
\label{eq:AKLT_U}
    U_{\Lambda^*} =  h\left(\frac{1}{3}\right).
\end{equation}
By combining Eqs.~(\ref{eq:AKLT_L1}) and (\ref{eq:AKLT_U}), we can exactly evaluate the work deficit density as
\begin{equation}
    \delta = h\left(\frac{1}{3}\right).
\end{equation}

\subsubsection{Cluster state} \label{sss:cluster_state}
We next calculate the work deficit density in the cluster state \cite{briegel2001persistent} under open boundary condition. 
The cluster state is the ground state of the cluster Hamiltonian
\begin{equation*}
    H_{\rm cluster} \equiv \sum_{i=1}^N \sigma^z_{n} \sigma^x_{n+1} \sigma^z_{n+2},
\end{equation*}
where $\sigma_n^x$ and $\sigma_n^z$ represent the pauli $x$ and $z$ matrices on $n$-th qubit. It is represented in an MPS form with bond dimension $D_B=2$ with the following set of matrices:
\begin{equation}
\label{eq:MPS_cluster}
    A^0 = \begin{pmatrix}
        0 & 0 \\
        1 & 1
    \end{pmatrix}, 
    A^1 = \begin{pmatrix}
        1 & -1 \\ 
        0 & 0 
    \end{pmatrix}.
\end{equation}

While the direct calculation is not applicable in the thermodynamic limit, we can calculate the lower bounds $L_2$ by performing the optimization over $2\times 2\times 2$ dimensional system. This optimization can be performed analytically in this case, and the lower bound is obtained as
\begin{equation}
    \label{eq:cluster_L2}
    L_2 \geq \frac{1}{2}\ln 2.
\end{equation}
Since the lower bound for $l=1$ is calculated as $L_1=0$ only providing the trivial bound, we can see that the bound is significantly improved by increasing locality parameter $l$.

We can also derive the upper bound $U_{\Lambda^*}$ with the ansatz protocol $\Lambda^*$ where the projective measurement on $\{\ket{0},\ket{1}\}$ and $\{\ket{+},\ket{-}\}$ are alternately performed on each site. Here, basis $\ket{\pm}$ are defined as $\ket{\pm} \equiv \frac{1}{\sqrt{2}}(\ket{0}\pm\ket{1})$. This upper bound is calculated as 
\begin{equation}
    \label{eq:cluster_U}
    U_{\Lambda^*} = \frac{1}{2}\ln 2.
\end{equation}
By combining Eqs.~(\ref{eq:cluster_L2}) and (\ref{eq:cluster_U}), we can exactly evaluate the work deficit density in this case as follows:
\begin{equation}
\label{eq:WDD_cluster}
    \delta = \frac{1}{2}\ln 2.
\end{equation}

\subsubsection{MPS family under quantum phase transition} \label{sss:calc_QPT}
\begin{figure}[]
    \centering
    \includegraphics[width=0.5\textwidth]{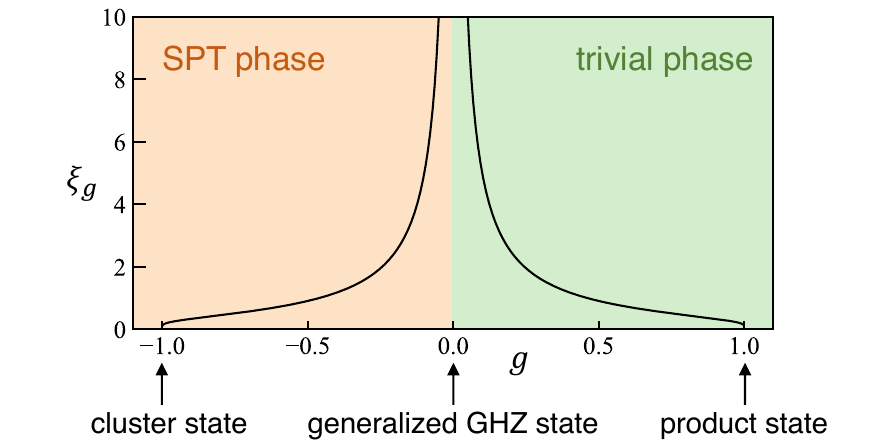}
    \caption{Correlation length as a function of the system parameter $g$ for the MPS family (Eq.~(\ref{eq:MPSfam})). The correlation length diverges at the quantum phase transition point $g=0$, indicating a transition from the SPT phase in $g<0$ to the trivial phase in $g>0$.}
    \label{fig:MPSfam_prev}
\end{figure}

We next numerically calculate the work deficit density in an MPS family introduced in Ref. \cite{wolf2006quantum}, by utlizing our calculation method. 
The MPS family is represented with the following set of matrices with one parameter $g$:
\begin{equation}
\label{eq:MPSfam}
    A^0 =\begin{pmatrix}
        0 & 0 \\
        1 & 1 
    \end{pmatrix}, 
    A^1 =\begin{pmatrix}
        1 & g \\
        0 & 0 
    \end{pmatrix}.
\end{equation}
We consider the region $-1\leq g \leq 1$, and system under preodic boundary condition.
These MPS are the ground states of the parametrized Hamiltonian
\begin{equation}
\label{eq:MPSfam_Hami}
    H_g \equiv \sum_{n} 2(g^2-1)\sigma^z_{n}\sigma^z_{n+1} - (g+1)^2 \sigma^x_n + (g-1)^2 \sigma_n^z \sigma_{n+1}^x \sigma_{n+2}^z,
\end{equation} 
where $\sigma_n^x$ and $\sigma_n^z$ are the pauli $x$ and $z$ matrices on $n$-th site, respectively. 
As the special cases of this MPS family, the cluster state (\ref{eq:MPS_cluster}) is realized at $g=-1$, the generalized GHZ state $\frac{1}{\sqrt{2}}(\ket{00...0}+\ket{11...1})$ is realized at $g=0$, and the product state $\ket{++...+}$ is realized at $g=1$.

This MPS family is widely studied as a model that exhibits a quantum phase transition while maintaining a fixed bond dimension \( D_B = 2 \) \cite{smith2022crossing,jones2021skeleton,tantivasadakarn2023pivot,wei2023efficient,smith2024constant}. The phase transition occurs at \( g = 0 \), transitioning from a symmetry-protected topological phase (SPT phase) when \( g < 0 \) to a trivial phase when \( g > 0 \). The correlation length \( \xi_g \) of this MPS family diverges at the transition point as follows, depicted in Fig.~\ref{fig:MPSfam_prev}:
\begin{equation}
\label{eq:MPS_corr}
    \xi_g = \left\{ 
    \begin{alignedat}{2}
        \left(\ln \left|\frac{1-g}{1+g} \right|\right)^{-1} \ \ \ (g < 0) \\
        \left(\ln \left|\frac{1+g}{1-g}\right|\right)^{-1} \ \ \ (g > 0)
    \end{alignedat}
    \right.
    .
\end{equation}
We note that at \( g = 0 \), the system resides in a generalized GHZ state characterized by long-range, non-decaying correlation. This feature is distinct from conformal critical points, which are known for power-law decaying correlations. Previous studies have identified the point \( g = 0 \) in this MPS family as a multicritical point \cite{smith2022crossing,jones2021skeleton,tantivasadakarn2023pivot}.

\begin{figure}[]
    \centering
    \includegraphics[width=0.5\textwidth]{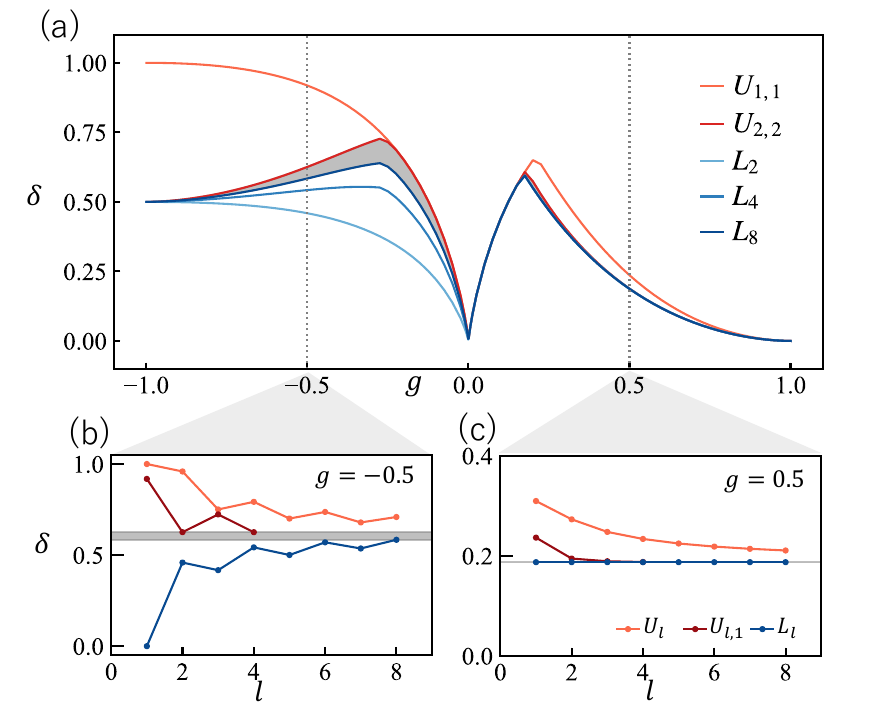}
    \caption{Numerical calculation of the work deficit density in the MPS family (Eq.~(\ref{eq:MPSfam})). (a) Work deficit density as a function of the system parameter $g$. The gray region, sandwiched by the lower and upper bounds, represents the range for the work deficit density $\delta$. The step size for $g$ is $0.025$, except near $g=0$, where the data points are more densely sampled at $g=\pm 0.025 \times 0.5^{n}$ for $n$ ranging from $1$ to $5$. (b,c) Lower and upper bounds of the work deficit density as a function of the locality parameter $l$ at fixed system parameters $g = -0.5$ and $0.5$, respectively. Systematic improvement of the bounds with increasing $l$ is observed. The data points for  $L_l$, $U_l$, and $U_{l,k}$ are taken as the best value among results from numerical optimization for random initializations.}
    \label{fig:MPSfam_WDcalc}
\end{figure}

The result of our numerical calculation is shown in Fig.~\ref{fig:MPSfam_WDcalc}.
In Fig.~\ref{fig:MPSfam_WDcalc}(a), the behavior of the work deficit density as a function of the system parameter $g$ is depicted, while in Figs.~\ref{fig:MPSfam_WDcalc}(b) and (c) the lower and upper bounds for fixed $g$ as a function of the locality parameter $l$ are displayed.
All of these figures illustrate that the lower and upper bounds are improved systematically with increasing locality parameter $l$, showcasing the validity of our calculation method.

From Fig.~\ref{fig:MPSfam_WDcalc}(a), we can observe the bimodal structure of $\delta$, with its peaks in the regions $-1 \leq g \leq 0$ and $0 \leq g \leq 1$, respectively.
The reasons for this bimodal structure can be attributed to two factors: long-range correlation around $g=0$ and the short-range correlation around $g=\pm 1$. 
At $g=1$, work deficit density $\delta$ becomes zero, since the system is in a product state and does not have inter-site correlation at all.
On the other hand, at $g=0$, $\delta$ drop to zero due to the non-decaying correlation. 
The generalized GHZ state realized at $g=0$, is the superposition of the completely aligned states $\ket{00...0}$ and $\ket{11...1}$, and therefore the multipartite work deficit remains the same value $\Delta = \ln 2$ independent of the system size $N$. In the thermodynamic limit $N\to \infty$, 
this leads to $\delta \equiv \frac{\Delta}{N} \to 0$.
Consequently, the peak of $\delta$ occurs between these two extremes, where the system has the intermediate-scale correlation length. The peak in the region $-1 \leq g \leq 0$ occurs in a similar mechanism.

We further discuss the kinks observed in Fig.~\ref{fig:MPSfam_WDcalc}(a). In this figure, the unsmooth change in $\delta$ is observed at the phase transition point $g=0$, reflecting the drastic change of the quantum correlation structure. 
Around this point $g=0$, the optimal measurement protocol $\Lambda^*$ is kept unchanged; it is to perform the measurement on the basis $\{\ket{0},\ket{1}\}$ for all sites.
On the other hand, the kink observed around $g = 0.175$ in Fig.~\ref{fig:MPSfam_WDcalc}(a) is due to a discontinuous change in the optimal measurement protocol at this point; in the region $g < 0.175$, the optimal protocol is to perform projective measurements on the basis $\{\ket{0},\ket{1}\}$ for all sites, while for $g \geq 0.175$, the optimal basis suddenly switches to $\{\ket{+},\ket{-}\}$, where $\ket{\pm} \equiv \frac{1}{\sqrt{2}}(\ket{0} \pm \ket{1})$.
Additionally, the kink may occur for the same reason in the region $\-1 \leq g\leq 0$, although it is not clearly observable in the current numerical calculation due to the gap in the lower and upper bounds. Such kinks, resulting from changes in the optimal measurement protocols, do not necessarily indicate a drastic change in the quantum correlation structure at those points.
To distinguish between such kinks and the unsmoothness due to the phase transition at $g=0$, we further introduce the framework of coarse-graining in Sec.~\ref{s:coarse_grain}, demonstrating that the kink around $g=0$ becomes sharper while the kinks at the other points shift or get dull under the coarse-graining.

\section{Coarse-graining}\label{s:coarse_grain}
\begin{figure}[]
    \centering
    \includegraphics[width=0.5\textwidth]{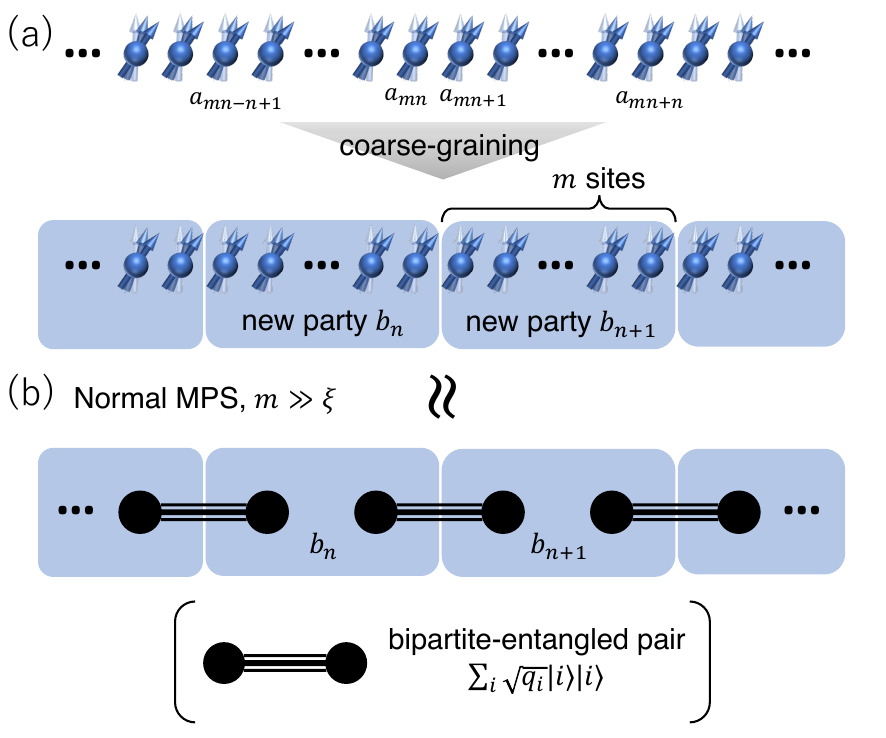}
    \caption{(a) Schematic for the coarse-graining. The neighbouring $m$ sites $\{a_{mn+1},a_{mn+2},\dots,a_{(m+1)n-1}\}$ are merged and regarded as a single party $b_{n+1}$. (b) Convergence of normal MPS to bipartite-entangled form under coarse-graining. By taking the party size $m$ much larger than the correlation length $\xi$, normal MPS converges to the form where bipartite-entangled pair $\sum_i \sqrt{q_i}\ket{i}\ket{i}$ is shared between the neighbouring parties.}
    \label{fig:coarse}
\end{figure}

In the foregoing sections, we have focused on the situation where each site is regarded as a single party.
In this section, we calculate the work deficit density defined in a coarse-grained picture.
Specifically, we consider the situation where $m$ sites $\{a_{mn+1},a_{mn+2},\dots,a_{(m+1)n-1}\}$ are merged and regarded as a single party $b_{n+1}$, as shown in Fig.~\ref{fig:coarse}(a).
To represent the work deficit density defined under such a partitioning, we introduce the notation $\delta_{m}$, where the subscript $m$ denotes the number of sites per a single party. In the case of $m=1$, the subscript may be omitted, as has been done so far.

In the framework of coarse-graining, we can get access to the structure of longer-range multipartite quantum correlation. 
For example, the work deficit density $\delta_m$, defined under the partitioning of Fig.~\ref{fig:coarse}(a), does not account for the short-range quantum correlation among $m$ sites $\{a_{mn+1},\dots,a_{(m+1)n-1}\}$, while it captures the structure of longer-range quantum correlation well. 
This is contrary to $\delta_1$, which accounts for the effects of local quantum correlations across all sites.
In this section, we analyze the long-range structure using the framework of coarse-graining, demonstrating that the quantum phase transition in the MPS family (\ref{eq:MPSfam}) can be effectively probed with this framework.

\subsection{Work deficit density for normal MPS under coarse-graining} \label{ss:coarse_EE}
In this subsection, we show that the work deficit density $\delta_m$ converges to half of the entanglement entropy by increasing number of sites per party $m$.
Especially, we focus on the translationally invariant normal MPS \cite{cirac2017matrix,cirac2021matrix} under periodic boundary condition.
We consider an $N$-site system divided into $m$-site parties, where the number of parties is denoted as $M\equiv \frac{N}{m}$.
In this setup, we can derive the following equality, which is the main result of this section: 
\begin{equation}
\label{eq:EE_coarse}
    \delta_{m} = \frac{S_{\text{EE}}}{2} + \Tilde{\mathcal{O}}(e^{-\frac{m}{2\xi}}),
\end{equation}
where \(\xi\) is the correlation length of this MPS, $S_{\text{EE}}$ is the entanglement entropy for subsystem much larger than \(\xi\), and the notation \(\Tilde{\mathcal{O}}(x)\) represents \(\mathcal{O}(x\ln x)\).
In the following, we provide an intuitive discussion of why Eq.~(\ref{eq:EE_coarse}) is satisfied, while the rigorous proof is presented in Supplemental Material as Theorem~\blue{4}.

The essence of the derivation of Eq.~(\ref{eq:EE_coarse}) is the fact that under coarse-graining, a normal MPS converges to the state where the bipartite-entangled pairs are shared between the neighbouring parties, as shown in Fig.~\ref{fig:coarse}(b).
This convergence is the consequence of the renormalization group transformation, as discussed e.g., in Refs.~\cite{cirac2017matrix,cirac2021matrix,piroli2021quantum,malz2024preparation}.
The state in Fig.~\ref{fig:coarse}(b) is represented as
\begin{equation}
\label{eq:coarse_approx_state}
\ket{\psi} \simeq \sum_{i_1,\dots,i_{M}} \sqrt{q_{i_1}q_{i_2}\dots q_{i_{M}}} \ket{i_1 i_2}_{1}\ket{i_2i_3}_{2} \dots \ket{i_M i_1}_{M},
\end{equation}
where \(\{\ket{jk}_{n}\}_{j,k=1}^{D_B}\) denotes the orthonormal basis of $n$-th party \(b_n\). 
In this representation, each $m$-site party is effectively regarded as a $D_B^2$-dimensional system due to the fixed bond dimension of $\ket{\psi}$.
Since Eq.~(\ref{eq:coarse_approx_state}) is a generalized version of the Bell-pair-shared state in Fig.~\ref{fig:bipartite_div}(b), where the Bell pair is replaced with the bipartite-entangled pair \(\sum_{i} \sqrt{q_{i}} \ket{i}_{n}\ket{i}_{n+1}\), the work deficit density for this state is reduced to the bipartite contribution as follows:
\begin{equation}
\label{eq:coarse_H}
    \delta_m  \simeq \frac{1}{M}\sum_{n=1}^{M}H(\{q_{i}\}) = H(\{q_{i}\}).
\end{equation}
Here, the optimal measurement protocol for $\delta_m$ is to perform the projective measurement with basis \(\{\ket{jk}_{n}\}_{j,k}\) for all parties. 
Since $S_{\text{EE}} = 2 H(\{q_{i}\})$ as shown in Fig.~\ref{fig:coarse}(b), we can derive Eq.~(\ref{eq:EE_coarse}). The error term $\Tilde{\mathcal{O}}(e^{-\frac{m}{2\xi}})$ in Eq.~(\ref{eq:EE_coarse}) is the consequence of the error in the approximation in Eq.~(\ref{eq:coarse_approx_state}).

\subsection{Detection of quantum phase transition with work deficit density} \label{ss:coarse_QPT}
\begin{figure*}[]
    \centering
    \includegraphics[width=\textwidth]{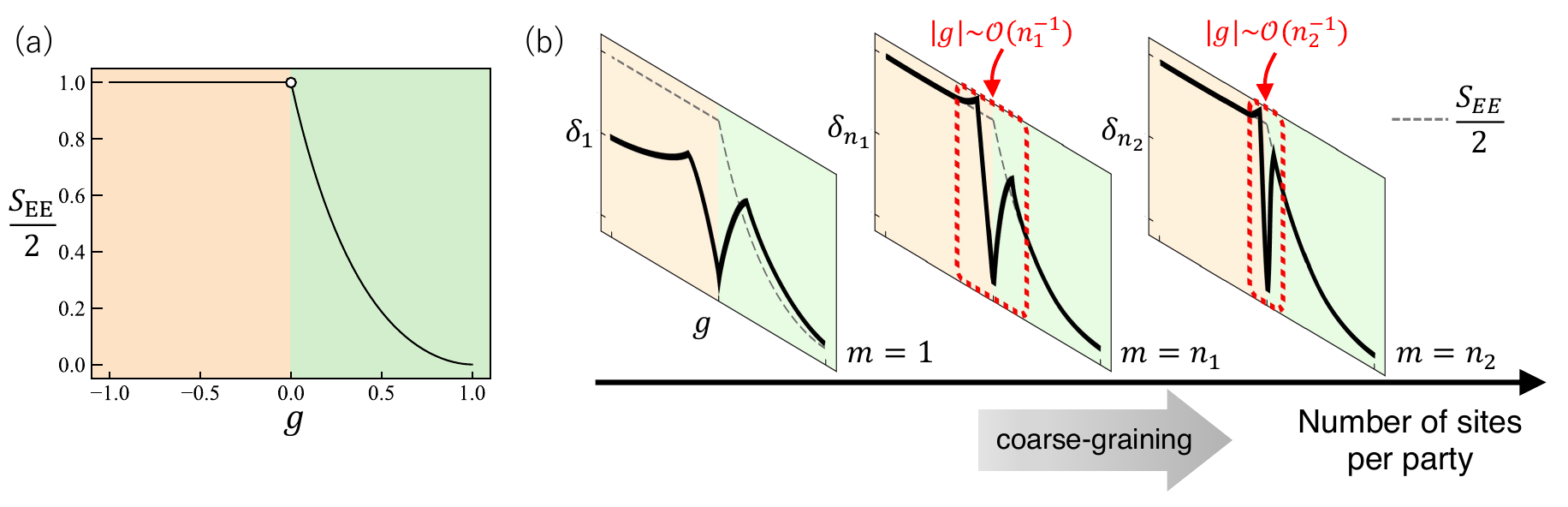}
    \caption{(a) Entanglement entropy for the MPS family (Eq.~(\ref{eq:MPSfam})). Here, $S_{\rm EE}$ is defined as the entanglement entropy of a subsystem whose size is much larger than the correlation length $\xi_g$, for each parameter $g$. Therefore, $S_{\rm EE}$ is ill-defined at $g=0$ due to the divergence of the correlation length. (b) Rough sketch of the behavior of work deficit density $\delta_m$ in the MPS family under coarse-graining. In the region $|g|\gg\frac{1}{m}$, $\delta_m$ converges to half of the entanglement entropy depicted in (a), while in the region $|g|\ll\frac{1}{m}$, $\delta_m$ is close to zero. Based on these analyses, we can conclude that $\delta_m$ undergoes drastic changes in the parameter region $|g| \sim \mathcal{O}(m^{-1})$. It is important to note that this figure is a rough sketch derived from analytical calculations in the regions $|g|\ll\frac{1}{m}$ and $|g|\gg\frac{1}{m}$, and is not based on the calculations in the entire parameter range $-1 \leq g \leq 1$.}
    \label{fig:MPSfam_coarse}
\end{figure*}

We apply the framework of coarse-graining to the MPS family defined in Eq.~(\ref{eq:MPSfam}), to reveal its long-range quantum correlation structure. We consider the system under periodic boundary condition in the thermodynamic limit, in the same way as Sec.~\ref{sss:calc_QPT}.
Instead of conducting numerical calculations across the entire parameter range $-1\leq g \leq 1$ in a coarse-grained framework, we here perform the analysis solely within the regions $|g| \gg \frac{1}{m}$ and $|g| \ll \frac{1}{m}$ to grasp a rough sketch for the behavior of the work deficit density.
Given that the correlation length (\ref{eq:MPS_corr}) is approximately $\xi_g \simeq \frac{1}{2|g|}$ for $|g| \ll 1$, the regions $|g| \gg \frac{1}{m}$ and $|g| \ll \frac{1}{m}$ correspond to the situations of $\xi_g\ll m$ and $\xi_g \gg m$, respectively. 

In the region $|g| \gg \frac{1}{m}$, we can utilize Eq.~(\ref{eq:EE_coarse}), since $\xi_g \ll m$ is satisfied.
Therefore, the work deficit density converges to half of the entanglement entropy of subsystem much larger than the correlation length $\xi_g$, which can be calculated as follows:
\begin{equation}
\label{eq:MPSfam_EE}
    \frac{S_{\text{EE}}}{2} = \left\{
    \begin{alignedat}{2}
        & \ln 2 \ \  &(g < 0) \\
        & h\left( \frac{1}{2}-\frac{\sqrt{g}}{1+g} \right) \ \ &(g > 0) 
    \end{alignedat}
    \right.
    ,
\end{equation}
where $h(p)\equiv -p\ln p - (1-p) \ln (1-p)$ denotes the binary entropy function. It is depicted in Fig.~\ref{fig:MPSfam_coarse}(a).

On the other hand, in the region $|g| \ll \frac{1}{m}$, work deficit density deviates from the entanglement entropy since the party size $m$ is smaller than the correlation length $\xi_g$.
Therefore, in this region, we take a different strategy to evaluate $\delta_m$. To be specific, we can derive the heuristic upper bound 
\begin{equation}
\label{eq:MPSfam_UB}
    \delta_m \leq \Tilde{\mathcal{O}}(m|g|),
\end{equation}
by utilizing a certain ansatz measurement protocol \cite{SM}.
From this inequality, we can see that $\delta_m$ is close to zero in the region $|g| \ll \frac{1}{m}$.
Such a behavior of work deficit density can be understood as the consequence of the long-range correlation around $g= 0$.
For example, in the generalized GHZ state at $g=0$, the total work deficit remains the same value $\ln 2$ independent of the number of parties $M$, resulting in $\delta_m =0$ in the thermodynamic limit $M \to \infty$.

Figure \ref{fig:MPSfam_coarse}(b) shows the rough sketch of the behavior of work deficit density, reflecting these analytical calculations. 
This figure illustrates that $\delta_m$ converges to half of the entanglement entropy in the region $|g| \gg \frac{1}{m}$, while it rapidly approaches to zero in the region $|g| \ll \frac{1}{m}$. Therefore, by changing the size of the party $m$, the unsmoothness of the work deficit density around $g=0$ becomes sharper, showcasing the drastic change in the long-range structure of multipartite quantum correlation.
On the other hand, the other kinks observed in Fig.~\ref{fig:MPSfam_WDcalc}(c), such as the one at $g= 0.175$, shift or become dull under sufficient coarse-graining, since $\delta_m$ converges to $\frac{S_{\rm EE}}{2}$, which is smooth at these points as depicted in Fig.~\ref{fig:MPSfam_coarse}(a).
This illuminates that these kinks are regarded as artifacts of optimization.
We remark that Fig.~\ref{fig:MPSfam_coarse}(b) roughly predicts the behavior of the work deficit density based on the analysis in the regions $|g| \gg \frac{1}{m}$ and $|g| \ll \frac{1}{m}$. The behaviors in other parameter regions are merely interpolations from these extremes.

\section{Multipartite one-way work deficit} \label{s:adaptive_WD}
In the foregoing sections, we focused on the multipartite work deficit in non-adaptive situation for simplicity. 
In this section, we discuss the work deficit with adaptive protocols. Specifically, we introduce the multipartite one-way work deficit, defined with the LOCC class called one-way closed LOCC, where fixed one-way classical communication and local adaptive measurements are allowed. Due to the restriction in communication direction, we can derive its expression exactly calculable in general few-body systems. Furthermore, we show that the efficient calculation method introduced in Sec.~\ref{s:calc_many_body} is applicable also to this measure, thus significantly reducing numerical cost, while there remains an open question regarding the accuracy of this method.

\subsection{Definition of multipartite one-way work deficit} \label{ss:def_oWD}
\begin{figure}[]
    \centering
    \includegraphics[width=0.5\textwidth]{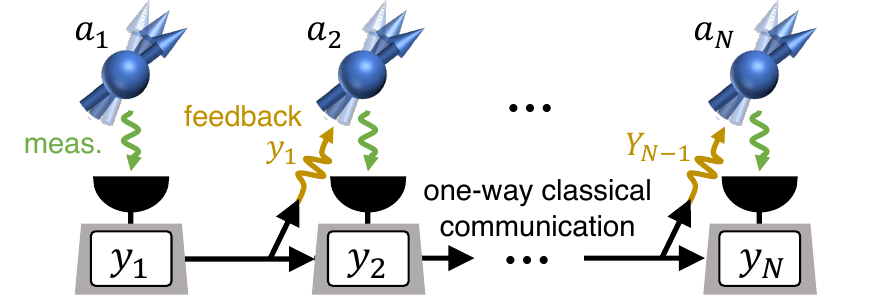}
    \caption{Schematic of one-way closed LOCC. The direction of classical communication is fixed as $a_1 \to a_2 \to \dots \to a_N$. At each party $a_n$, adaptive local projective measurement is allowed, which can depend on the communicated outcomes denoted as $Y_{n-1} \equiv (y_1, y_2, \dots, y_{n-1})$.}
    \label{fig:definition_oWD}
\end{figure}

We first introduce the multipartite one-way work deficit, which is defined as
\begin{equation}
\label{eq:Delta^to}
    \Delta^\to \equiv W_g - W_l^\to,
\end{equation}
where $W_g$ and $W_l^\to$ are extractable works through global operations and one-way closed LOCC, respectively. Since $W_g$ is represented as
\begin{math}
    W_g = N \ln d - S(\rho),
\end{math}
as discussed in Eq.~(\ref{eq:m_global_ext_work}), we here evaluate $W_l^\to$ to derive the expression for $\Delta^\to$.

In one-way closed LOCC, adaptive local operations dependent on the communicated measurement outcomes are allowed, which significantly differs from the zero-way case. Particularly, we consider the situation where the direction of allowed one-way communication is fixed as $a_1 \to a_2 \to \dots \to a_N$, as shown in Fig.~\ref{fig:definition_oWD}.
In this situation, the work extractable through a certain protocol $\Lambda$ can be represented as
\begin{equation}
\label{eq:work_lambda_ow}
    W_{\Lambda} = \sum_{n=1}^N \left\{ \ln d - \sum_{Y_{n}} P_\Lambda [Y_n] S(\widetilde{\rho}_n^{\Lambda,Y_n}) \right\} - H_{\Lambda}(Y_N),
\end{equation}
where $\widetilde{\rho}_n^{\Lambda,Y_n}$ denotes the conditional density operator of the $n$-th party for the measurement outcomes $Y_n$, and $P_\Lambda [Y_n]$ is the corresponding measurement probability. Similarly to Eq.~(\ref{eq:work_lambda}), the terms in the curly brackets represent the extractable work from each party, and the Shannon entropy term $H_{\Lambda}(Y_N)$ represents the work required for the memory erasure.

The extractable work through one-way closed LOCC $W_l^\to$ can be quantified by optimizing $W_\Lambda$ defined in Eq.~(\ref{eq:work_lambda_ow}) over the allowed set of measurement protocols. By performing this optimization, we can derive the representation
\begin{equation}
\label{eq:Wl_one-way}
    W_l^\to = N \ln d - \min_{\Lambda\in \mathcal{M}_N^\to} H_{\Lambda}(Y_N).
\end{equation}
Here, $\mathcal{M}_N^\to$ denotes the set of measurement protocols whose operators are denoted as $\{\Pi_{y_1}\otimes\Pi_{y_2}^{y_1}\otimes \dots \Pi_{y_N}^{Y_{N-1}}\}_{Y_N}$, where $\{\Pi_{y_n}^{Y_{n-1}}\}_{y_n}$ is the set of the rank-one orthogonal projectors on the $n$-th party depending on the measurement outcomes $Y_{n-1}$. The derivation of $W_l^\to$ in Eq.~(\ref{eq:Wl_one-way}) is essentially the same as that of $W_l$ in Eq.~(\ref{eq:Wl_zero-way}), which is explained in detail in Appendix~\ref{apps:expressions}.

By combining Eqs.~(\ref{eq:Delta^to}) and (\ref{eq:Wl_one-way}), we can derive an expression for the one-way work deficit density as
\begin{equation}
\label{eq:oWD_Sh_def}
    \Delta^\to = \min_{\Lambda \in \mathcal{M}_N^\to} H_{\Lambda}(Y_N) - S(\rho).
\end{equation}
This expression shows that the multipartite one-way work deficit is calculated through the optimization of the local measurement bases $\{\Pi_{y_1}\}, \{\Pi_{y_2}^{y_1}\}, \dots \{\Pi_{y_N}^{Y_{N-1}}\}$, which can be performed exactly for general few-body systems.
One can also show that Eq.~(\ref{eq:oWD_Sh_def}) is rewritten using the quantum relative entropy as
\begin{equation}
\label{eq:oWD_one_way_c}
    \Delta^\to = \min_{\sigma \in \mathcal{C}^\to_N} S(\rho\|\sigma),
\end{equation}
where $\mathcal{C}^\to_N$ denotes the set of one-way classically correlated states described as
\begin{equation}
    \sigma \equiv \sum_{Y_N} q(Y_N) \Pi_{y_1} \otimes \Pi_{y_2}^{y_1} \dots \otimes \Pi_{y_N}^{Y_{N-1}}.
\end{equation}
Since the set of one-way classically correlated states $\mathcal{C}^\to_N$ includes the set of classically correlated states $\mathcal{C}_N$, we can show the following inequality:
\begin{equation}
\label{eq:ozWD_size}
    \Delta \geq \Delta^\to.
\end{equation}

\subsection{Efficient calculation method for multipartite one-way work deficit} \label{ss:calc_oWD}

We show that the efficient calculation method introduced in Sec.~\ref{s:calc_many_body} is also applicable to the one-way work deficit, although there remains an open question in the error bound. 
As the leading term of $\Delta^\to$ in many-body systems, we define the one-way work deficit density as 
\begin{equation}
    \delta^\to \equiv \frac{\Delta^\to}{N},
\end{equation}
where $N$ denotes the number of parties. The numerical cost for directly calculating $\delta^\to$ explodes in $e^{e^{\mathcal{O}(N)}}$, since we need to optimize
all the measurement bases $\{\Pi_{y_1}\}_{y_1}, \{\Pi_{y_2}^{y_1}\}_{y_1,y_2}, \dots \{\Pi_{y_N}^{Y_{N-1}}\}_{Y_N}$, as shown Eq.~(\ref{eq:oWD_Sh_def}).
In the efficient calculation method, we evaluate $\delta^\to$ by utilizing efficiently computable lower and upper bounds:
\begin{equation}
    L_l^\to \leq \delta^\to \leq U_l^\to.
\end{equation}
For systems described with MPS, the numerical costs for these bounds $L_l^\to$ and $ U_l^\to$ are proportional to the system size. Moreover, for translationally invariant normal MPS in the thermodynamic limit, these bounds can be calculated with constant numerical cost. 

The validity of these bounds can be demonstrated through calculations in some examples. For instance, in the AKLT state and the cluster state in the thermodynamic limit, we can exactly calculate the one-way work deficit density using these bounds as follows \cite{SM}:
\begin{alignat}{2}
    \delta^\to &= h\left(\frac{1}{3}\right) \quad &(\text{for AKLT state}), \\
    \delta^\to &= \frac{1}{2}\ln 2 \quad &(\text{for cluster state}).
\end{alignat}

However, it remains an open question how the tightness of the bounds $L_l^\to$ and $U_l^\to$ improves with increasing the locality parameter $l$ in general MPS. 
This is in contrary to the zero-way case, where the systematic improvement of the error bound is guaranteed by Eq.~(\ref{eq:WD_calc_presecion}).
To establish such a guarantee in one-way case remains an open question.

\section{Summary and discussion} \label{s:sum_outlook}
In this work, we introduced the multipartite work deficit, a measure that quantifies multipartite quantum correlation based on work extraction. It is defined as the difference in extractable works through global operations and non-adaptive LOCC. We obtained its expression (\ref{eq:zWD_Sh_def}) that is exactly calculable for general few-body states and computed its values in simple examples.

Importantly, we introduced an efficient calculation method for this measure in many-body systems described with MPS, in Secs.~\ref{ss:overview_calc} and \ref{ss:upp_low_bounds}. While the numerical cost for directly calculating the multipartite work deficit grows exponentially with system size $N$, the cost for our efficient method is linear with respect to $N$, enabling significant cost reduction (Table~\ref{tab:num_cost}). This method is applicable not only to the multipartite work deficit but also to another measure called global quantum discord \cite{rulli2011global}, which highlights its potential applicability to various measures. In Sec.~\ref{ss:demo_in_MPS}, we demonstrated this method in several MPS examples, including the AKLT state, the cluster state, and an MPS family~\cite{wolf2006quantum}. 
We also introduced the framework of coarse-graining of the multipartite work deficit in Sec.~\ref{s:coarse_grain}. 
By utilizing this framework in the MPS family, we observed the drastic change of multipartite quantum correlaiton structure at the quantum phase transition point.

A remaining issue of this work is to investigate whether our efficient calculation method for MPS can be applied to broader classes of many-body states. 
Such extension may be straightforward for the systems described with projected-entangled pair states (PEPS), matrix product operators (MPO), and projected entangled pair operators (PEPO), which are the generalizations of MPS to higher-dimensional and/or mixed states \cite{cirac2021matrix,cirac2017matrix}. 
In fact, these states can be represented with fixed bond dimensions independent of the system size, which is essential for our calculation method. On the other hand, extension to states that exhibit volume-law entanglement entropy is much more nontrivial, since the bond dimension for these states grows exponentially with the system size. It is an important open question to develop novel approaches for evaluating multipartite work deficit in such states.

Another future task is to study the multipartite work deficit defined with fully adaptive LOCC. 
In this work, we introduced the work deficits defined with the LOCC classes called zero-way and one-way closed LOCC, where the set of allowed operations are reasonably restricted. This restriction is crucial for deriving their closed formula which can be calculated exactly for general few-body states, while maintaining clear operational meanings based on work extraction.
On the other hand, the investigation of the measure defined with fully adaptive LOCC, in which the adaptive local operations and arbitrary classical communication are allowed, poses a greater challenge due to the extensive range of allowable protocols. 
Since studies on such a case have so far been limited to demonstrations in a few examples \cite{oppenheim2002thermodynamical,horodecki2005local}, it remains a future task to establish calculation methods for general few-body and many-body systems.

It is also important to explore how the multipartite work deficit reveals the structures of quantum correlations in quantum many-body systems. 
While the quantum correlations of many-body systems at zero and finite temperature have been extensively studied in the previous works \cite{alicki2010thermal,kuwahara2022exponential,wolf2008area,kuwahara2021improved,anshu2020entanglement,cirac2011entanglement,schuch2013topological,kato2019locality,anshu2022entanglement,kuwahara2020clustering,kato2019quantum}, they primarily focused on bipartite or tripartite measures, leaving the structure of multipartite quantum correlation largely unexplored. 
We expect that the multipartite work deficit introduced in this work would serve as a probe for elucidating such multipartite structures.
Particularly, its convergence to entanglement entropy in normal MPS, as discussed in Sec.~\ref{ss:coarse_EE}, represents a significant step forward. 
Clarifying the conditions under which this convergence to bipartite entanglement occurs poses an open question.

Finally, we discuss the relevance of our work to experiments.
Recently, single-site measurement and manipulation of quantum many-body systems \cite{bakr2009quantum, weitenberg2011single}, as well as thermodynamic work extraction from quantum systems \cite{cottet2017observing, masuyama2018information, naghiloo2018information}, are becoming possible on real experimental platforms, such as ultracold atoms and superconducting qubits.
The multipartite work deficit has fully operational meaning, as it is defined based on the work extraction through site-wise operations, which would be feasible with these current or near-future quantum technologies. 
Therefore, the multipartite work deficit is expected to serve as an experimental probe of multipartite structure of quantum correlations in many-body quantum systems.
It would also be an interesting perspective to investigate the role of quantum correlations in thermodynamics in such experimental platforms.

\begin{acknowledgments}
We thank Tsuyoshi Okubo and Yohei Fuji for fruitful discussions.
T.Y. is supported by World-leading Innovative Graduate Study Program for Materials Research, Information, and Technology (MERIT-WINGS) of the University of Tokyo. T.Y. is also supported by JSPS KAKENHI Grant No. JP23KJ0672.
N.Y. wishes to thank JST PRESTO No. JPMJPR2119, JST Grant Number JPMJPF2221, JST CREST Grant Number JPMJCR23I4, IBM Quantum, and JST ERATO Grant Number JPMJER2302, Japan.
T.S. is supported by JST ERATO-FS Grant Number JPMJER2204, JST ERATO Grant Number JPMJER2302, Japan, JSPS KAKENHI Grant Number JP19H05796, and JST CREST Grant Number JPMJCR20C1. N.Y. and T.S. are also supported by Institute of AI and Beyond of the University of Tokyo.
\end{acknowledgments}

\appendix

\section{Relationship to other measures of multipartite quantum correlations}
We discuss the relationship between the multipartite work deficit introduced in this work and other existing measures of multipartite quantum correlations, especially in terms of the numerical cost. 
Many existing measures, such as the relative entropy of entanglement \cite{plenio2001bounds,wei2004connections}, geometric entanglement \cite{barnum2001monotones,wei2003geometric,shimony1995degree}, and global quantum discord \cite{rulli2011global}, involve optimization parameters proportional to the system size, similarly to the multipartite work deficit. 
This increase of the optimization parameters makes it difficult to calculate these measures in many-body systems. 
While our efficient calculation method can reduce the numerical cost for the multipartite work deficit and global quantum discord, extending this method to other measures remains as a future issue. 

On the other hand, some of multipartite entanglement measures, such as quantum Fisher information \cite{hyllus2012fisher,toth2012multipartite,frowis2012measures,hauke2016measuring}, do not require such extensive optimization. This is because these measures are designed to witness specific types of multipartite entanglement rather than to quantify entanglement of general states. Therefore, the values of these measures can change under local unitary transformations, which distinguishes them from the previously mentioned measures \cite{plenio2001bounds,wei2004connections,barnum2001monotones,wei2003geometric,shimony1995degree,rulli2011global}.

\section{Expressions of multipartite work deficit}\label{apps:expressions}
\subsection{Expression of $W_l$} \label{apps:rank_one_oWD}
We here show that extractable work through zero-way closed LOCC is expressed as
\begin{equation*}
    W_l = N\ln d - \min_{\Lambda\in \mathcal{M}_N} H_{\Lambda}(Y_N). \tag{\ref{eq:Wl_zero-way}}
\end{equation*}
To derive this expression from Eq.~(\ref{eq:work_lambda}) in the main text, we need to show that optimal measurement protocol is always rank-one.
In the following, we prove that for any measurement protocol $\Lambda$ which includes higher-rank projectors, there exists a certain protocol $\Lambda'$ consisting solely of rank-one projectors, such that $W_{\Lambda'} \geq W_{\Lambda}$ is satisfied. 

In the zero-way closed LOCC, the allowed local measurement on each party is not arbitrary POVM but \textit{projective} measurement. Therefore, the measurement protocol $\Lambda$ is described with the sets of orthogonal projectors, represented as $\{\Pi_{y_1}\}, \{\Pi_{y_2}\}, \dots, \{\Pi_{y_N}\}$. In this case, the conditional density operator on the $n$-th party is defined as
\begin{equation*}
p_{\Lambda} (y_n) \widetilde{\rho}_n^{\Lambda,y_n} \equiv \Pi_{y_n} \rho_n \Pi_{y_n},
\end{equation*}
where $p_{\Lambda} (y_n)$ represents the probability of the measurement outcome $y_n$.
The rank-one measurement protocol $\Lambda'$, which satisfies $W_{\Lambda'} \geq W_{\Lambda}$, can be constructed by replacing each higher-rank projector $\Pi_{y_n}$ in $\Lambda$ with its fine-grained rank-one projectors $\{\Pi_{x_n}^{y_n} \Pi_{y_n}\}_{x_n}$, where $\{\Pi_{x_n}^{y_n}\}_{x_n}$ is the projectors on the diagonal basis of the conditional density operator defined as $\widetilde{\rho}_n^{\Lambda,y_n} = \sum_{x_n} q^{y_n}(x_n) \Pi_{x_n}^{y_n}$.
By representing the measurement outcomes for the new protocol $\Lambda'$ as $(Y_n, X_n) \equiv (y_1, x_1, y_2, x_2, \dots, y_n, x_n)$, we can derive the inequality $W_{\Lambda'} \geq W_{\Lambda}$ as follows:
\begin{align*}
    W_{\Lambda'} &= N\ln d - H_{\Lambda'}(Y_{N}, X_{N}) \\
    &= N\ln d - \sum_{n=1}^N H_{\Lambda'}(y_n, x_n|Y_{n-1}, X_{n-1}) \\
    &\geq N\ln d - \sum_{n=1}^N  H_{\Lambda'}(y_n, x_n|Y_{n-1}) \\
    &= N\ln d -\left\{\sum_{n=1}^N  H_{\Lambda'}(x_n|Y_{n}) + H_{\Lambda'}(y_n|Y_{n-1})\right\} \\
    &\geq N\ln d -\left\{\sum_{n=1}^N  H_{\Lambda'}(x_n|y_{n}) + H_{\Lambda'}(y_n|Y_{n-1})\right\} \\
    &= N\ln d - \sum_{n=1}^N \left\{\sum_{y_{n}} p_{\Lambda} (y_n) S(\widetilde{\rho}_n^{\Lambda,y_n})\right\} - H_{\Lambda}(Y_N)\\
    &= W_{\Lambda}.
\end{align*}

\subsection{Expression with quantum relative entropy}
We here show the equivalence of the two expressions of multipartite work deficit, namely Eqs.~(\ref{eq:zWD_Sh_def}) and~(\ref{eq:zWD_zero_way_c}).
Equation~(\ref{eq:zWD_Sh_def}) can be transformed into the representation with quantum relative entropy as
\begin{equation}
\label{appeq:zWD_sh_trans}
    \begin{split}
        \min_{\Lambda \in \mathcal{M}_N } H_{\Lambda}(Y_N) - S(\rho) &= \min_{\Lambda \in \mathcal{M}_N} S\left(\mathcal{D}_\Lambda (\rho)\right) - S(\rho) \\
        &= \min_{\Lambda \in \mathcal{M}_N } S\left(\rho \| \mathcal{D}_\Lambda (\rho)\right),
    \end{split}
\end{equation}
where $\mathcal{D}_\Lambda$ denotes the dephasing map with respect to the measurement basis of $\Lambda$. 
Since $\mathcal{D}_\Lambda (\rho) \in \mathcal{C}_N$, the following inequality holds:
\begin{equation}
\label{appeq:zWD_equiv1}
    \min_{\Lambda \in \mathcal{M}_N } S\left(\rho \| \mathcal{D}_\Lambda (\rho)\right) \geq \min_{\sigma \in \mathcal{C}_N} S(\rho\|\sigma).
\end{equation}

Furthermore, quantum relative entropy of $\rho$ with respect to a classically correlated state $\sigma = \sum_{Y_N}q(Y_N) \Pi_{y_1} \otimes \Pi_{y_2} \dots\otimes \Pi_{y_N}$ can be decomposed as 
\begin{equation*}
    S(\rho \| \sigma) = S\left(\rho \| \mathcal{D}_\sigma (\rho)\right) + S\left(\mathcal{D}_\sigma (\rho)\| \sigma \right),
\end{equation*}
where $\mathcal{D}_\sigma$ denotes the dephasing map with respect to the diagonal basis of $\sigma$.
Therefore, we can derive the following inequality:
\begin{equation}
\label{appeq:zWD_equiv2}
    \begin{split}
        \min_{\sigma \in \mathcal{C}_N}S(\rho \| \sigma) &= \min_{\sigma \in \mathcal{C}_N}S\left(\rho \| \mathcal{D}_\sigma (\rho)\right) + S\left(\mathcal{D}_\sigma (\rho)\| \sigma \right) \\
        &\geq \min_{\Lambda \in \mathcal{M}_N } S\left(\rho \| \mathcal{D}_\Lambda (\rho)\right).
    \end{split}
\end{equation}
From Eqs.~(\ref{appeq:zWD_sh_trans}), (\ref{appeq:zWD_equiv1}), and (\ref{appeq:zWD_equiv2}), we can derive the equivalence of Eqs.~(\ref{eq:zWD_Sh_def}) and (\ref{eq:zWD_zero_way_c}) as follows:
\begin{equation}
    \min_{\Lambda \in \mathcal{M}_N } H_{\Lambda}(Y_N) - S(\rho) = \min_{\sigma \in \mathcal{C}_N}S(\rho \| \sigma).
\end{equation}

\section{Relationship between multipartite work deficit and bipartite quantum correlation} \label{apps:decomp_WD}
We here derive the inequality
\begin{equation}
    \Delta \geq  \sum_{n=1}^{N-1} D_{(1,\dots,n);n+1}, \tag{\ref{eq:zWD_biD}}
\end{equation}
which indicates the relationship between multipartite work deficit and quantum discord $D_{(1,\dots,n);n+1}$, a well-known measure of bipartite quantum correlation \cite{ollivier2001quantum,henderson2001classical}.
The quantum discord for a bipartite state $\rho_{1,2}$ is defined as 
\begin{equation}
    D_{1;2} \equiv \min_{\Lambda \in \mathcal{M}_1} \sum_{y_1} p_\Lambda (y_1) S(\rho_2^{\Lambda,y_1}) - S_{2|1} (\rho_{1,2}),
\end{equation}
where $\mathcal{M}_1$ is the set of projective measurements on party $a_1$, $p_\Lambda (y_1)$ is the probability of obtaining measurement outcome $y_1$, $\rho_2^{\Lambda,y_1}$ is the conditional density operator on party $a_2$ given the outcome $y_1$, and $S_{2|1} (\rho_{1,2}) \equiv S(\rho_{1,2})-S(\rho_1)$ denotes the conditional von Neumann entropy.
The inequality (\ref{eq:zWD_biD}) can be derived as follows:
\begin{equation}
    \begin{split}
        &\Delta^ = \min_{\Lambda \in \mathcal{M}_{N}} \left\{ \sum_{n=1}^{N-1}H_{\Lambda}(y_{n+1}|Y_{n}) +H_{\Lambda}(y_1) \right\}-S(\rho_{1,\dots,N})  \\
    &\geq  \sum_{n=1}^{N-1} \left\{\min_{\Lambda \in \mathcal{M}_{n+1}} H_{\Lambda}(y_{n+1}|Y_{n}) \right\} +S(\rho_1) -S(\rho_{1,\dots,N})\\
    &\geq  \sum_{n=1}^{N-1} \left\{\min_{\Lambda \in \mathcal{M}_{n+1}} H_{\Lambda}(y_{n+1}|Y_{n})  -S_{n+1|(1,\dots,n)}(\rho_{1,\dots,n+1})\right\} \\
    &\geq  \sum_{n =1}^{N-1}  D_{(1,\dots,n);n+1}.
    \end{split}
\end{equation}

\section{Efficient calculation method for work deficit density $\delta$} \label{apps:calc_method}
\subsection{Upper bound $U_{l,k}$}
We here introduce the definition of the upper bound $U_{l,k}$ of work deficit density $\delta$, which is defined for the translationally invariant state with periodic boundary condition. This is regarded as a generalization of the upper bound $U_{l}$, which is defined as follows:
\begin{equation*}
    U_l \equiv \frac{1}{N}\left\{\sum_{n \in I_l} \min_{\Lambda\in \mathcal{M}_{n:l}} H_{\Lambda}(Y_{n:l})-S(\rho)\right\}, \tag{\ref{eq:U^phi_l}}
\end{equation*}
where $\mathcal{M}_{n:l}$ describes the set of measurement protocols whose measurement operators are denoted as
\begin{equation*}
    \{\Pi_{y_{n+1}}\otimes\Pi_{y_{n+2}}\dots\otimes\Pi_{y_{n+l}}\}_{Y_{n:l}}.
\end{equation*}
Here, $\{\Pi_{y_{n+i}}\}$ represents the set of orthogonal rank-one projectors on party $a_{n+i}$.
For the translationally invariant system under periodic boundary condition, $U_l$ is represented as
\begin{equation}
\label{appeq:Ul_trans_perod}
    U_l = \frac{1}{l}\min_{\Lambda\in \mathcal{M}_{n:l}} H_{\Lambda}(Y_{n:l}) - \frac{S(\rho)}{N}, 
\end{equation}
since all the terms in the summation $\left\{\min_{\Lambda\in \mathcal{M}_{n:l}}H_{\Lambda}(Y_{n:l})\right\}_{n \in I_l}$ are identical.

For such states, the upper bound $U_{l,k}$ is defined as follows, which is reduced to Eq.~(\ref{appeq:Ul_trans_perod}) when $k=0$:
\begin{equation}
\label{appeq:U^phi_l,k}
    U_{l,k} \equiv \frac{kl}{N}\ln d + \frac{1}{l}\min_{\Lambda \in \mathcal{M}_{l}}H_{\Lambda^{\otimes(k+1)}} (Y_{n:l}|Y_{n-kl:kl}) - \frac{S(\rho)}{N} ,
\end{equation}
where $\mathcal{M}_{l}$ is the set of local projective measurements on the $l$-site segment, and the measurement protocol $\Lambda^{\otimes(k+1)}$ is to perform the identical measurement $\Lambda$ on $(k+1)$ sequential $l$-site segments, specifically $\{a_i\}_{i=n-kl+1}^{n-kl+l},\dots,\{a_i\}_{i=n+1}^{n+l}$. We note that in the thermodynamic limit $N\to \infty$, the first term in Eq.~(\ref{appeq:U^phi_l,k}) is negligible.
We can derive the inequality $\delta \leq U_{l,k}$ as follows:
\begin{align*}
    \delta &= \frac{1}{N}\min_{\Lambda \in \mathcal{M}_{N}} \left\{\sum_{n \in I_l} H_{\Lambda}(Y_{n:l}|Y_{n})\right\}  - \frac{S(\rho)}{N}\\
    &\leq \frac{1}{N}\min_{\Lambda \in \mathcal{M}_{N}} \left\{\sum_{n \in I_l}H_{\Lambda}(Y_{n:l}|Y_{n-kl:kl}) \right\} - \frac{S(\rho)}{N}\\
    &\leq \frac{1}{N}\min_{\Lambda \in \mathcal{M}_{l}} \left\{\sum_{n \in I_l} H_{\Lambda^{\otimes(k+1)}} (Y_{n:l}|Y_{n-kl:kl})\right\} - \frac{S(\rho)}{N} \\
    &\leq \frac{1}{N}\min_{\Lambda \in \mathcal{M}_{l}} \left\{\sum_{\substack{n \in I_l \\ n \geq kl}} H_{\Lambda^{\otimes(k+1)}} (Y_{n:l}|Y_{n-kl:kl})\right\}\\
    &\quad \quad \quad \quad \quad \quad + \frac{kl \ln d}{N}- \frac{S(\rho)}{N} \\
    &= \left(\frac{1}{l}-\frac{k}{N}\right) \min_{\Lambda\in \mathcal{M}_{l}} H_{\Lambda^{\otimes(k+1)}} (Y_{n:l}|Y_{n-kl:kl})\\
    & \quad \quad \quad \quad \quad \quad + \frac{kl \ln d}{N}- \frac{S(\rho)}{N} \\
    &\leq U_{l,k}.
\end{align*}
The equality in the fifth line follows from the fact that all the terms in the summation $\{H_{\Lambda^{\otimes(k+1)}} (Y_{n:l}|Y_{n-kl:kl})\}_{n \in I_l,n \geq kl}$ are identical due to the translation invariance of the system and measurement protocol.

\subsection{Reduction of the optimization region for the lower bounds} 
\begin{figure}[]
    \centering
    \includegraphics[width=0.5\textwidth]{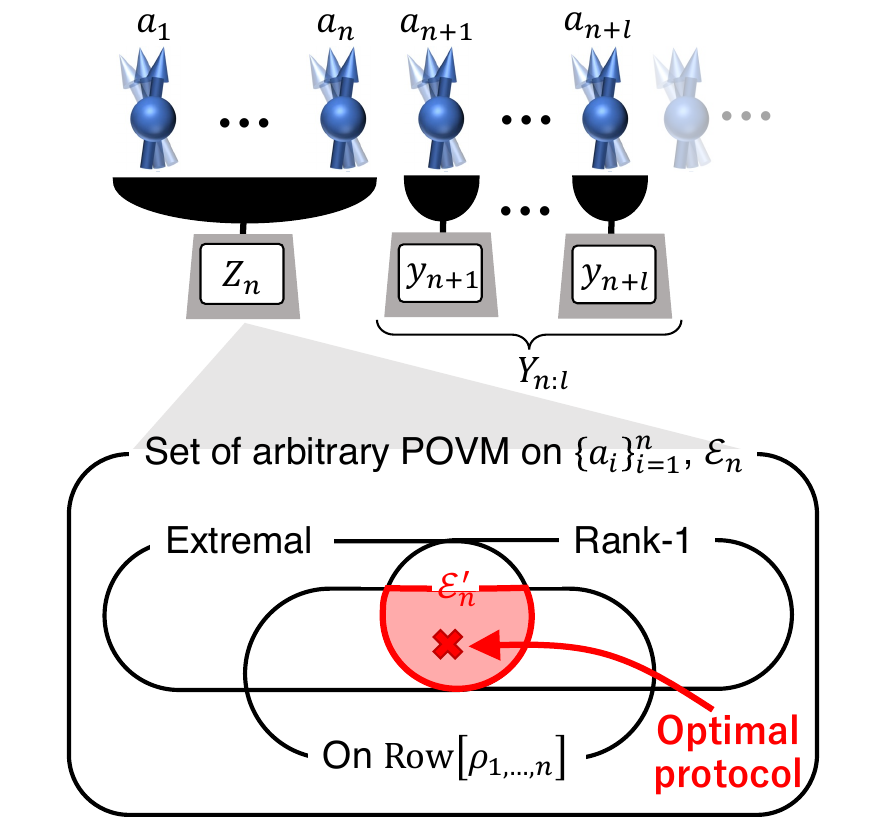}
    \caption{Reduction of the optimization region in the calculation of lower bound $L_l$. Although the lower bound $L_l$ includes the optimization over the set of arbitrary POVMs on the parties $\{a_i\}_{i=1}^{n}$, denoted as $\mathcal{E}_{n}$, this optimization region can be effectively reduced to the set of extremal rank-one POVMs on $\mathrm{Row}[\rho_{1,\dots,n}]$, denoted as $\mathcal{E}^\prime_{n}$ (Theorem \ref{Thm1}). Especially for the MPS with fixed bond dimension $D_B$, optimization over the reduced region $\mathcal{E}^\prime_{n}$ can be performed at a constant cost independent of $n$.}
    \label{fig:LB_opt_region}
\end{figure}

We provide a detailed discussion on the optimization involved in the definition of the lower bound $L_l$.
The lower bound is defined as
\begin{equation}
\label{appeq:L^phi_l_mini}
L_l \equiv \frac{1}{N} \left\{\sum_{n \in I_l} \min_{\mu \in \mathcal{E}_{n}} \min_{\pi \in \mathcal{M}_{n:l}} H_{(\mu,\pi)}(Y_{n:l}|Z_n) -S(\rho) \right\},
\end{equation}
where $\mathcal{E}_n$ denotes the set of arbitrary POVMs on the parties $\{a_i\}_{i=1}^{n}$, and $\mathcal{M}_{n:l}$ represents the set of local projective measurements on the neighbouring $l$ parties $\{a_i\}_{i=n+1}^{n+l}$. Equation~(\ref{appeq:L^phi_l_mini}) is equivalent to the representation in Eq.~(\ref{eq:L^phi_l}), since the measurement protocol $\Lambda \in \mathcal{N}_{n:l}$ is divided into $\mu \in \mathcal{E}_{n}$ and $\pi \in \mathcal{M}_{n:l}$ as $\Lambda = (\mu, \pi)$. The measurement operators for $(\mu,\pi)$ are given by
\begin{equation*}
    \{E_{Z_{n}}\otimes\Pi_{y_{n+1}}\otimes\Pi_{y_{n+2}}\dots\otimes\Pi_{y_{n+l}}\}_{Z_n,Y_{n:l}},
\end{equation*}
where $\{E_{Z_{n}}\}$ is the set of POVM operators for $\mu$, and $\{\Pi_{y_{n+i}}\}$ is the set of orthogonal projectors on party $a_{n+i}$.
Since the Hilbert space dimension of the subsystem $\{a_i\}_{i=1}^{n}$ is $d^{n}$, the optimization cost for $\mu \in \mathcal{E}_{n}$ appears to grow exponentially with respect to $n$, at first glance.

However, we here show that this cost can be reduced to constant independent of $n$, for MPS with fixed bond dimension $D_B$. 
While we have only provided an intuitive explanation for this cost reduction in the main text, suggesting that the subsystem $\{a_i\}_{i=1}^{n}$ is regarded as a fixed-dimensional system, we present a more detailed discussion in this subsection.
Specifically, we show that the optimization region $\mathcal{E}_{n}$ for the POVM $\mu$ can be reduced to the set of rank-one extremal POVMs on $\mathrm{Row}[\rho_{1,\dots,n}]$ denoted as $\mathcal{E}_{n}^\prime$, as illustrated in Fig.~\ref{fig:LB_opt_region}. 
Here, $\mathrm{Row}[\rho_{1,\dots,n}]$ is the row space of the reduced density operator $\rho_{1,\dots,n}$, namely, the subspace spanned by the row vectors of $\rho_{1,\dots,n}$. For an MPS with bond dimension $D_B$, the dimension for $\mathrm{Row}[\rho_{1,\dots,n}]$ is fixed ($D_B$ for open boundary condition, $D_B^2$ for periodic boundary condition).
A rank-one POVM is a class of POVM defined by rank-one operators $\{E_{Z_{n}}\}$, and an extremal POVM is one that cannot be described as a convex combination of other POVMs~\cite{d2005classical,haapasalo2012quantum}.
This reduction of optimization region is stated as the following theorem:
\begin{theorem} \label{Thm1}
    Let $\mathcal{E}_{n}$ be the set of arbitrary POVMs on parties $\{a_i\}_{i=1}^{n}$, and $\mathcal{E}_{n}^\prime$ be the set of rank-one extremal POVMs on $\mathrm{Row}[\rho_{1,\dots,n}]$. Then, the following equality is satisfied:
    \begin{equation}
    \begin{split}
        L_l &\equiv \frac{1}{N} \left\{\sum_{n \in I_l} \min_{\mu \in \mathcal{E}_{n}} \min_{\pi \in \mathcal{M}_{n:l}} H_{(\mu,\pi)}(Y_{n:l}|Z_n) -S(\rho) \right\} \\
        &= \frac{1}{N} \left\{\sum_{n \in I_l} \min_{\mu \in \mathcal{E}_{n}^\prime} \min_{\pi \in \mathcal{M}_{n:l}} H_{(\mu,\pi)}(Y_{n:l}|Z_n) -S(\rho) \right\}.
    \end{split}
    \end{equation} 
\end{theorem}
The proof of this theorem is provided in the Supplemental Material \cite{SM}. 
Since the rank-one extremal POVM on $x$-dimensional system is described with fewer than $x^2$ rank-one measurement operators $\{E_{Z_{n}}\}$~\cite{d2005classical,haapasalo2012quantum}, the optimization of $\mu$ within $\mathcal{E}_{n}^\prime$ can be performed at a constant cost independent of $n$.
This is crucial for the lower bound $L_l$ to be calculated with linear cost with respect to the system size $N$ (Table~\ref{tab:num_cost}).

\subsection{Lower and upper bounds for states \textit{approximated} by MPS}
While we have primarily focused on systems exactly represented by MPS, we here explain that our efficient calculation method is applicable also for states \textit{approximated} by MPS.
Specifically, we consider the state $\rho$ which is approximated with an MPS $\rho^\prime$ with fixed bond dimension.
To evaluate the lower bound $L_l(\rho)$, we utilize the fact that it is well-approximated with $L_l(\rho^\prime)$, which is efficiently computable as the consequence of Theorem~\ref{Thm1}.
To be more precise, we derive the following theorem:
\begin{theorem} \label{Thm2}
Let $\rho$ and $\rho^\prime$ be the states satisfying $\|\rho -\rho^\prime \|_{\rm tr} \leq \nu$, where $\|X\|_{\rm tr} \equiv \tr[|X|]$ denotes the trace norm. Then, the following inequalities are satisfied:
\begin{align}
    | U_l(\rho) -U_l (\rho^\prime) | &\leq \Tilde{\mathcal{O}}(\nu), \\
    | L_l(\rho) -L_l (\rho^\prime) | &\leq \Tilde{\mathcal{O}}(\nu). \label{appeq:approx_LB}
\end{align}
\end{theorem}
The proof of this theorem is provided in the Supplemental Material \cite{SM}. 
Since Eq.~(\ref{appeq:approx_LB}) implies the inequality
\begin{equation*}
    \delta(\rho) \geq L_l(\rho) \geq L_l (\rho^\prime) - \Tilde{\mathcal{O}}(\nu),
\end{equation*}
we can obtain the efficiently computable lower bound of $\delta (\rho)$ through the calculation in the MPS $\rho^\prime$.

\subsection{Tightness of the lower and upper bounds}
We here derive the systematic improvement of our lower and upper bounds by increasing the locality parameter $l$.
Especially for MPS with fixed bond dimension  $D_B$, we can derive the inequality
\begin{equation}
    U_l - L_l \leq \mathcal{O}\left(\frac{1}{l}\right). \tag{\ref{eq:WD_calc_presecion}}
\end{equation}
This inequality shows that we can achieve arbitrary accuracy $\varepsilon$ by increasing the locality $l$.
In the derivation, we utilize the measurement protocol $\Lambda^*_{n:l} \in \mathcal{N}_{n:l}$, which achieves the minimum for $\min_{\Lambda \in \mathcal{N}_{n:l}} H_\Lambda(Y_{n:l}|Z_{n})$.
With this notation, we can derive the inequality (\ref{eq:WD_calc_presecion}) as follows:
\begin{align*}
    U_l - L_l &\leq \frac{1}{N}\left\{\sum_{n \in I_l} H_{\Lambda^*_{n:l}}(Y_{n:l}) - H_{\Lambda^*_{n:l}}(Y_{n:l}|Z_{n}) \right\} \\
    &= \frac{1}{N}\sum_{n \in I_l} H_{\Lambda^*_{n:l}} (Y_{n:l}:Z_{n}) \\
    &\leq \frac{1}{N}\sum_{n \in I_l} I_{\{a_i\}_{i=1}^{n}:\{a_i\}_{i=n+1}^{n+l}}(\rho) \\
    &\leq \frac{1}{N}\left\{\sum_{n \in I_l} S(\rho_{1,\dots,n}) + S(\rho_{n+1,\dots,n+l})\right\} \\
    &\leq \frac{1}{l} 4 \ln D_B = \mathcal{O}\left(\frac{1}{l}\right).
\end{align*}
Here, $H_{\Lambda}(y:z)\equiv H_{\Lambda}(y) + H_{\Lambda}(z) - H_{\Lambda}(y,z)$ represents the classical mutual information between measurement outcomes $y$ and $z$ under the measurement protocol $\Lambda$, and $I_{a_1:a_2}(\rho)\equiv S(\rho_1) +S(\rho_2) -S(\rho_{1,2})$ represents the quantum mutual information between parties $a_1$ and $a_2$.
The inequality in the third line follows from the fact that quantum mutual information decreases monotonically under local CPTP map, which is the consequence of the monotonicity of quantum relative entropy \cite{petz2003monotonicity}. 
In the derivation of the final line, we utilize that the system is described with an MPS with a fixed bond dimension $D_B$.

Additionally, we can derive the following inequalities, which are satisfied in general, not only in MPS:
\begin{equation}
    L_l \leq L_{kl}, \quad U_{kl} \leq U_{l}. \tag{\ref{eq:LUb_improve}}
\end{equation}
The inequality $L_l \leq L_{kl}$ can be derived as
\begin{equation}
\label{appeq:sys_imp_LB}
    \begin{split}
        &L_{kl} = \frac{1}{N} \left\{\sum_{n \in I_{kl}} \min_{\Lambda \in \mathcal{N}_{n:kl}}  H_{\Lambda}(Y_{n:kl}|Z_{n}) - S(\rho) \right\}\\
    &= \frac{1}{N} \left\{\sum_{n \in I_{kl}} \min_{\Lambda \in \mathcal{N}_{n:kl}}  \sum_{j=0}^{k-1} H_{\Lambda}(Y_{n+jl:l}|Y_{n:jl},Z_{n}) - S(\rho) \right\} \\
    &\geq  \frac{1}{N} \left\{\sum_{n \in I_{kl}} \sum_{j=0}^{k-1} \min_{\Lambda \in \mathcal{N}_{n:(j+1)l}} H_{\Lambda}(Y_{n+jl:l}|Y_{n:jl},Z_{n}) - S(\rho) \right\}\\
    &\geq \frac{1}{N} \left\{\sum_{m \in I_{l}}  \min_{\Lambda \in \mathcal{N}_{m:l}}   H_{\Lambda}(Y_{m:l}|Z_{m})  - S(\rho) \right\}  \\
    &= L_l,
    \end{split}
\end{equation}
where subscript $m$ in the fourth line corresponds to $n+jl$ in the third line, and therefore the inequality in the fourth line follows from the inclusion relation $\mathcal{N}_{n:(j+1)l} \subset \mathcal{N}_{n+jl:l}$. 
The inequality $U_{kl} \leq U_{l}$ can be derived as follows:
\begin{equation}
\label{appeq:sys_imp_UB}
    \begin{split}
        U_{kl} &= \frac{1}{N} \left\{\sum_{n \in I_{kl}} \min_{\Lambda \in \mathcal{M}_{n:kl}}  H_{\Lambda}(Y_{n:kl}) - S(\rho) \right\}\\
    &= \frac{1}{N}\left\{\sum_{n \in I_{kl}} \min_{\Lambda \in \mathcal{M}_{n:kl}} \sum_{j=0}^{k-1} H_{\Lambda}(Y_{n+jl:l}|Y_{n:jl}) - S(\rho) \right\} \\
    &\leq \frac{1}{N}\left\{\sum_{n \in I_{kl}} \min_{\Lambda \in \mathcal{M}_{n:kl}} \sum_{j=0}^{k-1} H_{\Lambda}(Y_{n+jl:l}) - S(\rho) \right\} \\
    &\leq \frac{1}{N}\left\{\sum_{m \in I_{l}}  \min_{\Lambda \in \mathcal{M}_{m:l}}  H_{\Lambda}(Y_{m:l}) - S(\rho) \right\}\\
    &= U_l,
    \end{split}
\end{equation}
where tne subscript $m$ in the fourth line corresponds to $n+jl$ in the third line.

\section{Efficient calculation method for one-way work deficit density $\delta^\to$} \label{apps:calc_method_oWD}
\subsection{Upper bounds}
We here introduce the upper bound $U_l^\to$ for the one-way work deficit density $\delta^\to$.
The bound $U_l^\to$ is defined as
\begin{equation}
    U_l^\to \equiv \frac{1}{N}\left\{\sum_{n \in I_l} \min_{\Lambda\in \mathcal{M}_{n:l}^{\to}} H_{\Lambda}(Y_{n:l}) -S(\rho)\right\},\label{appeq:U^to_l}
\end{equation}
where $\mathcal{M}_{n:l}^{\to}$ is the set of measurement protocols whose operators are described as 
\begin{equation*}
    \{\Pi_{y_{n+1}}\otimes\Pi_{y_{n+2}}^{y_{n+1}}\dots\otimes\Pi_{y_{n+l}}^{Y_{n:l-1}}\}_{Y_{n:l}}.
\end{equation*}
Here, $\{\Pi_{y_{n+k+1}}^{Y_{n:k}}\}_{Y_{n:k+1}}$ represents the local projective measurement on $a_{n+k+1}$ under the condition that the measurement outcomes on $\{a_i\}_{i=n+1}^{n+k}$ are $Y_{n:k}$. 
Since each term in the summation of Eq.~(\ref{appeq:U^to_l}), $\min_{\Lambda\in \mathcal{M}_{n:l}^{\to}} H_{\Lambda}(Y_{n:l})$, can be calculated at a constant numerical cost independent of system size $N$, the bound $U_l^\to$ is calculated with the cost proportional to $N$.

The upper bound $U_l^\to$ is derived as follows:
\begin{align*}
    \delta^\to &= \frac{1}{N}\left\{ \min_{\Lambda \in \mathcal{M}^\to_{N}}\sum_{n \in I_l} H_{\Lambda}(Y_{n:l}|Y_{n})-S(\rho) \right\}\\
    &\leq \frac{1}{N}\left\{ \min_{\Lambda \in \mathcal{M}^\to_{N}}\sum_{n \in I_l} H_{\Lambda}(Y_{n:l}) -S(\rho) \right\} \\
    &\leq \frac{1}{N}\left\{ \sum_{n \in I_l} \min_{\Lambda \in \mathcal{M}^{\to}_{n:l}}  H_{\Lambda}(Y_{n:l}) -S(\rho) \right\} \\
    &= U_l^\to.
\end{align*}
Furthermore, we can show the inequality
\begin{equation}
\label{appeq:sys_imp_U^to}
    U_l^\to \geq U_{kl}^\to,
\end{equation}
for an arbitrary natural number $k$, which implies that the bound $U_l^\to$ is systematically improved by increasing the locality parameter $l$. The derivation of Eq.~(\ref{appeq:sys_imp_L^to}) is essentially the same as Eq.~(\ref{appeq:sys_imp_UB}).

\subsection{Lower bounds}
We here introduce the lower bound $L_l^\to$ for one-way work deficit density $\delta^\to$. 
The lower bound is defined as 
\begin{equation}
    L_l^\to \equiv  \frac{1}{N}\left\{\sum_{n \in I_l} \min_{\Lambda \in\mathcal{N}^{\to}_{n:l} }  H_{\Lambda}(Y_{n:l}|Z_{n}) -S(\rho) \right\}. \label{appeq:L^to_l}
\end{equation}
where $\mathcal{N}_{n:l}^{\to}$ is the set of measurements whose operators are represented as
\begin{equation*}
    \{E_{Z_{n}}\otimes\Pi_{y_{n+1}}^{Z_{n}}\otimes\Pi_{y_{n+2}}^{Z_{n},y_{n+1}}\dots\otimes\Pi_{y_{n+l}}^{Z_{n},Y_{n:l-1}} \}_{Z_{n},Y_{n:l}}.
\end{equation*}
Here, the local measurement on party $a_{n+k+1}$ depends on the past measurement outcomes on parties $\{a_{i}\}_{i=1}^{n+k}$ (i.e., $Z_{n}$ and $Y_{n:k}$). 
Similarly to $L_l$, the optimization region of the POVM $\{E_{Z_{n}}\}$ can be reduced as shown in Fig.~\ref{fig:LB_opt_region} for systems described with MPS.
Furthermore, the continuity of the lower bound can also be proved, similarly to Eq.~(\ref{appeq:approx_LB}).
Since each term in the summation of Eq.~(\ref{appeq:L^to_l}), $\min_{\Lambda \in\mathcal{N}^{\to}_{n:l} }  H_{\Lambda}(Y_{n:l}|Z_{n})$, can be calculated with the constant numerical cost for systems described with MPS, the bound $L_l^\to$ can be calculated with the cost proportional to the system size $N$.

The lower bound $L_l^\to$ is derived as follows:
\begin{align*}
    \delta^\to &= \frac{1}{N}\min_{\Lambda \in \mathcal{M}^\to_{N}} \left\{\sum_{n \in I_l} H_{\Lambda}(Y_{n:l}|Y_{n}) -S(\rho)\right\}  \\
    &\geq \frac{1}{N}\left\{\sum_{n \in I_l} \min_{\Lambda \in \mathcal{M}^\to_{n+l}}  H_{\Lambda}(Y_{n:l}|Y_{n}) -S(\rho)\right\}  \\
    &\geq  \frac{1}{N}\left\{\sum_{n \in I_l} \min_{\Lambda \in \mathcal{N}^{\to}_{n:l}}  H_{\Lambda}(Y_{n:l}|Z_{n}) -S(\rho)\right\}\\
    &= L_l^\to.
\end{align*}
Here, we use the inclusion relation of the measurement protocols $\mathcal{M}^\to_{n+l} \subset \mathcal{N}^{\to}_{n:l}$ to derive the inequality in the third line.
Furthermore, we can show the inequality
\begin{equation}
\label{appeq:sys_imp_L^to}
    L_l^\to \leq L_{kl}^\to,
\end{equation}
for an arbitrary natural number $k$, which implies that the bound $L_l^\to$ is systematically improved by increasing the locality parameter $l$. The derivation of Eq.~(\ref{appeq:sys_imp_L^to}) is essentially the same as Eq.~(\ref{appeq:sys_imp_LB}).

\bibliography{biblio.bib}

\onecolumngrid

\end{document}



\onecolumngrid

\begin{center}
	\Large
	\textbf{Supplemental Material for ``Measuring multipartite quantum correlations by thermodynamic work extraction"}
\end{center}

\setcounter{section}{0}
\setcounter{equation}{0}
\setcounter{figure}{0}
\setcounter{table}{0}
\setcounter{page}{1}
\renewcommand{\thesection}{S\arabic{section}}
\renewcommand{\theequation}{S\arabic{equation}}
\renewcommand{\thefigure}{S\arabic{figure}}
\renewcommand{\thetable}{S\arabic{table}}
\renewcommand{\bibnumfmt}[1]{[S#1]}
\renewcommand{\citenumfont}[1]{S#1}

\tableofcontents

\section{Efficient calculation method for global quantum discord}
In this section, we show that our efficient calculation method is applicable to a measure of multipartite quantum correlation called global quantum discord \cite{rulli2011global}. In Sec.~\ref{Sss:Ov_GQD}, we introduce the definition of global quantum discord and overview its efficient calculation method. In Sec.~\ref{Sss:LUB_GQD}, we derive the lower and upper bounds, and their systematic improvement in trade-off with the numerical cost.

\subsection{Overview of efficient calculation method} \label{Sss:Ov_GQD}
The global quantum discord is a measure of multipartite quantum correlation introduced in Ref.~\cite{rulli2011global}.
For a state $\rho$ in an $N$-partite system composed of parties $\{a_i\}_{i=1}^N$, the global quantum discord is defined as
\begin{equation}
    \label{appeq:gQD}
    D^{(G)} \equiv  \min_{\Lambda \in \mathcal{M}_N} \left\{ H_\Lambda (Y_N) -S(\rho) - \sum_{n=1}^N \left( H_\Lambda (y_n) -S(\rho_n) \right) \right\},
\end{equation}
where $\rho_n$ is the reduced density operator for party $a_n$, $\mathcal{M}_N$ is the set of measurement protocols composed of the local projective measurements on all sites, $y_n$ is the measurement outcome obtained from party $a_n$, and $Y_N\equiv (y_1,y_2\dots,y_N)$ denotes all the measurement outcomes.
The leading term of this measure is defined as
\begin{equation}
    \label{appeq:gQD_dens}
    d^{(G)} \equiv \frac{D^{(G)}}{N},
\end{equation}
which we call global quantum discord density.

While the previous numerical studies of global quantum discord required exponentially increasing computational costs with respect to the system size $N$, our method reduces the cost to the linear growth with $N$, when the state is described with MPS. 
In this calculation method, similar to the work deficit densities, we introduce the upper and lower bounds for $d^{(G)}$ as
\begin{equation}
\label{appeq:gQD_calc}
L_l^G \leq d^{(G)} \leq U_l^G,
\end{equation}
where $l$ is a natural number representing the locality of these bounds. The bounds $L_l^G$ and $U_l^G$ are calculated by dividing the entire system into $l$-site segments and performing optimization within each segment.

We can further show that these bounds can be systematically improved in trade-off with the numeraical cost by increasing the locality parameter $l$.
Especially for the states well-approximated with MPS, we can derive the following inequality which guarantees that arbitrary error bound is achieved with this systematic improvement:
\begin{equation}
\label{appeq:gQD_UL_tightness}
U_l^G- L_l^G \leq \mathcal{O} \left(\frac{1}{l}\right).
\end{equation}
In the next subsection, we introduce the specific expressions for these bounds $L_l^G$ and $U_l^G$, and derive Eqs.~(\ref{appeq:gQD_calc}) and (\ref{appeq:gQD_UL_tightness}).

\subsection{Lower and upper bounds}\label{Sss:LUB_GQD}
The lower and upper bounds of the global quantum discord density $d^{(G)}$ are defined as follows:
\begin{align}
    U_l^G &\equiv \frac{1}{N}\sum_{n \in I_l} \min_{\Lambda \in\mathcal{M}_{n:l} }  \left\{H_{\Lambda}(Y_{n:l}) -\sum_{i=n+1}^{n+l} H_{\Lambda}(y_i) \right\} - \frac{1}{N} \left(S(\rho)-\sum_{n=1}^N S(\rho_n)\right), \\
    L_l^G &\equiv \frac{1}{N}\sum_{n \in I_l} \min_{\Lambda \in\mathcal{N}_{n:l} }  \left\{H_{\Lambda}(Y_{n:l}|Z_{n}) -\sum_{i=n+1}^{n+l} H_{\Lambda}(y_i) \right\}- \frac{1}{N} \left(S(\rho)-\sum_{n=1}^N S(\rho_n)\right),
\end{align}
where $\mathcal{M}_{n:l}$ is the set of measurement protocols composed of local projective measurements on $\{a_i\}_{i=n+1}^{n+l}$, $\mathcal{N}_{n:l}$ is the set of measurement protocols composed of a POVM on $\{a_i\}_{i=1}^{n}$ and local projective measurements on $\{a_i\}_{i=n+1}^{n+l}$, $Z_{n}$ denotes the measurement outcome of the POVM on $\{a_i\}_{i=1}^{n}$, and $Y_{n:l} \equiv (y_{n+1}, \dots, y_{n+l})$ denotes the outcomes of parties $\{a_i\}_{i=n+1}^{n+l}$.

The inequality $d^{(G)} \leq U_l^G$ in Eq.~(\ref{appeq:gQD_calc}) can be derived as follows:
\begin{align*}
    d^{(G)} &= \frac{1}{N} \min_{\Lambda \in\mathcal{M}_{N}}\sum_{n \in I_l}  \left\{ H_{\Lambda}(Y_{n:l}|Y_n) -\sum_{i=n+1}^{n+l} H_{\Lambda}(y_i) \right\} - \frac{1}{N} \left(S(\rho)-\sum_{n=1}^N S(\rho_n)\right)\\
    &\leq \frac{1}{N} \min_{\Lambda \in\mathcal{M}_{N}} \sum_{n \in I_l} \left\{H_{\Lambda}(Y_{n:l}) -\sum_{i=n+1}^{n+l} H_{\Lambda}(y_i) \right\} - \frac{1}{N} \left(S(\rho)-\sum_{n=1}^N S(\rho_n)\right)\\
    &= \frac{1}{N}\sum_{n \in I_l} \min_{\Lambda \in\mathcal{M}_{n:l} }  \left\{H_{\Lambda}(Y_{n:l}) -\sum_{i=n+1}^{n+l} H_{\Lambda}(y_i) \right\} - \frac{1}{N} \left(S(\rho)-\sum_{n=1}^N S(\rho_n)\right) \\
    &=U_l^G.
\end{align*}
The inequality for the lower bound $d^{(G)} \geq L_l^G$ in Eq.~(\ref{appeq:gQD_calc}) is derived as
\begin{align*}
    d^{(G)} &= \frac{1}{N} \min_{\Lambda \in\mathcal{M}_{N}}\sum_{n \in I_l} \left\{  H_{\Lambda}(Y_{n:l}|Y_n) -\sum_{i=n+1}^{n+l} H_{\Lambda}(y_i) \right\} - \frac{1}{N} \left(S(\rho)-\sum_{n=1}^N S(\rho_n)\right) \\
    &\geq \frac{1}{N}\sum_{n \in I_l} \min_{\Lambda \in\mathcal{M}_{n+l}}  \left\{H_{\Lambda}(Y_{n:l}|Y_n) -\sum_{i=n+1}^{n+l} H_{\Lambda}(y_i) \right\} - \frac{1}{N} \left(S(\rho)-\sum_{n=1}^N S(\rho_n)\right) \\
    &\geq \frac{1}{N}\sum_{n \in I_l} \min_{\Lambda \in \mathcal{N}_{n:l}}  \left\{H_{\Lambda}(Y_{n:l}|Z_{n}) -\sum_{i=n+1}^{n+l} H_{\Lambda}(y_i) \right\} - \frac{1}{N} \left(S(\rho)-\sum_{n=1}^N S(\rho_n)\right) \\
    &=L_l^G,
\end{align*}
where the inequality in the third line follows from the inclusion relationship $\mathcal{M}_{n+l}\subset \mathcal{N}_{n:l}$.
By denoting the optimal protocol which achieves the minimum value of $\min_{\Lambda \in\mathcal{N}_{n:l} } \left\{ H_{\Lambda}(Y_{n:l}|Z_{n}) -\sum_{i=n+1}^{n+l} H_{\Lambda}(y_i) \right\}$ as $\Lambda^*_{n:l} \in \mathcal{N}_{n:l}$, Eq.~(\ref{appeq:gQD_UL_tightness}) is derived as follows:
\begin{align*}
    U_l^G -L_l^G &=\frac{1}{N}\sum_{n \in I_l} \min_{\Lambda \in\mathcal{M}_{n:l} }  \left\{H_{\Lambda}(Y_{n:l}) -\sum_{i=n+1}^{n+l} H_{\Lambda}(y_i) \right\} -\frac{1}{N}\sum_{n \in I_l} \min_{\Lambda \in\mathcal{N}_{n:l} } \left\{ H_{\Lambda}(Y_{n:l}|Z_{n}) -\sum_{i=n+1}^{n+l} H_{\Lambda}(y_i) \right\} \\
    &\leq \frac{1}{N}\sum_{n \in I_l} \left\{ H_{\Lambda^*_{n:l}}(Y_{n:l}) -\sum_{i=n+1}^{n+l} H_{\Lambda^*_{n:l}}(y_i) \right\} -\frac{1}{N}\sum_{n \in I_l}  \left\{H_{\Lambda^*_{n:l}}(Y_{n:l}|Z_{n}) -\sum_{i=n+1}^{n+l} H_{\Lambda^*_{n:l}}(y_i) \right\} \\
    &= \frac{1}{N} \sum_{n \in I_l}  H_{\Lambda^*_{n:l}}(Y_{n:l}:Z_{n}) \\
    &\leq \frac{1}{N} \sum_{n \in I_l}  I_{\{a_i\}_{i=1}^n:\{a_i\}_{i=n+1}^{n+l}} (\rho) \\
    &\leq \frac{1}{N} \sum_{n \in I_l}  S (\rho_{1,\dots,n}) + S(\rho_{n+1,\dots,n+l}) \\
    &\leq \frac{4 \ln D_B}{l} = \mathcal{O} \left(\frac{1}{l}\right).
\end{align*}
Here, $H_{\Lambda}(y:z)\equiv H_{\Lambda}(y) + H_{\Lambda}(z) - H_{\Lambda}(y,z)$ represents the classical mutual information between measurement outcomes $y$ and $z$ under the measurement protocol $\Lambda$, and $I_{a_1:a_2}(\rho)\equiv S(\rho_1) +S(\rho_2) -S(\rho_{1,2})$ represents the quantum mutual information between parties $a_1$ and $a_2$.
In the derivation of the fourth line, we use the monotonicity of the mutual information under local CPTP map.

\section{Reduction of optimization region involved in the lower bounds} 
In this section, we prove that we can effectively reduce the range of optimization involved in the definitions of the lower bounds $L_l$, $L_l^\to$ and $L_l^G$.
For simplicity, we here focus on $L_l$, while the discussion can be directly extended to the other lower bounds $L_l^\to$ and $L_l^G$.
It is defined as 
\begin{equation}
    L_l \equiv \frac{1}{N} \left\{\sum_{n \in I_l} \min_{\mu \in \mathcal{E}_{n}} \min_{\pi \in \mathcal{M}_{n:l}} H_{(\mu,\pi)}(Y_{n:l}|Z_n) -S(\rho) \right\} ,\label{Seq:L^phi_l_mini}
\end{equation}
where $\mathcal{E}_n$ denotes the set of arbitrary POVMs on parties $\{a_i\}_{i=1}^{n}$, and $\mathcal{M}_{n:l}$ is the set of local projective measurements on parties $\{a_i\}_{i=n+1}^{n+l}$. 

We show that the optimization region $\mathcal{E}_n$ in Eq.~(\ref{Seq:L^phi_l_mini}) can be effectively reduced to the set of rank-one extremal POVMs on $\mathrm{Row}[\rho_{1,\dots,n}]$ denoted as $\mathcal{E}_n^\prime$, as illustrated in Fig.~\blue{12} in the main text. The three conditions (rank-one, extremal, and on $\mathrm{Row}[\rho_{1,\dots,n}]$) for $\mu \in \mathcal{E}_n^\prime$ are detailed as follows:
\begin{itemize}
    \item A rank-one POVM is a class of POVM whose operators $\{E_{Z_{n}}\}_{Z_n}$ are all of rank one. 
    \item An extremal POVM is a class of POVM which cannot be described as a convex combination of other POVMs. For instance, POVM $\vartheta$ represented with the operators $\{\frac{1}{2}\ket{0}\bra{0}, \frac{1}{2}\ket{1}\bra{1}, \frac{1}{2}\ket{+}\bra{+}, \frac{1}{2}\ket{-}\bra{-}\}$ is non-extremal since it can be decomposed as $\frac{1}{2} \vartheta_1+ \frac{1}{2} \vartheta_2$, where $\vartheta_1$ is POVM with operators $\{\ket{0}\bra{0}, \ket{1}\bra{1}\}$ and $\vartheta_2$ is POVM with operators $\{\ket{+}\bra{+},\ket{-}\bra{-}\}$. This decomposition shows that $\vartheta$ is equivalent to performing $\vartheta_1$ and $\vartheta_2$ with probability $\frac{1}{2}$, respectively. On the other hand, POVMs $\vartheta_1$ and $\vartheta_2$ are extremal, since they are not decomposable.
    \item $\mathrm{Row}[\rho_{1,\dots,n}]$ represents the row space of the reduced density operator $\rho_{1,\dots,n}$. In other words, $\mathrm{Row}[\rho_{1,\dots,n}]$ is spanned by the eigenvectors of $\rho_{1,\dots,n}$, and its dimension becomes $\mathrm{rank}[\rho_{1,\dots,n}]$. For an MPS with fixed bond dimension $D_B$, the dimension for $\mathrm{Row}[\rho_{1,\dots,n}]$ is also fixed ($D_B$ for open boundary condition, $D_B^2$ for periodic boundary condition).
\end{itemize}
Since rank-one extremal POVMs on $x$-dimensional system are described with fewer than $x^2$ POVM operators \cite{d2005classical,haapasalo2012quantum}, the optimization cost within $\mathcal{E}_n^\prime$ become constant, independent of $n$, for MPS with fixed bond dimension $D_B$.

Such a reduction of the optimization region is stated in the following theorem:
\begin{theorem} (restatement of Theorem \blue{1} in Appendix~\blue{D2}) \label{Sthm:red_opt_Ll}
Let $\mathcal{E}_{n}$ be the set of arbitrary POVMs on parties $\{a_i\}_{i=1}^{n}$, $\mathcal{M}_{n:l}$ be the set of local projective measurements on $\{a_i\}_{i=n+1}^{n+l}$, and $\mathcal{E}_{n}^\prime$ be the set of rank-one extremal POVMs on $\mathrm{Row}[\rho_{1,\dots,n}]$. Then, the following equality is satisfied:
\begin{equation}
    \label{Seq:Thm1_Ll}
    L_l \equiv \frac{1}{N} \left\{\sum_{n \in I_l} \min_{\mu \in \mathcal{E}_{n}} \min_{\pi \in \mathcal{M}_{n:l}} H_{(\mu,\pi)}(Y_{n:l}|Z_n) -S(\rho) \right\} 
    = \frac{1}{N} \left\{\sum_{n \in I_l} \min_{\mu \in \mathcal{E}_{n}^\prime} \min_{\pi \in \mathcal{M}_{n:l}} H_{(\mu,\pi)}(Y_{n:l}|Z_n) -S(\rho) \right\}.
\end{equation}
\end{theorem}
\begin{proof}
By utilizing Lemma~\ref{Slem:red_opt_cond} derived later in this section, we can show the following inequality:
\begin{equation}
\label{Seq:red_Ll_lower}
    \begin{split}
        \frac{1}{N} \left\{\sum_{n \in I_l} \min_{\mu \in \mathcal{E}_{n}} \min_{\pi \in \mathcal{M}_{n:l}} H_{(\mu,\pi)}(Y_{n:l}|Z_n) -S(\rho) \right\} &= \frac{1}{N} \left\{\sum_{n \in I_l}  H_{(\mu^*_n,\pi^*_n)}(Y_{n:l}|Z_n) -S(\rho) \right\} \\
        &\geq \frac{1}{N} \left\{\sum_{n \in I_l} \min_{\mu \in \mathcal{E}_{n}^\prime} \min_{\pi \in \mathcal{M}_{n:l}} H_{(\mu,\pi)}(Y_{n:l}|Z_n) -S(\rho) \right\},
    \end{split}
\end{equation}
where $\mu^*_n\in \mathcal{E}_{n}$ and $\pi^*_n \in \mathcal{M}_{n:l}$ denote the measurement protocols which achieve the minimum of $\min_{\mu\in \mathcal{E}_{n}} \min_{\pi\in \mathcal{M}_{n:l}} H_{(\mu,\pi)}(Y_{n:l}|Z_n)$. The inequality in the second line follows from Lemma~\ref{Slem:red_opt_cond}. 

Additionally, the following inequality directly follows from the inclusion relationship $\mathcal{E}_{n}^\prime \subset  \mathcal{E}_{n}$:
\begin{equation}
\label{Seq:red_Ll_upper}
    \frac{1}{N} \left\{\sum_{n \in I_l} \min_{\mu \in \mathcal{E}_{n}} \min_{\pi \in \mathcal{M}_{n:l}} H_{(\mu,\pi)}(Y_{n:l}|Z_n) -S(\rho) \right\} 
        \leq \frac{1}{N} \left\{\sum_{n \in I_l} \min_{\mu \in \mathcal{E}_{n}^\prime} \min_{\pi \in \mathcal{M}_{n:l}} H_{(\mu,\pi)}(Y_{n:l}|Z_n) -S(\rho) \right\}.
\end{equation}
By combining Eqs.~(\ref{Seq:red_Ll_lower}) and (\ref{Seq:red_Ll_upper}), we can derive Eq.~(\ref{Seq:Thm1_Ll}).
\end{proof}

We finish this section by proving Lemma~\ref{Slem:red_opt_cond}, which is utilized in the proof of Theorem \ref{Sthm:red_opt_Ll}.
\begin{lemma} \label{Slem:red_opt_cond}
For an arbitrary POVM $\mu \in \mathcal{E}_{n}$, and local projective measurement on $l$ sites $\pi \in \mathcal{M}_{n:l}$, there exists a rank-one extremal POVM on $\mathrm{Row}[\rho_{1,\dots,n}]$, $\mu^\prime \in \mathcal{E}_{n}^\prime$, which satisfies 
\begin{equation}
    \label{Seq:Thm1}
    H_{(\mu,\pi)}(Y_{n:l}|Z_n) \geq H_{(\mu^\prime,\pi)}(Y_{n:l}|Z_n).
\end{equation}
\end{lemma}

\begin{proof}
We first show that for an arbitrary POVM $\mu$, there exists a POVM on $\mathrm{Row}[\rho_{1,\dots,n}]$, denoted as $\mu^\prime$, which satisfies Eq.~(\ref{Seq:Thm1}). Such a POVM $\mu^\prime$ is constructed with a set of POVM operators $\{\Pi_{\rm Row} E_{Z_{n}} \Pi_{\rm Row}\}_{Z_n}$, where $\{E_{Z_{n}}\}_{Z_n}$ is the set of POVM operators for $\mu$, and $\Pi_{\rm Row}$ denotes the projector onto the subspace $\mathrm{Row}[\rho_{1,\dots,n}]$. 
For these POVMs, $\mu$ and $\mu^\prime$, we can derive the equality $H_{(\mu,\pi)}(Y_{n:l}|Z_n) = H_{(\mu^\prime,\pi)}(Y_{n:l}|Z_n)$ from the equivalence of the probability distributions as shown below:
\begin{equation}
\label{Seq:Thm1_row}
    \begin{split}
        P_{(\mu,\pi)}[Z_n, Y_{n:l}] &= \tr[E_{Z_{n}} \otimes \Pi_{y_{n+1}} \otimes \dots \Pi_{y_{n+l}} \rho] \\
        &= \tr[E_{Z_{n}} \otimes \Pi_{y_{n+1}} \otimes \dots \Pi_{y_{n+l}} (\Pi_{\rm Row} + \Pi_{\rm Nul}) \otimes \mathbbm{1}_{d^{N-n}} \rho (\Pi_{\rm Row} + \Pi_{\rm Nul}) \otimes \mathbbm{1}_{d^{N-n}}] \\
        &= \tr[\Pi_{\rm Row} E_{Z_{n}} \Pi_{\rm Row} \otimes \Pi_{y_{n+1}} \otimes \dots \Pi_{y_{n+l}} \rho] \\
        &= P_{(\mu^\prime, \pi)}[Z_n, Y_{n:l}],
    \end{split}
\end{equation}
where $\{\Pi_{y_{n+1}} \otimes \dots \Pi_{y_{n+l}}\}$ represents the set of rank-one projectors for the measurement $\pi \in \mathcal{M}_{n:l}$. Here, $\Pi_{\rm Nul}$ represents the projector onto the null space of $\rho_{1,\dots,n}$, which is the orthogonal complement of $\mathrm{Row}[\rho_{1,\dots,n}]$, and satisfies the equality $\Pi_{\rm Row} + \Pi_{\rm Nul} = \mathbbm{1}_{d^n}$. Therefore, the second line of Eq.~(\ref{Seq:Thm1_row}) follows from simple insertions of identities on both sides of $\rho$. The equality in the third line follows from $\Pi_{\rm Nul} \rho = \rho \Pi_{\rm Nul} = 0$.

We next prove that for any higher-rank POVM $\mu$, there exists a rank-one POVM $\mu^\prime$, which satisfies Eq.~(\ref{Seq:Thm1}).
Such a POVM $\mu^\prime$ can be introduced as a fine-grained version of $\mu$. Specifically, $\mu^\prime$ is defined by replacing the higher-rank POVM operator $E_{Z_n}$ of $\mu$ with its fine-grained rank-one operators $\{E^\prime_{Z_n,W_n}\}_{W_n}$, where the relationship $E_{Z_n} = \sum_{W_n} E^\prime_{Z_n,W_n}$ holds. 
Since the equality $P_{(\mu,\pi)}[Z_n,Y_{n:l}] = \sum_{W_n} P_{(\mu^\prime,\pi)}[Z_n,W_n,Y_{n:l}]$ directly follows from this relationship, we can derive the following inequality:
\begin{equation}
\label{Seq:Thm1_rank1}
    \begin{split}
        H_{(\mu,\pi)}(Y_{n:l}|Z_n) &= H_{(\mu^\prime,\pi)}(Y_{n:l}|Z_n) \\
        &\geq H_{(\mu^\prime,\pi)}(Y_{n:l}|Z_n,W_n).
    \end{split}
\end{equation}
This inequality corresponds to Eq.~(\ref{Seq:Thm1}), as the set of labels $(Z_n,W_n)$ represents the measurement outcomes of the POVM $\mu^\prime$.

Finally, we show that for any non-extremal POVM $\mu$, there exists a corresponding extremal POVM $\mu^\prime$, which satisfies Eq.~(\ref{Seq:Thm1}). 
A non-extremal POVM $\mu$ can be expressed as a convex combination of extremal POVMs $\{\mu^k\}_k$, formulated as $\mu = \sum_k p_k \mu^k$ \cite{d2005classical,haapasalo2012quantum}. In this case, the relationship 
\begin{math}
    E_{Z_{n}} = \sum_{k} p_k E_{Z_{n}}^k
\end{math}
is satisfied, where $\{E_{Z_{n}}\}_{Z_{n}}$ and $\{E_{Z_{n}}^k\}_{Z_{n}}$ represent the sets of POVM operators for $\mu$ and $\mu^k$ , respectively.
Since the equality 
\begin{math}
    P_{(\mu,\pi)} [Z_n,Y_{n:l}] = \sum_k p_k P_{(\mu^k,\pi)} [Z_n,Y_{n:l}]
\end{math}
directly follows from this relationship, we can derive the following inequality:
\begin{equation}
\label{Seq:Thm1_extremal}
    \begin{split}
        H_{(\mu,\pi)}(Y_{n:l}|Z_n) &\geq \sum_k p_k H_{(\mu^k,\pi)}(Y_{n:l}|Z_n) \\
        &\geq \min_k H_{(\mu^k,\pi)}(Y_{n:l}|Z_n).
    \end{split}
\end{equation}

By combining Eqs.~(\ref{Seq:Thm1_row}), (\ref{Seq:Thm1_rank1}), and (\ref{Seq:Thm1_extremal}), we conclude that for an arbitrary POVM $\mu \in \mathcal{E}_{n}$, there exists a rank-one extremal POVM on $\mathrm{Row}[\rho_{1,\dots,n}]$, $\mu^\prime \in \in \mathcal{E}_{n}^\prime$, which satisfies Eq.~(\ref{Seq:Thm1}).
\end{proof}

\section{Continuity of the lower and upper bounds with respect to the quantum state}
In this section, we show the continuity of the lower and upper bounds $L_l$, $U_l$, $L_l^\to$, $U_l^\to$, $L_l^G$, and $U_l^G$, with respect to the quantum state. For simplicity, we here focus on the bounds $U_l$ and $L_l$, although the proof can be directly extended to the other bounds.
By explicitly showing the dependence on the quantum state $\rho$, the bounds $U_l$ and $L_l$ can be expressed as follows:
\begin{align}
    U_l (\rho) &\equiv \frac{1}{N} \left\{\sum_{n \in I_l} \min_{\Lambda \in \mathcal{M}_{n:l}} H_{\Lambda;\rho} (Y_{n:l}) - S(\rho)\right\}, \\
    L_l (\rho) &\equiv \frac{1}{N} \left\{\sum_{n \in I_l} \min_{\Lambda \in \mathcal{N}_{n:l}} H_{\Lambda;\rho} (Y_{n:l}|Z_{n}) - S(\rho)\right\}.
\end{align}
Here, $H_{\Lambda;\rho} (Y_{n:l})$ represents the Shannon entropy of the measurement outcomes $Y_{n:l}$ when the measurement $\Lambda \in \mathcal{M}_{n:l}$ is performed on the quantum state $\rho$. Similarly, $H_{\Lambda;\rho} (Y_{n:l}|Z_{n})$ denotes the conditional Shannon entropy when the measurement $\Lambda \in \mathcal{N}_{n:l}$ is performed on the state $\rho$. 
The dependence on the state $\rho$ can be abbreviated in the obvious cases, as has been done in the main text.
For these bounds, we can prove the following theorem:

\begin{theorem} (restatement of Theorem \blue{2} in Appendix~\blue{D3}) \label{Sthm:cont_UlLl}
Let $\rho$ and $\rho^\prime$ be $d$-dimensional $N$-partite states, which satisfy
\begin{math}
    \|\rho-\rho^\prime\|_{\rm tr} \leq \nu \leq \frac{1}{2},
\end{math}
where $\|X\|_{\rm tr}\equiv\tr[|X|]$ denotes the trace norm.
Then, the following inequalities hold: 
\begin{align}
    |U_l(\rho) - U_l(\rho^\prime)| &\leq \frac{\nu}{2} \ln d  + \frac{h(\frac{\nu}{2})}{l} = \Tilde{\mathcal{O}} (\nu), \label{Seq:U^phi_approx} \\
    |L_l(\rho) - L_l(\rho^\prime)| &\leq 4\nu \ln d  + \frac{2h(\nu)}{l} = \Tilde{\mathcal{O}} (\nu), \label{Seq:L^phi_approx}
\end{align}
where $h(p) \equiv -p \ln p -(1-p)\ln (1-p)$ is the binary entropy function and $\Tilde{\mathcal{O}} (x)$ represents $\mathcal{O} (x \ln x)$.
\end{theorem}

\begin{proof}
By utilizing Lemma~\ref{Slem:cont_ent}, which is derived later in this section, we can prove Eq.~(\ref{Seq:U^phi_approx}) as follows:
\begin{equation}
    \begin{split}
        |U_l(\rho) -U_l(\rho^\prime)| &\leq \frac{1}{N} \sum_{n \in I_l} \left| \min_{\Lambda \in \mathcal{M}_{n:l}} H_{\Lambda;\rho} (Y_{n:l}) - \min_{\Lambda \in \mathcal{M}_{n:l}} H_{\Lambda;\rho^\prime} (Y_{n:l}) \right| \\
        &\leq \frac{\nu}{2} \ln d  + \frac{h(\frac{\nu}{2})}{l},
    \end{split}
\end{equation}
where the inequality in the first line follows from the triangle inequality, and the inequality in the second line is the consequence of Lemma~\ref{Slem:cont_ent}.

Similarly, by utilizing Lemma~\ref{Slem:cont_cond} discussed later, we can derive Eq.~(\ref{Seq:L^phi_approx}) as follows:
\begin{equation}
    \begin{split}
        |L_l(\rho) -L_l(\rho^\prime)| &\leq \frac{1}{N} \sum_{n \in I_l} \left| \min_{\Lambda \in \mathcal{N}_{n:l}} H_{\Lambda;\rho} (Y_{n:l}|Z_{n}) - \min_{\Lambda \in \mathcal{N}_{n:l}} H_{\Lambda;\rho^\prime} (Y_{n:l}|Z_{n})\right| \\
        &\leq 4 \nu \ln d + \frac{2h(\nu)}{l},
    \end{split}
\end{equation}
where the inequality in the first line follows from the triangle inequality, and the inequality in the second line is the consequence of Lemma~\ref{Slem:cont_cond}.
\end{proof}

We finish this section by providing the statements and proofs of Lemma~\ref{Slem:cont_ent} and~\ref{Slem:cont_cond} used to prove Theorem~\ref{Sthm:cont_UlLl}.
\begin{lemma} \label{Slem:cont_ent}
Let $\rho$ and $\rho^\prime$ be $d$-dimensional $N$-partite states, which satisfy
\begin{math}
    \|\rho-\rho^\prime\|_{\rm tr} \leq \nu \leq \frac{1}{2}.
\end{math}
Then, the following inequality is satisfied: 
\begin{equation}
\left|\min_{\Lambda \in \mathcal{M}_{n:l}} H_{\Lambda;\rho} (Y_{n:l}) - \min_{\Lambda \in \mathcal{M}_{n:l}} H_{\Lambda;\rho^\prime} (Y_{n:l})\right| \leq f_l(\nu) = \Tilde{\mathcal{O}} (\nu), \label{Seq:cond_sh_approx}
\end{equation}
where $f_l$ is defined as $f_l(\nu) \equiv  \frac{l\nu}{2} \ln d  + h \left(\frac{\nu}{2}\right)$.
\end{lemma}

\begin{proof}
We here suppose $\min_{\Lambda \in \mathcal{M}_{n:l}} H_{\Lambda;\rho} (Y_{n:l}) - \min_{\Lambda \in \mathcal{M}_{n:l}} H_{\Lambda;\rho^\prime} (Y_{n:l}) \geq 0$, without loss of generality. By denoting the measurement protocol which realizes the minimum of $\min_{\Lambda \in \mathcal{M}_{n:l}} H_{\Lambda;\rho} (Y_{n:l})$ as  $\Lambda^*\in \mathcal{M}_{n:l}$, we can derive the following inequality:
\begin{equation}
\begin{split}
\left|\min_{\Lambda \in \mathcal{M}_{n:l}} H_{\Lambda;\rho} (Y_{n:l}) - \min_{\Lambda \in \mathcal{M}_{n:l}} H_{\Lambda;\rho^\prime} (Y_{n:l})\right|
&\leq H_{\Lambda^*;\rho} (Y_{n:l}) - H_{\Lambda^*;\rho^\prime} (Y_{n:l}) \\
&= S(\mathcal{D}_{\Lambda^*}(\rho))-S(\mathcal{D}_{\Lambda^*}(\rho^\prime)).
\end{split} 
\end{equation}
Here, $\mathcal{D}_{\Lambda^*}$ is the CPTP (completely positive and trace preserving) map corresponding to the measurement protocol $\Lambda^* \in \mathcal{M}_{n:l}$, which is defined as follows:
\begin{equation*}
    \mathcal{D}_{\Lambda^*}(\rho)  \equiv \sum_{Y_{n:l}} \tr[\Pi_{y_{n+1}}\otimes \Pi_{y_{n+2}}\dots \otimes \Pi_{y_{n+l}}\rho] \Pi_{y_{n+1}}\otimes \Pi_{y_{n+2}}\dots \otimes \Pi_{y_{n+l}},
\end{equation*}
where $\{\Pi_{y_{n+1}}\otimes \Pi_{y_{n+2}}\dots \otimes \Pi_{y_{n+l}}\}_{Y_{n:l}}$ is the set of measurement operators for $\Lambda^*$.

Therefore, by introducing the notation $\nu^\prime \equiv \|\mathcal{D}_{\Lambda^*}(\rho)-\mathcal{D}_{\Lambda^*}(\rho^\prime)\|_{\rm tr}$, we can derive the following inequality:
\begin{equation}
\begin{split}
\left|\min_{\Lambda \in \mathcal{M}_{n:l}} H_{\Lambda;\rho} (Y_{n:l}) - \min_{\Lambda \in \mathcal{M}_{n:l}} H_{\Lambda;\rho^\prime} (Y_{n:l})\right|
&\leq S(\mathcal{D}_{\Lambda^*}(\rho))-S(\mathcal{D}_{\Lambda^*}(\rho^\prime)) \\
&\leq f_l (\nu^\prime) \\
&\leq f_l (\nu) = \Tilde{\mathcal{O}} (\nu).
\end{split} 
\end{equation}
Here, the inequality in the second line directly follows from the Audenaert-Fannes inequality \cite{audenaert2007sharp}, which states the continuity of the von Neumann entropy $S(\rho)$ with respect to the quantum state $\rho$.  
The third line is derived from the fact that $f_l(x)$ is the monotonically increasing function in the range $x\leq 1$ and the inequality $\nu^\prime \leq \nu \leq\frac{1}{2}$. 
This inequality is derived as $\nu^\prime \equiv \|\mathcal{D}_{\Lambda^*}(\rho)-\mathcal{D}_{\Lambda^*}(\rho^\prime)\|_{\rm tr} \leq \|\rho-\rho^\prime\|_{\rm tr} \leq \nu$ by using the monotonicity of the trace distance under the CPTP map $\mathcal{D}_{\Lambda^*}$.
\end{proof}

\begin{lemma} \label{Slem:cont_cond}
Let $\rho$ and $\rho^\prime$ be $d$-dimensional $N$-partite states, which satisfy
\begin{math}
\|\rho-\rho^\prime\|_{\rm tr} \leq \nu \leq \frac{1}{2}.
\end{math}
Then, the following inequality is satisfied: 
\begin{equation}
\left|\min_{\Lambda \in \mathcal{N}_{n:l}} H_{\Lambda;\rho} (Y_{n:l}|Z_{n}) - \min_{\Lambda \in \mathcal{N}_{n:l}} H_{\Lambda;\rho^\prime} (Y_{n:l}|Z_{n})\right| \leq g_l(\nu) = \Tilde{\mathcal{O}} (\nu), \label{Seq:cond_sh_approx}
\end{equation}
where $g_l$ is defined as $g_l(\nu) \equiv  4l \nu \ln d  + 2h(\nu)$.
\end{lemma}

\begin{proof}
We here suppose $\min_{\Lambda \in \mathcal{N}_{n:l}} H_{\Lambda;\rho} (Y_{n:l}|Z_{n}) -\min_{\Lambda \in \mathcal{N}_{n:l}} H_{\Lambda;\rho^\prime} (Y_{n:l}|Z_{n}) \geq 0$, without loss of generality. By denoting the optimal measurement protocol for the state $\rho$ as $\Lambda^*\in \mathcal{N}_{n:l}$, we can derive the following inequality:
\begin{equation}
\begin{split}
\left|\min_{\Lambda \in \mathcal{N}_{n:l}} H_{\Lambda;\rho} (Y_{n:l}|Z_{n}) - \min_{\Lambda \in \mathcal{N}_{n:l}} H_{\Lambda;\rho^\prime} (Y_{n:l}|Z_{n})\right|
&\leq H_{\Lambda^*;\rho} (Y_{n:l}|Z_{n}) - H_{\Lambda^*;\rho^\prime} (Y_{n:l}|Z_{n}) \\
&= S_{n+1,\dots,n+l|1,\dots,n}(\mathcal{D}_{\Lambda^*}(\rho))-S_{n+1,\dots,n+l|1,\dots,n}(\mathcal{D}_{\Lambda^*}(\rho^\prime)).
\end{split} 
\end{equation}
Here, $S_{2|1}(\rho) \equiv S(\rho_{1,2}) -S(\rho_1)$ represents the conditional von Neumann entropy and $\mathcal{D}_{\Lambda^*}$ is the CPTP map corresponding to the measurement protocol $\Lambda^* \in \mathcal{N}_{n:l}$, which is defined as follows:
\begin{equation*}
    \mathcal{D}_{\Lambda^*}(\rho)  \equiv \sum_{Z_{n},Y_{n:l}} \tr[E_{Z_{n}} \otimes \Pi_{y_{n+1}}\dots \otimes \Pi_{y_{n+l}}\rho] \ \Pi_{Z_{n}} \otimes \Pi_{y_{n+1}}\dots \otimes \Pi_{y_{n+l}},
\end{equation*}
where $\{E_{Z_{n}} \otimes \Pi_{y_{n+1}}\dots \otimes \Pi_{y_{n+l}}\}_{Z_{n},Y_{n:l}}$ is the set of measurement operators for $\Lambda^*$, and $\{\Pi_{Z_{n}}\}_{Z_n}$ is the set of orthogonal rank-one projectors on $\{a_i\}_{i=1}^n$.

Therefore, by introducing the notation $\nu^\prime \equiv \|\mathcal{D}_{\Lambda^*}(\rho)-\mathcal{D}_{\Lambda^*}(\rho^\prime)\|_{\rm tr}$, we can derive the following inequality:
\begin{equation}
\begin{split}
\left|\min_{\Lambda \in \mathcal{N}_{n:l}} H_{\Lambda;\rho} (Y_{n:l}|Z_{n}) - \min_{\Lambda \in \mathcal{N}_{n:l}} H_{\Lambda;\rho^\prime} (Y_{n:l}|Z_{n})\right|
&\leq S_{n+1,\dots,n+l|1,\dots,n}(\mathcal{D}_{\Lambda^*}(\rho))-S_{n+1,\dots,n+l|1,\dots,n}(\mathcal{D}_{\Lambda^*}(\rho^\prime)) \\
&\leq  g_l(\nu^\prime) \\
&\leq g_l(\nu) = \Tilde{\mathcal{O}} (\nu).
\end{split} 
\end{equation}
Here, the inequality in the second line follows from the Alicki-Fannes inequality \cite{alicki2004continuity}, which states that conditional von Neumann entropy $S_{2|1}(\rho)$ is continuous with respect to the quantum state $\rho$. 
The third line follows from the fact that $g_l(x)$ is monotonically increasing function in $x\leq\frac{1}{2}$ and the inequality $\nu^\prime \leq \nu \leq\frac{1}{2}$. This inequality is derived as $\nu^\prime \equiv \|\mathcal{D}_{\Lambda^*}(\rho)-\mathcal{D}_{\Lambda^*}(\rho^\prime)\|_{\rm tr} \leq \|\rho-\rho^\prime\|_{\rm tr} \leq \nu$ by using the monotonicity of the trace distance under the CPTP map $\mathcal{D}_{\Lambda^*}$.
\end{proof}

\section{Lower and upper bounds in thermodynamic limit} \label{Ss:inf_MPS}
In this section, we discuss the lower and upper bounds $L_l$, $U_l$, $L_l^\to$, $U_l^\to$, $L_l^G$, and $U_l^G$ in the thermodynamic limit.
Since the numerical costs for calculating these bounds grow linearly with system size $N$, their calculations might appear impossible in the thermodynamic limit, in the naive expectation.
However, we here show that for translationally invariant normal MPS, these bounds can be calculated with constant costs in the limit of $N\to \infty$.
We note that in this section, we focus on the bounds $L_l$ and $U_l$, although the discussion can be directly extended to the other bounds as well.

In Sec.~\ref{Sss:approx_fixed_point}, we review the concept of normal MPS. In Sec.~\ref{Sss:red_bounds_normalMPS}, we prove the main theorem of this section, Theorem~\ref{SThm3:thermo_limit}, which implies that $L_l$ and $U_l$ can be calculated with constant costs for normal MPS in the thermodynamic limit.
While Secs.~\ref{Sss:approx_fixed_point} and~\ref{Sss:red_bounds_normalMPS} focus on open boundary conditions, we show that Theorem~\ref{SThm3:thermo_limit} can be extended to cases with periodic boundary conditions in Sec.~\ref{Sss:PBC_inf_normal}.

\subsection{Approximation of normal MPS with the fixed point tensor} \label{Sss:approx_fixed_point}
\begin{figure*}[]
    \centering
    \includegraphics[width=1\textwidth]{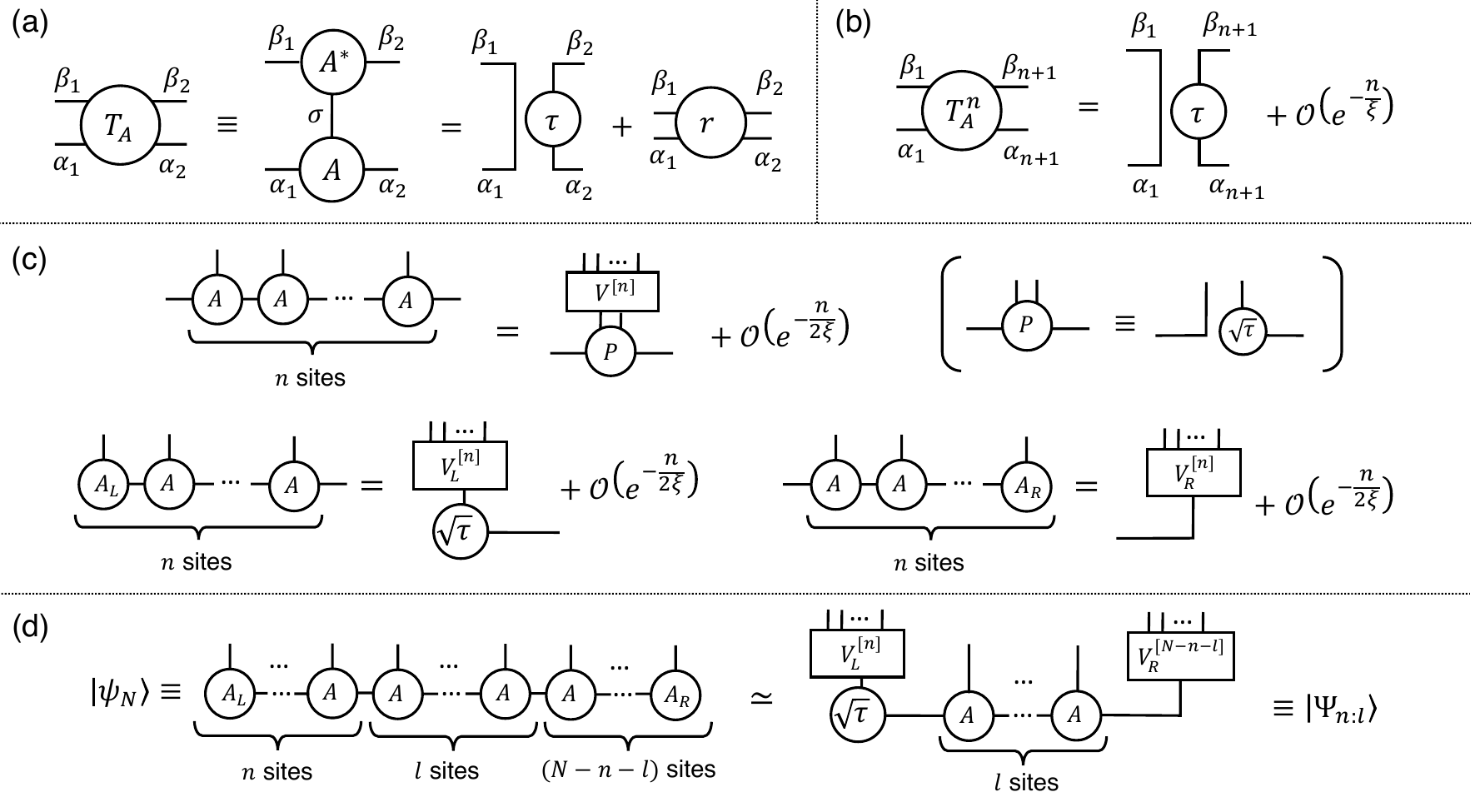}
    \caption{Tensor network diagrams for normal MPS and corresponding transfer matrix.
    (a) Transfer matrix $T_A$ for tensor $A$. When the MPS generated with tensor $A$ is normal and in the right-canonical form, the unique maximum eigenvalue of the transfer matrix $T_A$ becomes $\lambda_1 = 1$, and its corresponding right and left eigenvectors are $\mathbbm{1}_{D_B}$ and $\tau$, respectively. Here, $\tau$ is positive definite Hermitian matrix with trace $1$. The tensor $r$ represents the remaining Jordan blocks of $T_A$, and its spectral radius $|\lambda_2|$ is less than $1$. 
    (b) The $n$-fold product of the transfer matrix $T_A$ approaches the fixed point $\mathbbm{1}_{D_B} \otimes \tau$ by increasing $n$. The size of the remaining part becomes $\mathcal{O}(e^{-\frac{n}{\xi}})$, where $\xi \equiv (-\ln |\lambda_2|)^{-1}$ is the correlation length of this MPS. 
    (c) Approximation of $n$-site tensor with renormalization fixed point tensor $P \equiv \mathbbm{1}_{D_B} \otimes \sqrt{\tau}$. Here, $V^{[n]}$, $V_L^{[n]}$, and $V_R^{[n]}$ are the isometries from dimension $D_B^2$ (or $D_B$) to $d^n$.
    (d) Approximation of an $N$-site normal MPS $\ket{\psi_N}$ with the state $\ket{\Psi_{n:l}}$. The tensors for $n$-site segment $\{a_i\}_{i=1}^n$ and $(N-n-l)$-site segment $\{a_i\}_{i=n+l+1}^N$ are approximated by utilizing the relationships in (c).
    }
    \label{SMfig:coarse_basic}
\end{figure*}

In this subsection, we review the definition and properties of normal MPS following the previous works such as Refs.~\cite{perez2006matrix,cirac2017matrix,cirac2021matrix,piroli2021quantum,malz2024preparation}.
We consider an $N$-site MPS defined as 
\begin{equation}
    \ket{\psi_N} \equiv \sum_{\sigma_1,\dots,\sigma_N} A_L^{\sigma_1} A^{\sigma_2} \dots A^{\sigma_{N-1}} A_R^{\sigma_N} \ket{\sigma_1,\sigma_2,\dots \sigma_{N-1},\sigma_N},
\end{equation}
where $\{A^{\sigma}\}_{\sigma}$ is a set of $D_B$-dimensional square matrices, $\{A_L^{\sigma}\}_\sigma$ is a set of $D_B$-dimensional row vectors, and $\{A_R^{\sigma}\}_\sigma$ is a set of $D_B$-dimensional column vectors.
To fix the redundancy of gauge degree of freedom, we consider the right-canonical form, where the condition $\sum_{\sigma} A^\sigma A^{\sigma \dag} = \mathbbm{1}_{D_B}$ is satisfied. The normal MPS we address here is the class of MPS that satisfies the following two conditions: $(\mathrm{i})$ it is irreducible (i.e., $A^\sigma$ does not have a nontrivial common invariant subspace), $(\mathrm{ii})$ the transfer matrix $T_A$ has a unique largest eigenvalue $\lambda_1 = 1$. The transfer matrix of tensor $A^\sigma$ is defined as
\begin{equation}
    [T_A]_{(\alpha_1,\beta_1);(\alpha_2,\beta_2)} \equiv \sum_{\sigma} (A^{\sigma}_{\beta_1,\beta_2})^* A^{\sigma}_{\alpha_1,\alpha_2},
\end{equation}
whose diagram is depicted in Fig.~\ref{SMfig:coarse_basic}(a). Due to the above-mentioned conditions, the transfer matrix for normal MPS can be described as \cite{perez2006matrix}
\begin{equation}
\label{Seq:transfer_matrix}
    [T_A]_{(\alpha_1,\beta_1);(\alpha_2,\beta_2)} = \delta_{\alpha_1,\beta_1} \tau_{\alpha_2,\beta_2} + r_{(\alpha_1,\beta_1);(\alpha_2,\beta_2)},
\end{equation}
where $\delta$ and $\tau$ are the leading right and left eigenvectors associated with $\lambda_1 = 1$, respectively, and $r$ is the remaining Jordan blocks associated with the subleading eigenvalues. 
It is also noteworthy that when we regard tensor $\tau_{\alpha_2,\beta_2}$ as a $D_B$-dimensional square matrix, it is shown to be always Hermitian and positive definite.

We next consider the situation where multiple sites are blocked. Since the transfer matrix $T_A$ is represented as Eq.~(\ref{Seq:transfer_matrix}), its $n$-th power is described as follows, as depicted in Fig.~\ref{SMfig:coarse_basic}(b) \cite{cirac2017matrix,cirac2021matrix}:
\begin{equation}
\label{Seq:TE_n_fold_approx}
    [T_A^n]_{(\alpha_1,\beta_1);(\alpha_{n+1},\beta_{n+1})} = \delta_{\alpha_1,\beta_1} \tau_{\alpha_{n+1},\beta_{n+1}} + \mathcal{O}(e^{-\frac{n}{\xi}}),
\end{equation}
where the correlation length $\xi$ is defined as $\xi \equiv \frac{1}{\ln |\lambda_2|}$, and $\lambda_2$ is the second largest eigenvalue of $T_A$.
Furthermore, it can be shown from Eq.~(\ref{Seq:TE_n_fold_approx}) that the tensors of $n$-site segments are approximated as shown in Fig.~\ref{SMfig:coarse_basic}(c) \cite{piroli2021quantum,malz2024preparation}. 
Here, $V^{[n]}$, $V_L^{[n]}$, and $V_R^{[n]}$ are isometries mapping $D_B^2$(or $D_B$)-dimensional virtual system to $d^n$-dimensional physical system, satisfying $V^{[n]\dag} V^{[n]} = \mathbbm{1}_{D_B^2}$, $V_L^{[n]\dag} V_L^{[n]} = \mathbbm{1}_{D_B}$, and $V_R^{[n]\dag} V_R^{[n]} = \mathbbm{1}_{D_B}$, respectively.
The tensor $P \equiv \mathbbm{1}_{D_B} \otimes \sqrt{\tau}$ is called renormalization fixed point tensor.
By utilizing these relationships in Fig.~\ref{SMfig:coarse_basic}(c), the following lemma can be derived:
\begin{lemma}\label{Slem:approx_normalMPS}
For an $N$-site MPS with open boundary condition generated by normal tensor $A$, denoted as $\ket{\psi_N}$, the following inequality is satisfied:
\begin{equation}
\label{SMeq:approx_Psi_nl}
    \gamma_{n:l} \equiv \|\ket{\psi_N}\bra{\psi_N} - \ket{\Psi_{n:l}}\bra{\Psi_{n:l}}\|_{\rm tr} \leq \mathcal{O}(e^{-\frac{n}{2\xi}}) + \mathcal{O}(e^{-\frac{N-n}{2\xi}}).
\end{equation}
Here, $\ket{\Psi_{n:l}}$ is the state defined as shown in Fig.~\ref{SMfig:coarse_basic}(d). It can be represented as $\ket{\Psi_{n:l}} = V_L^{[n]}\otimes \mathbbm{1}_{d^l} \otimes V_R^{[N-n-l]} \ket{\Psi^l}$, where $\ket{\Psi^l}$ is defined as Fig.~\ref{SMfig:inf_MPS_approx}(a), and $V_L^{[n]}$ and $V_R^{[N-n-l]}$ are the isometries from dimension $D_B$ to $d^n$ and $d^{N-n-l}$, respectively.
\end{lemma}
We abbreviate the proof for this lemma, since essentially the same discussion can be found e.g., in Refs.~\cite{piroli2021quantum,malz2024preparation}.
The essence of the proof is the approximations shown in Fig.~\ref{SMfig:coarse_basic}(c).

\subsection{Convergence of the lower and upper bounds to calculable quantities in thermodynamic limit} \label{Sss:red_bounds_normalMPS}
\begin{figure*}[]
    \centering
    \includegraphics[width=1\textwidth]{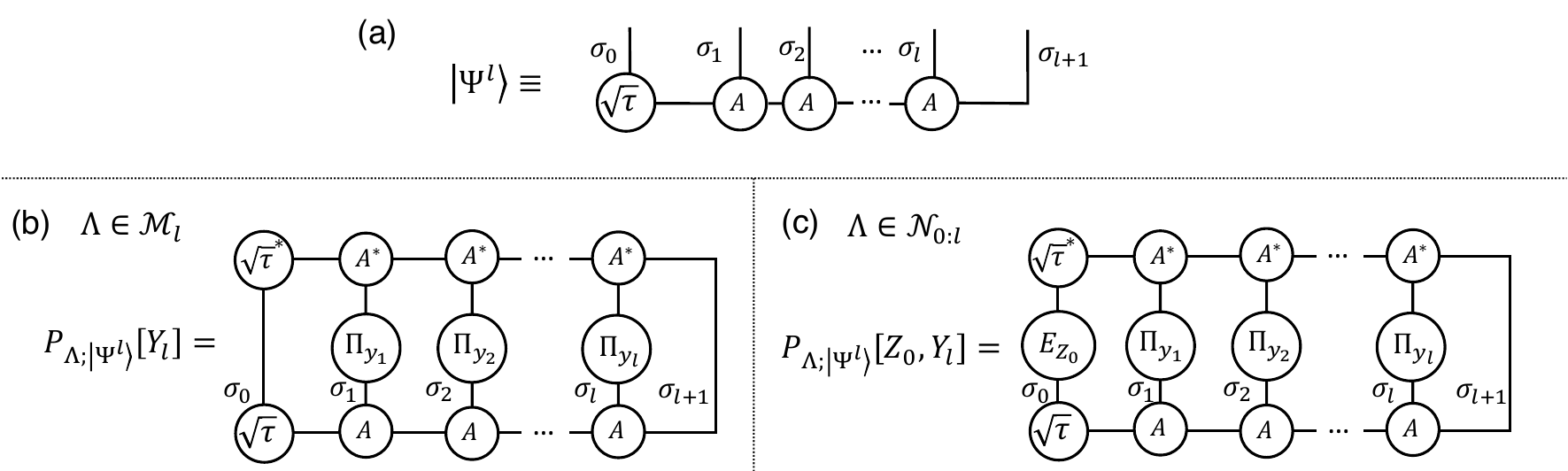}
    \caption{Calculations of $\overline{U}_l(\ket{\Psi^l})$ and $\overline{L}_l(\ket{\Psi^l})$ (i.e., Eqs.~(\ref{SMeq:U_l^*}) and (\ref{SMeq:L_l^*})). 
    (a) Diagram for the state $\ket{\Psi^l}$. 
    (b) Diagram for the probability $P_{\Lambda;\ket{\Psi^l}}[Y_l]$ with $\Lambda \in \mathcal{M}_l$. We can calculate $\overline{U}_l(\ket{\Psi^l})$ from this probability distribution. The tensor $\Pi_{y_n}$ represents the rank-one projector on the index $\sigma_n$.
    (c) Diagram for the probability $P_{\Lambda;\ket{\Psi^l}}[Z_0,Y_l]$ with $\Lambda \in \mathcal{N}_{0:l}$. We can calculate $\overline{L}_l(\ket{\Psi^l})$ from this probability distribution. In this diagram, $E_{Z_0}$ represents a rank-one POVM operator on the index $\sigma_0$, and $Z_0$ is the corresponding measurement outcome.}
    \label{SMfig:inf_MPS_approx}
\end{figure*}

In this section, we prove Theorem~\ref{SThm3:thermo_limit}, which states that for translationally invariant normal MPS, the lower and upper bounds $L_l$ and $U_l$ converge to the quantities $\overline{L}_l$ and $\overline{U}_l$ (defined later), respectively, in the thermodynamic limit. 
Both of these limits $\overline{L}_l$ and $\overline{U}_l$ are calculated with costs of $e^{\mathcal{O}(l)}$, thereby eliminating the numerical cost divergence of $L_l$ and $U_l$ in the naive expectation.
The precise statement of Theorem~\ref{SThm3:thermo_limit} is as follows:
\begin{theorem} \label{SThm3:thermo_limit}
Let $\ket{\psi_N}$ be an $N$-site state under open boundary condition described with translationally invariant normal MPS, and $U_l(\ket{\psi_N})$ and $L_l(\ket{\psi_N})$ be the upper and lower bounds for this state represented as 
\begin{align*}
    U_l (\ket{\psi_N}) &= \frac{1}{N}\sum_{n \in I_l} \min_{\Lambda \in \mathcal{M}_{n:l}} H_{\Lambda;\ket{\psi_N}} (Y_{n:l}) , \\
    L_l (\ket{\psi_N}) &= \frac{1}{N}\sum_{n \in I_l} \min_{\Lambda \in \mathcal{N}_{n:l}} H_{\Lambda;\ket{\psi_N}} (Y_{n:l}|Z_{n}).
\end{align*}
Then, for any infinitesimally small constant $\eta > 0$ there exists a certain $N_\eta$ such that for all $N\geq N_\eta$ the following inequalities hold:
\begin{align}
    \left| U_l (\ket{\psi_N}) - \overline{U}_l(\ket{\Psi^l}) \right| &\leq \eta, \label{Seq:Thm3_U} \\
    \left| L_l (\ket{\psi_N}) - \overline{L}_l(\ket{\Psi^l}) \right| &\leq \eta, \label{Seq:Thm3_L}
\end{align}
where $\overline{U}_l(\ket{\Psi^l})$ and $\overline{L}_l(\ket{\Psi^l})$ are defined as follows:
\begin{align}
    \overline{U}_l(\ket{\Psi^l}) &\equiv \frac{1}{l} \min_{\Lambda \in \mathcal{M}_l} H_{\Lambda;\ket{\Psi^l}}(Y_l), \label{SMeq:U_l^*}\\
    \overline{L}_l(\ket{\Psi^l}) &\equiv \frac{1}{l} \min_{\Lambda \in \mathcal{N}_{0:l}} H_{\Lambda;\ket{\Psi^l}}(Y_l|Z_0).\label{SMeq:L_l^*}
\end{align}
Here, the state $\ket{\Psi^l}$ is defined as Fig.~\ref{SMfig:inf_MPS_approx}(a), $\mathcal{M}_l$ represents the set of measurement protocols where the projective measurements are performed on the indices $\{\sigma_i\}_{i=1}^l$ as shown in Fig.~\ref{SMfig:inf_MPS_approx}(b), $\mathcal{N}_{0:l}$ denotes the set of measurement protocols where a POVM is performed on the index $\sigma_0$ and projective measurements are performed on the indices $\{\sigma_i\}_{i=1}^l$ as shown in Fig.~\ref{SMfig:inf_MPS_approx}(c), and $Z_0$ is defined as the measurement outcome of the POVM on $\sigma_0$.
\end{theorem}

\begin{proof}
We first prove the convergence of the upper bound, described as Eq.~(\ref{Seq:Thm3_U}).
Since the state $\ket{\Psi^l}$ is equivalent to $\ket{\Psi_{n:l}}$ up to the unitary transformation $V_L^{[n]}\otimes \mathbbm{1}_{d^l} \otimes V_R^{[N-n-l]}$, as shown in Fig.~\ref{SMfig:coarse_basic}(d), $\overline{U}_l(\ket{\Psi^l})$ is reformulated as follows:
\begin{equation}
\label{Seq:Thm3_Ubar_equiv}
    \overline{U}_l(\ket{\Psi^l}) \equiv \frac{1}{l} \min_{\Lambda \in \mathcal{M}_l} H_{\Lambda;\ket{\Psi^l}}(Y_l) = \frac{1}{N} \sum_{n \in I_l} \min_{\Lambda \in \mathcal{M}_{n:l}} H_{\Lambda;\ket{\Psi_{n:l}}} (Y_{n:l}).
\end{equation}
By utilizing Eq.~(\ref{Seq:Thm3_Ubar_equiv}) and Lemma~\ref{Slem:cont_ent}, we can derive the following inequality:
\begin{equation}
\label{Seq:Thm3_prf_U}
    \begin{split}
        \left| U_l (\ket{\psi_N}) - \overline{U}_l(\ket{\Psi^l}) \right| &= \frac{1}{N} \left|\sum_{n \in I_l} \min_{\Lambda \in \mathcal{M}_{n:l}} H_{\Lambda;\ket{\psi_N}} (Y_{n:l}) - \min_{\Lambda \in \mathcal{M}_{n:l}} H_{\Lambda;\ket{\Psi_{n:l}}} (Y_{n:l}) \right| \\
        &\leq \frac{1}{N} \sum_{n \in I_l} \left|\min_{\Lambda \in \mathcal{M}_{n:l}} H_{\Lambda;\ket{\psi_N}} (Y_{n:l}) - \min_{\Lambda \in \mathcal{M}_{n:l}} H_{\Lambda;\ket{\Psi_{n:l}}} (Y_{n:l}) \right| \\
        &\leq \frac{1}{N} \sum_{n \in I_l} f_l(\gamma_{n:l}).
    \end{split}
\end{equation}
The equality in the first line follows from Eq.~(\ref{Seq:Thm3_Ubar_equiv}), the inequality in the second line follows from the triangle inequality, and the inequality in the third line is the consequence of Lemma~\ref{Slem:cont_ent}. Here, the function $f_l$ is defined as $f_l(x) \equiv \frac{x}{2} l \ln d + h\left(\frac{x}{2}\right)$.

We next derive Eq.~(\ref{Seq:Thm3_U}) by utilizing Eq.~(\ref{Seq:Thm3_prf_U}) and Lemma~\ref{Slem:approx_normalMPS}.
Since the size of $\gamma_{n:l}$ is $\mathcal{O}\left(\exp\left(-\frac{\min \{n, N-n\} }{2\xi}\right)\right)$ from Lemma~\ref{Slem:approx_normalMPS}, the size of $f_l(\gamma_{n:l})$ becomes $\Tilde{\mathcal{O}}\left(\exp\left(-\frac{\min \{n, N-n\} }{2\xi}\right)\right)$. 
Therefore, we can take a constant $n_{\eta}$, which is determined by $l$, $\eta$ and $\xi$ and does not depend on $N$, such that for arbitrary $n$ in $n_{\eta} \leq n \leq N- n_{\eta}$, the inequality $f_l(\gamma_{n:l}) \leq \frac{l\eta}{2}$ is satisfied.
By combining this fact with Eq.~(\ref{Seq:Thm3_prf_U}), we can derive the following inequality:
\begin{equation}
\label{SMeq:TDL_UB_final}
\begin{split}
     \left| U_l (\ket{\psi_N}) - \overline{U}_l(\ket{\Psi^l}) \right| & \leq \frac{1}{N} \sum_{n \in I_l} f_l(\gamma_{n:l}) \\
     &= \frac{1}{N} \left\{ \sum_{\substack{n \in I_l, \\ n < n_\eta}} f_l(\gamma_{n:l}) + \sum_{\substack{n \in I_l, \\ n > N- n_\eta}} f_l(\gamma_{n:l})\right\} + \frac{1}{N}\sum_{\substack{n \in I_l, \\ n_\eta \leq n \leq N- n_\eta}} f_l(\gamma_{n:l}) \\
     & \leq \frac{2n_\eta}{Nl} C + \frac{\eta}{2},
\end{split} 
\end{equation}
where $C$ is a constant that satisfies $f_l(x) \leq C$ for arbitrary $x$.
Here, we divide the summation of $f_l(\gamma_{n:l})$ over $n\in I_l$ into the edge part (i.e., $n < n_\eta$ and $n > N- n_\eta$) and the bulk part (i.e., $ n_\eta \leq n \leq N- n_\eta$) in the second line, and upper-bound the both parts in the third line.
From Eq.~(\ref{SMeq:TDL_UB_final}), we can derive Eq.~(\ref{Seq:Thm3_U}) for any $N\geq N_\eta$ with $N_\eta \equiv \frac{4 C n_\eta}{l\eta}$.

The derivation of Eq.~(\ref{Seq:Thm3_L}) can be performed in essentially the same way as that of Eq.~(\ref{Seq:Thm3_U}).
Since $\overline{L}_l(\ket{\Psi^l})$ can be reformulated as
\begin{equation}
\label{Seq:Thm3_Lbar_equiv}
    \overline{L}_l(\ket{\Psi^l}) \equiv \frac{1}{l} \min_{\Lambda \in \mathcal{N}_{0:l}} H_{\Lambda;\ket{\Psi^l}}(Y_l|Z_0) = \frac{1}{N} \sum_{n \in I_l} \min_{\Lambda \in \mathcal{N}_{n:l}} H_{\Lambda;\ket{\Psi_{n:l}}} (Y_{n:l}|Z_n),
\end{equation}
we can derive the following inequality:
\begin{equation}
\label{Seq:Thm3_prf_L}
    \begin{split}
        \left| L_l (\ket{\psi_N}) - \overline{L}_l(\ket{\Psi^l}) \right| &= \frac{1}{N} \left|\sum_{n \in I_l} \min_{\Lambda \in \mathcal{N}_{n:l}} H_{\Lambda;\ket{\psi_N}} (Y_{n:l}|Z_n) - \min_{\Lambda \in \mathcal{N}_{n:l}} H_{\Lambda;\ket{\Psi_{n:l}}} (Y_{n:l}|Z_n) \right| \\
        &\leq \frac{1}{N} \sum_{n \in I_l} \left|\min_{\Lambda \in \mathcal{N}_{n:l}} H_{\Lambda;\ket{\psi_N}} (Y_{n:l}|Z_n) - \min_{\Lambda \in \mathcal{N}_{n:l}} H_{\Lambda;\ket{\Psi_{n:l}}} (Y_{n:l}|Z_n) \right| \\
        &\leq \frac{1}{N} \sum_{n \in I_l} g_l(\gamma_{n:l}),
    \end{split}
\end{equation}
where the inequality in the third line follows from Lemma~\ref{Slem:cont_cond}.
Since the size of $g_l(\gamma_{n:l})$ becomes $\Tilde{\mathcal{O}}\left(\exp\left(-\frac{\min \{n, N-n\} }{2\xi}\right)\right)$ from Lemma~\ref{Slem:approx_normalMPS}, we can take a constant $n_{\eta}^\prime$, which deos not depend on $N$, such that for arbitrary $n$ in $n_{\eta}^\prime \leq n \leq N- n_{\eta}^\prime$, the inequality $g_l(\gamma_{n:l}) \leq \frac{l\eta}{2}$ is satisfied.
By combining this fact with Eq.~(\ref{Seq:Thm3_prf_L}), we can derive the following inequality:
\begin{equation}
\label{SMeq:TDL_LB_final}
\begin{split}
     \left| L_l (\ket{\psi_N}) - \overline{L}_l(\ket{\Psi^l}) \right| & \leq \frac{1}{N} \sum_{n \in I_l} g_l(\gamma_{n:l}) \\
     &= \frac{1}{N} \left\{ \sum_{\substack{n \in I_l, \\ n < n_{\eta}^\prime}} g_l(\gamma_{n:l}) + \sum_{\substack{n \in I_l, \\ n > N- n_{\eta}^\prime}} g_l(\gamma_{n:l})\right\} + \frac{1}{N}\sum_{\substack{n \in I_l, \\ n_{\eta}^\prime \leq n \leq N- n_{\eta}^\prime}} g_l(\gamma_{n:l}) \\
     & \leq \frac{2n_{\eta}^\prime}{Nl} C^\prime + \frac{\eta}{2},
\end{split} 
\end{equation}
where $C^\prime$ is a constant that satisfies $g_l(x) \leq C^\prime$ for arbitrary $x$.
From Eq.~(\ref{SMeq:TDL_LB_final}), we can derive Eq.~(\ref{Seq:Thm3_L}) for any $N\geq N_\eta^\prime$ with $N_\eta^\prime \equiv \frac{4 C^\prime n_{\eta}^\prime}{l\eta}$.
\end{proof}

\subsection{Discussion for periodic boundary condition cases} \label{Sss:PBC_inf_normal}
\begin{figure*}[]
    \centering
    \includegraphics[width=1\textwidth]{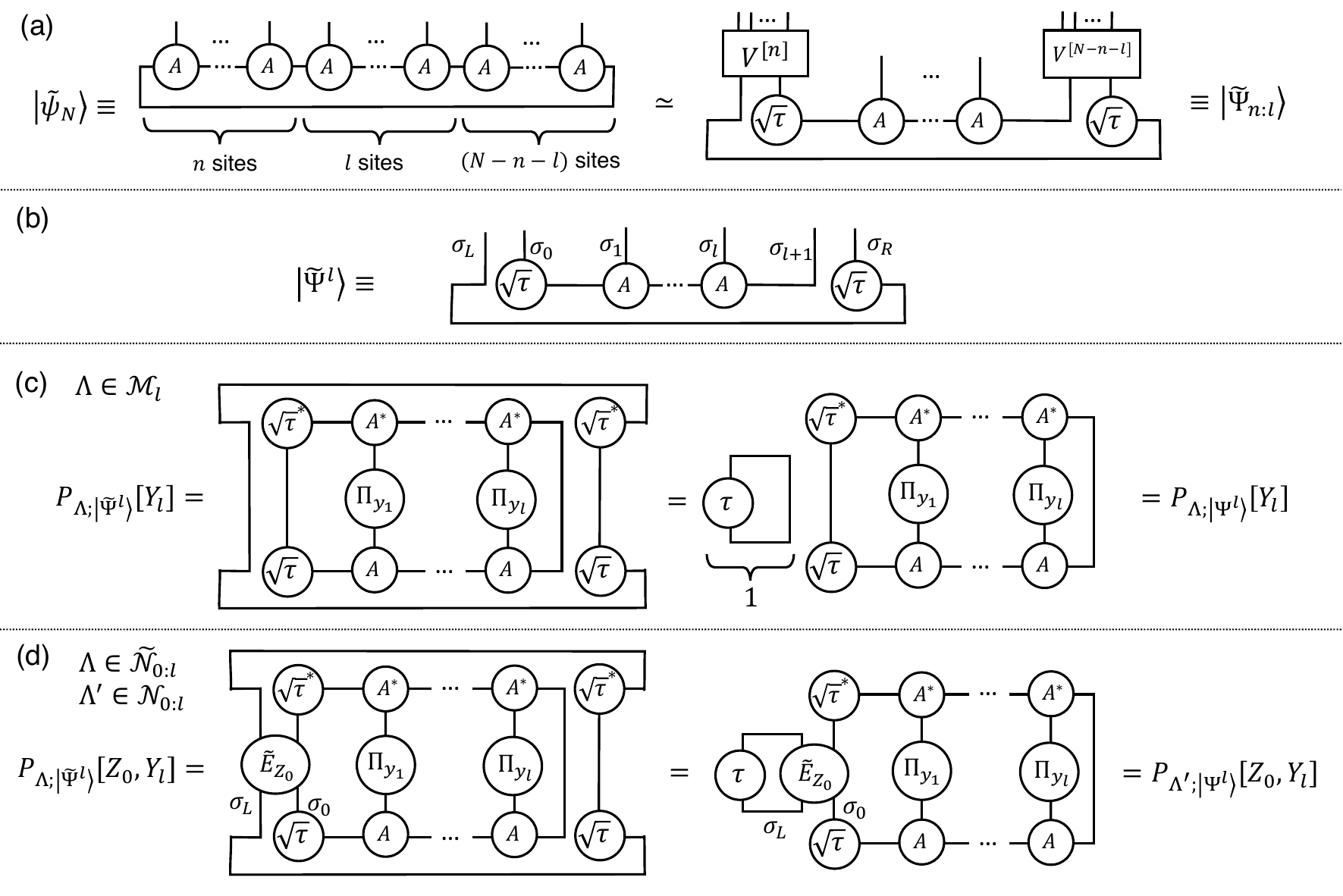}
    \caption{Calculation in normal MPS with periodic boundary conditions in the thermodynamic limit.
    (a) Approximation of the $N$-site MPS $\ket{\widetilde{\psi}_N}$ with the state $\ket{\widetilde{\Psi}_{n:l}}$. In this approximation, we utilize the relationship shown in Fig.~\ref{SMfig:coarse_basic}(c). Here, $V^{[n]}$ and $V^{[N-n-l]}$ are the isometries mapping $D_B$-dimensional virtual systems to $d^n$ and $d^{N-n-l}$-dimensional physical systems, respectively.
    (b) Diagram for the state $\ket{\widetilde{\Psi}^l}$. 
    (c) Derivation of the equality $P_{\Lambda;\ket{\widetilde{\Psi}^l}}[Y_l] = P_{\Lambda;\ket{\Psi^l}}[Y_l]$ for a measurement protocol $\Lambda \in \mathcal{M}_l$.
    (d) Derivation of the equality $P_{\Lambda;\ket{\widetilde{\Psi}^l}}[Z_0,Y_l] = P_{\Lambda^\prime;\ket{\Psi^l}}[Z_0,Y_l]$ for the measurement protocols $\Lambda \in \Tilde{\mathcal{N}}_{0:l}$ and $\Lambda^\prime \in \mathcal{N}_{0:l}$. In the protocol $\Lambda$, a POVM with operators $\{\widetilde{E}_{Z_0}\}$ is performed on the indices $(\sigma_L,\sigma_0)$. Correspondingly, in the protocol $\Lambda^\prime$, the POVM with operators $\{E_{Z_0} \equiv \tr_{\sigma_L}[\tau \widetilde{E}_{Z_0}]\}$ is performed on the index $\sigma_0$.}
    \label{SMfig:inf_MPS_PBC}
\end{figure*}

While Secs.~\ref{Sss:approx_fixed_point} and~\ref{Sss:red_bounds_normalMPS} focus on open boundary conditions, we here briefly explain that the discussion can be directly extended to periodic boundary conditions.
For that purpose, we first derive the following inequality, which is the extension of Lemma~\ref{Slem:approx_normalMPS} to the periodic boundary conditions:
\begin{equation}
\label{SMeq:approx_Til_Psi_nl}
 \|\ket{\widetilde{\psi}_N}\bra{\widetilde{\psi}_N} - \ket{\widetilde{\Psi}_{n:l}}\bra{\widetilde{\Psi}_{n:l}}\|_{\rm tr}  \leq \mathcal{O}(e^{-\frac{n}{2\xi}}) + \mathcal{O}(e^{-\frac{N-n}{2\xi}}),
\end{equation}
where $\ket{\widetilde{\psi}_N}$ is an $N$-site normal MPS with periodic boundary condition and $\ket{\widetilde{\Psi}_{n:l}}$ is the state defined as Fig.~\ref{SMfig:inf_MPS_PBC}(a). Here, $\ket{\widetilde{\Psi}_{n:l}}$ is represented as $\ket{\widetilde{\Psi}_{n:l}} = V^{[n]}\otimes \mathbbm{1}_{d^l} \otimes V^{[N-n-l]} \ket{\widetilde{\Psi}^l}$ with the state $\ket{\widetilde{\Psi}^l}$ defined in Fig.~\ref{SMfig:inf_MPS_PBC}(b) and the isometries $V^{[n]}$ and $V^{[N-n-l]}$ from dimension $D_B^2$ to $d^n$ and $d^{N-n-l}$, respectively. The essence for the derivation of Eq.~(\ref{SMeq:approx_Til_Psi_nl}) is the relationship shown in Fig.~\ref{SMfig:coarse_basic}(c).
By utilizing Eq.~(\ref{SMeq:approx_Til_Psi_nl}), we can extend Theorem~\ref{SThm3:thermo_limit} to the periodic boundary conditions.
To be specific, the following equations are derived:
\begin{align}
    \lim_{N\to\infty} U_l(\ket{\widetilde{\psi}_N}) &= \overline{U}_l (\ket{\widetilde{\Psi}^l}), \label{SMeq:Til_U_l^*}\\
    \lim_{N\to\infty} L_l(\ket{\widetilde{\psi}_N}) &= \overline{L}_l (\ket{\widetilde{\Psi}^l}) ,\label{SMeq:Til_L_l^*}
\end{align}
where $\overline{U}_l (\ket{\widetilde{\Psi}^l})$ and $\overline{L}_l (\ket{\widetilde{\Psi}^l})$ are defined as 
\begin{align}
    \overline{U}_l (\ket{\widetilde{\Psi}^l}) &\equiv \frac{1}{l} \min_{\Lambda \in \mathcal{M}_l} H_{\Lambda;\ket{\widetilde{\Psi}^l}}(Y_l), \\
    \overline{L}_l (\ket{\widetilde{\Psi}^l}) &\equiv \frac{1}{l} \min_{\Lambda \in \Tilde{\mathcal{N}}_{0:l}} H_{\Lambda;\ket{\widetilde{\Psi}^l}}(Y_l|Z_0).
\end{align}
Here, $\Tilde{\mathcal{N}}_{0:l}$ denotes the set of measurement protocols on the state $\ket{\widetilde{\Psi}^l}$ defined in Fig.~\ref{SMfig:inf_MPS_PBC}(b), where a POVM on the indices $(\sigma_L,\sigma_0)$ and local projective measurements on the indices $\{\sigma_i\}_{i=1}^l$ are performed.

While the limits $\overline{U}_l (\ket{\widetilde{\Psi}^l})$ and $\overline{L}_l (\ket{\widetilde{\Psi}^l})$ are already calculable at costs of $e^{\mathcal{O}(l)}$, we can further show the following equalities:
\begin{align}
    \overline{U}_l (\ket{\widetilde{\Psi}^l}) &= \overline{U}_l (\ket{\Psi^l}), \label{Seq:U_equiv_OBC_PBC}\\
    \overline{L}_l (\ket{\widetilde{\Psi}^l}) &= \overline{L}_l (\ket{\Psi^l}). \label{Seq:L_equiv_OBC_PBC}
\end{align}
Equation~(\ref{Seq:U_equiv_OBC_PBC}) directly follows from the fact that for an arbitrary measurement protocol $\Lambda \in \mathcal{M}_l$, the equality $P_{\Lambda;\ket{\Psi^l}}[Y_l]=P_{\Lambda;\ket{\widetilde{\Psi}^l}}[Y_l]$ is satisfied as shown in Fig.~\ref{SMfig:inf_MPS_PBC}(c).
Here, $P_{\Lambda;\ket{\Psi^l}}[Y_l]$ represents the probability of the outcomes $Y_l$ when the measurement protocol $\Lambda \in \mathcal{M}_l$ is applied to the state $\ket{\Psi^l}$.
In the derivation of Eq.~(\ref{Seq:L_equiv_OBC_PBC}), we use the fact that the probability distribution $P_{\Lambda;\ket{\widetilde{\Psi}^l}}[Z_0,Y_l]$ for arbitrary $\Lambda \in \Tilde{\mathcal{N}}_{0:l}$ is reproduced with a certain measurement $\Lambda^\prime \in \mathcal{N}_{0:l}$ on $\ket{\Psi^l}$, resulting in the equality $P_{\Lambda;\ket{\widetilde{\Psi}^l}}[Z_0,Y_l] = P_{\Lambda^\prime ;\ket{\Psi^l}}[Z_0,Y_l]$.
This is illustrated in Fig.~\ref{SMfig:inf_MPS_PBC}(d). Here, the POVM of $\Lambda^\prime \in \mathcal{N}_{0:l}$ on the index $\sigma_0$ is defined with the operators $\{E_{Z_{0}} \equiv \tr_{\sigma_{L}} [\tau \widetilde{E}_{Z_{0}}]\}_{Z_{0}}$, where $\{\widetilde{E}_{Z_0}\}_{Z_0}$ is the operators for the POVM of $\Lambda \in \Tilde{\mathcal{N}}_{0:l}$ on the indices $(\sigma_L,\sigma_0)$.

By combining Eqs.~(\ref{SMeq:Til_U_l^*}), (\ref{SMeq:Til_L_l^*}), (\ref{Seq:U_equiv_OBC_PBC}) and (\ref{Seq:L_equiv_OBC_PBC}), we can derive the following equalities, which are utilized in the numerical calculation in the MPS family \cite{wolf2006quantum} discussed later in Sec.~\ref{Sss:MPSfam}:
\begin{align}
    \lim_{N\to\infty} U_l(\ket{\widetilde{\psi}_N}) &= \overline{U}_l (\ket{\Psi^l}), \label{SMeq:Til_U_l^*_PBC}\\
    \lim_{N\to\infty} L_l(\ket{\widetilde{\psi}_N}) &= \overline{L}_l (\ket{\Psi^l}). \label{SMeq:Til_L_l^*_PBC}
\end{align}

\section{Calculations in MPS examples in thermodynamic limit}
\begin{figure*}[]
    \centering
    \includegraphics[width=1\textwidth]{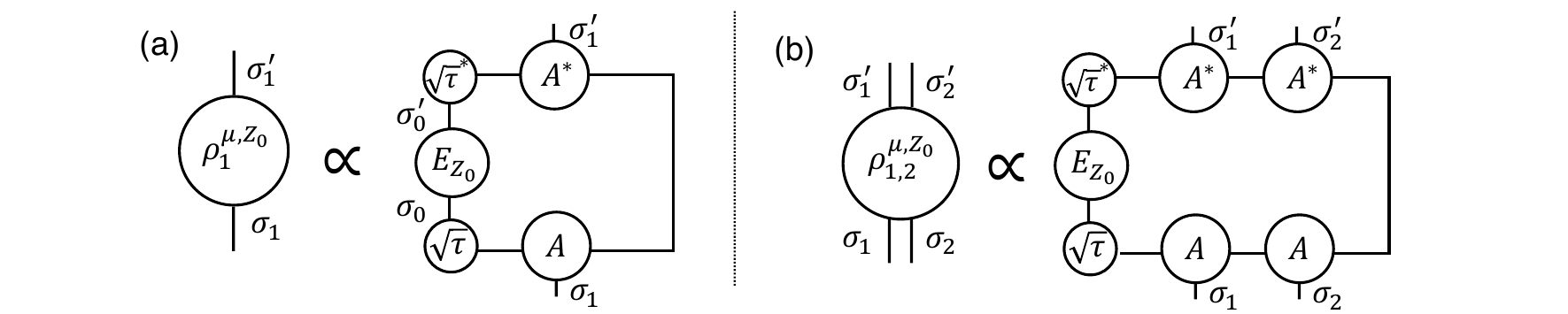}
    \caption{Diagrams for the conditional density operators (a) $\rho^{\mu,Z_{0}}_{1} \propto \tr_{\overline{1}}[E_{Z_{0}}\rho]$ and (b) $\rho^{\mu,Z_{0}}_{1,2} \propto \tr_{\overline{1,2}}[E_{Z_{0}}\rho]$. Here, $\{E_{Z_0}\}_{Z_0}$ is the set of operators for the POVM $\mu$ on the index $\sigma_0$.}
    \label{SMfig:MPS_AKLT_cluster}
\end{figure*}

In this section, we present the calculations of work deficit densities $\delta$ and $\delta^\to$ for MPS examples in the thermodynamic limit.
Since we consider translationally invariant normal MPS here, we can use Theorem~\ref{SThm3:thermo_limit} in the calculations of the lower and upper bounds.
Therefore, the bounds $L_l$, $U_l$, $L_l^\to$, and $U_l^\to$ can be calculated with the state $\ket{\Psi^l}$, as shown Fig.~\ref{SMfig:inf_MPS_approx}.
We provide the details of analytical calculation for the AKLT state in Sec.~\ref{Sss:AKLT} and that for the cluster state in Sec.~\ref{Sss:cluster}.
We present the details of numerical calculation in the MPS family \cite{wolf2006quantum} in Sec.~\ref{Sss:MPSfam}.

\subsection{AKLT state} \label{Sss:AKLT}
We calculate the work deficit densities $\delta$ and $\delta^\to$ in the AKLT state.
To evaluate the lower bound for $\delta^\to$, we utilize the following inequality:
\begin{equation}
\label{Seq:AKLT_LB}
    \begin{split}
        \delta^\to  &\geq L_1^\to \\
        &= \min_{\mu \in \mathcal{E}_{0}} \sum_{Z_{0}} P_\mu[Z_{0}] S(\rho^{\mu,Z_{0}}_{1}) \\
        &\geq \min_{\mu \in \mathcal{E}_{0}, Z_0}  S(\rho^{\mu,Z_{0}}_{1}),
    \end{split}
\end{equation}
where $\mathcal{E}_{0}$ is the set of POVMs on the index $\sigma_0$ of $\ket{\Psi^1}$, as illustrated in Fig.~\ref{SMfig:inf_MPS_approx}(c).
Here, probability $P_\mu[Z_{0}]$ and the conditional density operator $\rho^{\mu,Z_{0}}_{1}$ are defined as $ P_\mu[Z_{0}] \rho^{\mu,Z_{0}}_{1} \equiv \tr_{\sigma_0}[E_{Z_{0}}\rho]$, with $\{E_{Z_0}\}$ representing the set of the measurement operators for the POVM $\mu \in \mathcal{E}_{0}$.

To evaluate the final line of Eq.~(\ref{Seq:AKLT_LB}), we next calculate the conditional density operator $\rho^{\mu,Z_{0}}_{1}$, as shown in Fig.~\ref{SMfig:MPS_AKLT_cluster}(a).
The right-canonical tensors $\{A^\sigma\}$ and the fixed point tensor $\tau$ for the AKLT state are calculated as follows:
\begin{equation*}
    A^\uparrow = \sqrt{\frac{2}{3}}\begin{pmatrix}
        0 & 0 \\
        -1 & 0 
    \end{pmatrix}, \ 
    A^0 =\sqrt{\frac{1}{3}}\begin{pmatrix}
        1 & 0 \\
        0 & -1 
    \end{pmatrix},\ 
    A^\downarrow = \sqrt{\frac{2}{3}} \begin{pmatrix}
        0 & 1 \\
        0 & 0
    \end{pmatrix}, \ 
    \tau \equiv \frac{1}{2}\left(\begin{array}{cc}
    1 &  0\\
    0 & 1
    \end{array}\right).
\end{equation*}
Since a rank-one POVM operator $E_{Z_{0}}$ on two-dimensional system is represented as 
\begin{equation}
\label{Seq:rank_one_POVM_op}
    E_{Z_{0}} \propto \left(\begin{array}{cc}
    \sin^2 \theta &  \cos\theta \sin\theta e^{-i\phi}\\
    \cos\theta \sin\theta e^{i\phi} & \cos^2\theta
    \end{array}\right) ,
\end{equation}
with the parameters $\theta$ and $\phi$, the conditional density operator $\rho^{\mu,Z_{0}}_{1}$ is calculated as follows:
\begin{equation}
   \rho^{\mu,Z_{0}}_{1} = \frac{1}{3}\left(\begin{array}{ccc}
    2\cos^2\theta &  -\sqrt{2}\cos\theta\sin\theta e^{-i\phi} & 0 \\
    -\sqrt{2}\cos\theta\sin\theta e^{i\phi} & 1 & -\sqrt{2}\cos\theta\sin\theta e^{-i\phi} \\
    0 & -\sqrt{2}\cos\theta\sin\theta e^{i\phi} & 2 \sin^2\theta
    \end{array}\right).
\end{equation}
Here, the eigenvalues of this matrix are $\frac{2}{3}$, $\frac{1}{3}$, and $0$, independent of $\theta$ and $\phi$.

Therefore, the lower bound $L_1^\to$ can be evaluated as $L_1^\to \geq h\left(\frac{1}{3}\right)$, with the binary entropy function $h(p) \equiv -p\ln p -(1-p)\ln (1-p)$. 
Since the inequality $L_1\geq L_1^\to$ is always satisfied, we can derive the following inequality:
\begin{equation}
    L_1 \geq L_1^\to \geq h\left(\frac{1}{3}\right).
\end{equation}
Since the upper bound $U_{\Lambda^*} = h(\frac{1}{3})$ is also derived with the ansatz measurement protocol $\Lambda^*$, where local projective measurements with the basis $\{\ket{\uparrow},\ket{0},\ket{\downarrow}\}$ is performed on all parties, the work deficit densities are exactly calculated as
\begin{equation}
    \delta = \delta^\to =  h\left(\frac{1}{3}\right).
\end{equation}

\subsection{Cluster state} \label{Sss:cluster}
We calculate the work deficit densities $\delta$ and $\delta^\to$ for the cluster state.
To evaluate the lower bound of $\delta^\to$, we derive the following inequality:
\begin{equation}
\label{Seq:cluster_LB}
    \begin{split}
        \delta^\to  &\geq L_2^\to \\
        &\geq \min_{\mu \in \mathcal{E}_{0}} \frac{1}{2}\sum_{Z_{0}} P_\mu[Z_{0}] S(\rho^{\mu,Z_{0}}_{1,2}) \\
        &\geq \frac{1}{2} \min_{\mu \in \mathcal{E}_{0},Z_{0}}  S(\rho^{\mu,Z_{0}}_{1,2}),
    \end{split}
\end{equation}
where $\mathcal{E}_{0}$ is the set of POVMs on the index $\sigma_0$ of $\ket{\Psi^2}$, as illustrated in Fig.~\ref{SMfig:inf_MPS_approx}(c).
Here, probability $P_\mu[Z_{0}]$ and conditional density operator $\rho^{\mu,Z_{0}}_{1,2}$ are defined as $ P_\mu[Z_{0}] \rho^{\mu,Z_{0}}_{1,2} \equiv \tr_{\sigma_0}[E_{Z_{0}}\rho]$, with $\{E_{Z_0}\}$ representing the set of POVM operators for $\mu \in \mathcal{E}_{0}$.

To evaluate the final line of Eq.~(\ref{Seq:cluster_LB}), we calculate the conditional density operator $\rho^{\mu,Z_{0}}_{1,2}$, as shown in Fig.~\ref{SMfig:MPS_AKLT_cluster}(b). 
The right-canonical tensors $\{A^\sigma\}$ and the fixed point tensor $\tau$ for the cluster state are calculated as follows:
\begin{equation*}
    A^0 =\sqrt{\frac{1}{2}}\begin{pmatrix}
        0 & 0 \\
        1 & 1 
    \end{pmatrix},\ 
    A^1 =\sqrt{\frac{1}{2}}\begin{pmatrix}
        1 & -1 \\
        0 & 0 
    \end{pmatrix},\ 
    \tau \equiv \frac{1}{2}\left(\begin{array}{cc}
    1 &  0\\
    0 & 1
    \end{array}\right).
\end{equation*}
Since a rank-one POVM operator on two-dimensional system is designated as Eq.~(\ref{Seq:rank_one_POVM_op}), the conditional density operator can be calculated as
\begin{equation}
    \rho^{\mu,Z_{0}}_{1,2} = \frac{1}{2}\left(\begin{array}{cccc}
    \cos^2 \theta &0 & \sin\theta\cos\theta e^{-i\phi} &0 \\
    0 & \cos^2 \theta &0 & -\sin\theta\cos\theta e^{-i\phi} \\
    \sin\theta\cos\theta e^{i\phi} & 0 & \sin^2\theta &0 \\
    0 & -\sin\theta\cos\theta e^{i\phi} &0 &\sin^2\theta  \\
    \end{array}\right).
\end{equation}
The eigenvalues of this conditional density operator are $1,1,0,$ and $0$, independent of the parameters $\theta$ and $\phi$ of the POVM operator. 
Therefore, the lower bounds are evaluated as 
\begin{equation}
    L_2 \geq L_2^\to \geq \frac{1}{2}\ln 2.
\end{equation}

Since the upper bound $U_{\Lambda^*} = \frac{1}{2} \ln 2$ is also derived with the ansatz measurement protocol $\Lambda^*$, where the measurement on the basis $\{\ket{0},\ket{1}\}$ and $\{\frac{1}{\sqrt{2}} (\ket{0}+\ket{1}), \frac{1}{\sqrt{2}} (\ket{0}-\ket{1})\}$ are performed alternately to each party, we can exactly calculate the work deficit densities as
\begin{equation}
    \delta =\delta^\to= \frac{1}{2} \ln 2.
\end{equation}

\subsection{MPS family} \label{Sss:MPSfam}
We provide the detailed explanation on the numerical calculation in the MPS family~\cite{wolf2006quantum}. 
This MPS family is normal except at $g=0$, and its right-canonical form is derived as follows:
\begin{alignat}{2}
    &A^0 =  \sqrt{\frac{1}{1+|g|}}\begin{pmatrix}
        0 & 0 \\
        \sqrt{|g|} & 1 
    \end{pmatrix}, 
    A^1 = \sqrt{\frac{1}{1+|g|}} \begin{pmatrix}
        1 & -\sqrt{|g|} \\
        0 & 0 
    \end{pmatrix} \quad \quad &(g<0), \label{SMeq:MPS_fam_gnega}\\
    &A^0 = \sqrt{\frac{1}{1+g}} \begin{pmatrix}
        0 & 0 \\
        \sqrt{g} & 1 
    \end{pmatrix}, 
    A^1 = \sqrt{\frac{1}{1+g}}\begin{pmatrix}
        1 & \sqrt{g} \\
        0 & 0 
    \end{pmatrix} \quad \quad &(g>0).\label{SMeq:MPS_fam_gposi}
\end{alignat}
Furthermore, the fixed point tensor for this MPS family can be calculated as
\begin{equation}
\label{Seq:MPSfam_RDM}
    \tau \equiv \left\{
    \begin{alignedat}{2}
    &\left(
    \begin{array}{cc}
    \frac{1}{2} &  0\\
    0 & \frac{1}{2}
    \end{array}
    \right) \quad &(g < 0), \\
    &\left(
    \begin{array}{cc}
    \frac{1}{2} &  \frac{\sqrt{g}}{1+g}\\
    \frac{\sqrt{g}}{1+g}  & \frac{1}{2}
    \end{array}
    \right) \quad &(g > 0). \\
    \end{alignedat}
    \right.
\end{equation}

\begin{figure*}[]
    \centering
    \includegraphics[width=1\textwidth]{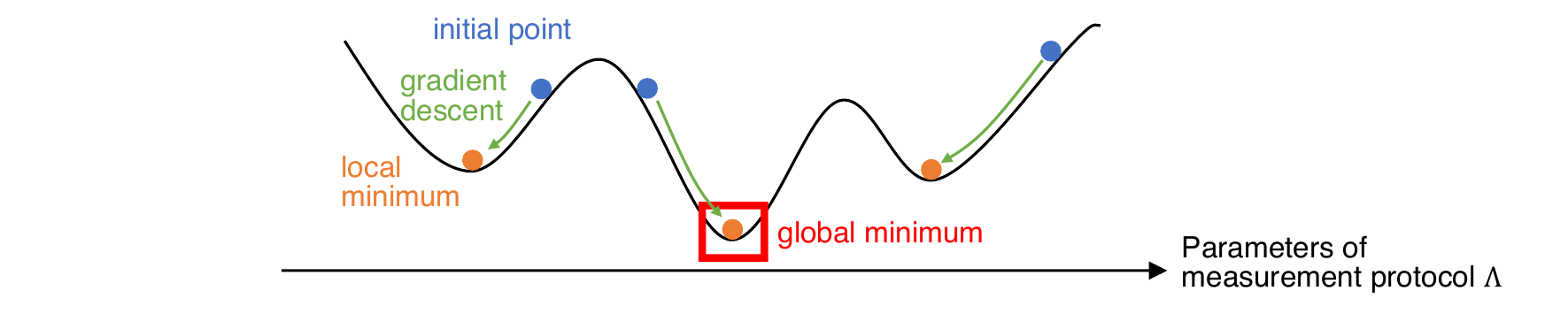}
    \caption{Schematic for the numerical optimizations of the measurement protocol $\Lambda$ involved in the lower and upper bounds. 
    For these optimizations, we perform the gradient descent method from random initial points for multiple times. We can reach the global minimum by running this process sufficiently many times.}
    \label{SMfig:num_calc_grad}
\end{figure*}

We numerically calculate the upper and lower bounds for this MPS family, by utilizing Eqs.~(\ref{SMeq:Til_U_l^*_PBC}) and (\ref{SMeq:Til_L_l^*_PBC}).
The upper bound $\overline{U}_l (\ket{\Psi^l})$ is computed by first calculating the cost function $H_{\Lambda;\ket{\Psi^l}}(Y_{l})$ for a certain measurement protocol $\Lambda \in \mathcal{M}_l$ as shown in Fig.~\ref{SMfig:inf_MPS_approx}(b), and then performing the optimization of $\Lambda$ over $\mathcal{M}_l$. The calculation of the lower bound $\overline{L}_l (\ket{\Psi^l})$ is performed in the similar manner.
In the optimizations of the measurement protocol $\Lambda$ over $\mathcal{M}_l$ and $\mathcal{N}_{0:l}$, we perform the gradient decent from random initial protocols for multiple times, as depicted in Fig.~\ref{SMfig:num_calc_grad}. 
By running this process for sufficiently many times and choosing the optimal protocol from all of these runs, we can avoid being trapped to the local minimum. 
It is also noteworthy that while we here adopt the gradient descent method in the optimizations of $\Lambda$, it is also possible to perform the brute-force optimizations over $\mathcal{M}_l$ and $\mathcal{N}_{0:l}$ at costs of $e^\mathcal{O}(l)$.

\section{Work deficit density under coarse-graining} \label{Ss:coarse_grain}
\begin{figure*}[]
    \centering
    \includegraphics[width=1\textwidth]{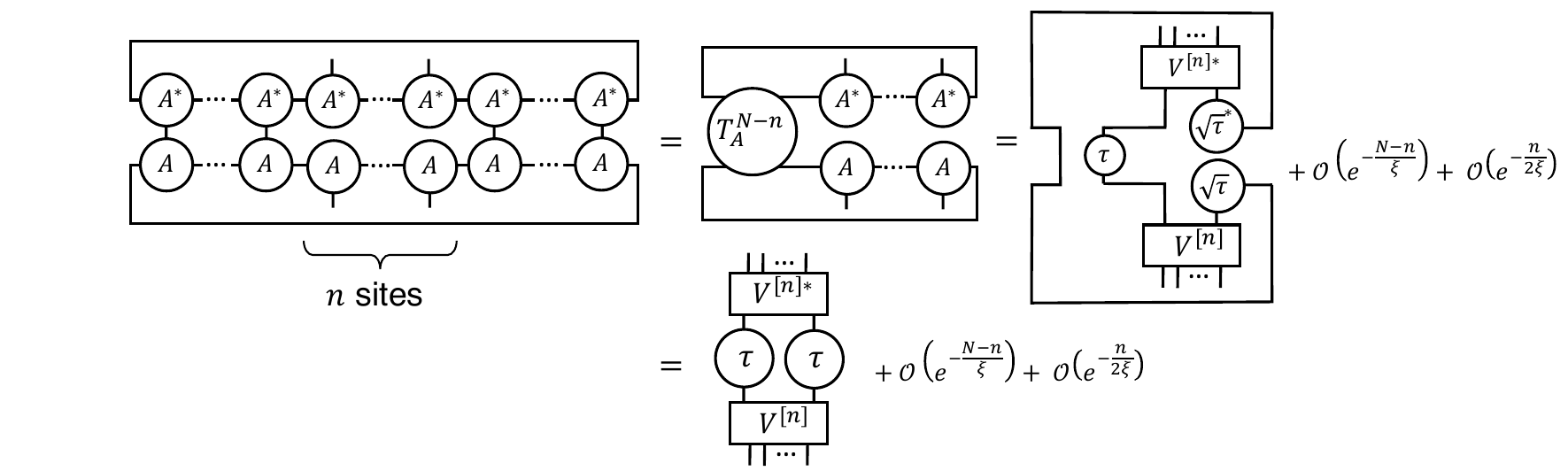}
    \caption{Reduced density operator for the $n$-site segment of an $N$-site normal MPS. Here, $T_A$ denote the transfer matrix for the tensor $A$ and $V^{[n]}$ represents the isometry which maps $D_B^2$-dimensional virtual system to $d^n$-dimensional physical system. In the derivation of the second equality, we utilize the approximations shown in Figs.~\ref{SMfig:coarse_basic}(b) and (c). In the limit of $\frac{n}{\xi}\to 0$ and $\frac{N-n}{\xi}\to 0$, the entanglement entropy of the subsystem converges to $S_{\rm EE} \equiv 2S(\tau)$.}
    \label{SMfig:coarse_MPS_fam}
\end{figure*}

In this section, we prove Theorem~\ref{SThm4:coarse-graining}, which shows the convergence of the work deficit density under coarse-graining as follows:
\begin{equation}
    \delta_m = \frac{S_{\text{EE}}}{2} + \Tilde{\mathcal{O}}(e^{-\frac{m}{2\xi}}).
\end{equation}
Here, $S_{\rm EE}$ denotes the entanglement entropy for the subsystem much larger than the correlation length $\xi$. 
It is known that $S_{\rm EE}$ is calculated as $S_{\rm EE} \equiv 2S(\tau)$ with the fixed point tensor $\tau$ \cite{cirac2017matrix,cirac2021matrix}, which directly follows from the tensor transformation shown in Fig.~\ref{SMfig:coarse_MPS_fam}.

In Sec.~\ref{Sss:coarse_normal_MPS}, we provide Lemma~\ref{Slem:coarse_approx}, which plays an important role in the derivation of Theorem~\ref{SThm4:coarse-graining}.
In Sec.~\ref{Sss:coarse_EE}, we prove the main result of this section, Theorem~\ref{SThm4:coarse-graining}, by utilizing this lemma.

\subsection{Coarse-graining of normal MPS} \label{Sss:coarse_normal_MPS}
\begin{figure*}[]
    \centering
    \includegraphics[width=1\textwidth]{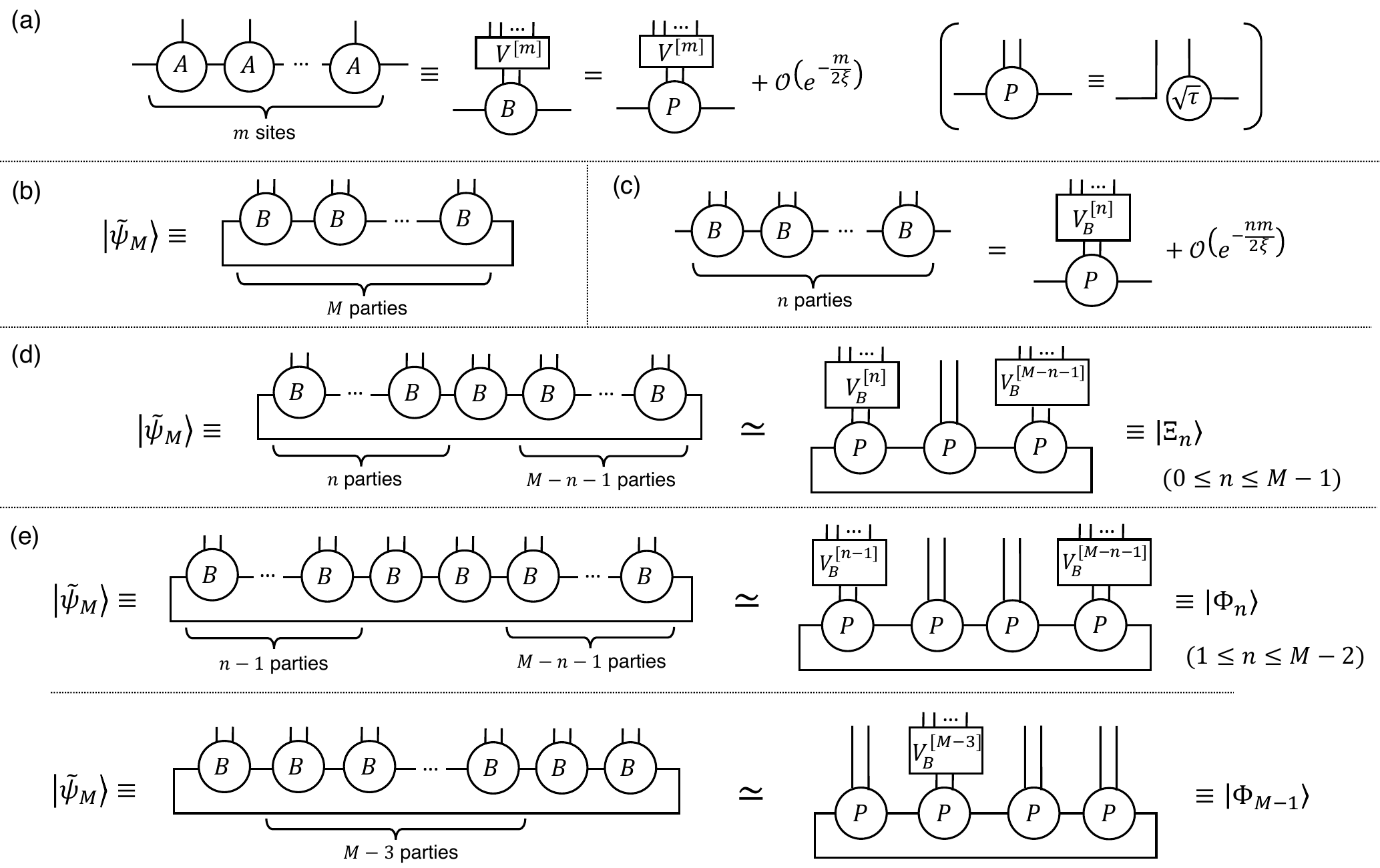}
    \caption{(a) Diagram for the $m$-site tensor $B$. Here, $V^{[m]}$ is the isometry from dimension $D_B^2$ to $d^m$, and $P\equiv \mathbbm{1}_{D_B}\otimes\sqrt{\tau}$ is the renormalization fixed point tensor. 
    (b) Definition of the $M$-partite state $\ket{\widetilde{\psi}_M}$. 
    (c) Approximation of the tensor corresponding to $n$ parties with the renormalization fixed point tensor $P$. Here, $V^{[n]}_{B}$ is the isometry from dimension $D_B^2$ to $D_B^{2n}$.
    (d) Approximation of the $M$-partite state $\ket{\widetilde{\psi}_M}$ with the state $\ket{\Xi_n}$. 
    (e) Approximation of the $M$-partite state $\ket{\widetilde{\psi}_M}$ with the state $\ket{\Phi_n}$. The approximations in (d) and (e) can be derived by utilizing the relationships shown in (a) and (c).}
    \label{SMfig:coarse_approx}
\end{figure*}

We here consider an $N$-site normal MPS under periodic boundary condition in the partitioning that $m$ sites are merged and regarded as a single party. 
In this setting, the number of parties is denoted as $M\equiv \frac{N}{m}$, $n$-th party is denoted as $b_n$, the tensor for $m$-site segment $B$ is defined as Fig.~\ref{SMfig:coarse_approx}(a), and the $M$-partite ($N$-site) state $\ket{\widetilde{\psi}_M}$ is defined as Fig.~\ref{SMfig:coarse_approx}(b).

Since the normal MPS under coarse-graining is well-approximated with the renormalization fixed point tensor, as illustrated in Figs.~\ref{SMfig:coarse_approx}(a) and (c), the following lemma can be derived:
\begin{lemma} \label{Slem:coarse_approx}
Let $\ket{\widetilde{\psi}_M}$ be $M$-partite state under periodic boundary condition, defined as Fig.~\ref{SMfig:coarse_approx}(b).
Then, the following inequalities are satisfied:
\begin{align}
    \gamma_n &\equiv \|\ket{\widetilde{\psi}_M}\bra{\Tilde{\psi}_M} - \ket{\Xi_n}\bra{\Xi_n}\|_{\rm tr} \leq \mathcal{O}(e^{-\frac{m}{2\xi}}), \\
    \gamma_n^\prime &\equiv \|\ket{\widetilde{\psi}_M}\bra{\Tilde{\psi}_M} - \ket{\Phi_n}\bra{\Phi_n}\|_{\rm tr} \leq \mathcal{O}(e^{-\frac{m}{2\xi}}),
\end{align}
where the states $\ket{\Xi_n}$ and $\ket{\Phi_n}$ are defined as shown in Figs.~\ref{SMfig:coarse_approx}(d) and (e), respectively, and $\xi$ is the correlation length of this state. 
\end{lemma}
We here abbreviate the proof of this lemma, since the essence of the proof is the same as Lemma~\ref{Slem:approx_normalMPS}.

\subsection{Convergence of work deficit density to entanglement entropy under coarse-graining} \label{Sss:coarse_EE}

By utilizing Lemma~\ref{Slem:coarse_approx}, we can derive the following theorem, which is the main result of this section:
\begin{theorem} \label{SThm4:coarse-graining}
Let $\ket{\widetilde{\psi}_M}$ be $M$-partite state under periodic boundary condition, defined as Fig.~\ref{SMfig:coarse_approx}(b).
Then, the following inequality is satisfied:
\begin{equation}
\label{SMeq:EE_converge}
    \left| \delta_m (\ket{\widetilde{\psi}_M}) - \frac{S_{\text{EE}}}{2}\right| \leq \Tilde{\mathcal{O}}(e^{-\frac{m}{2\xi}}),
\end{equation}
where $S_{\rm EE} \equiv 2S(\tau)$ is the entanglement entropy of sufficiently large subsystem, and $\xi$ is the correlation length of this state.
\end{theorem}

\begin{proof}
To derive Eq.~(\ref{SMeq:EE_converge}), we prove the following two inequalities:
\begin{align}
   \delta_m (\ket{\widetilde{\psi}_M}) &\geq \frac{S_{\text{EE}}}{2} -\Tilde{\mathcal{O}}(e^{-\frac{m}{2\xi}}), \label{SMeq:coarse_WDD_LB}\\
   \delta_m (\ket{\widetilde{\psi}_M}) &\leq \frac{S_{\text{EE}}}{2} +\Tilde{\mathcal{O}}(e^{-\frac{m}{2\xi}}). \label{SMeq:coarse_WDD_UB}
\end{align}
The inequality~(\ref{SMeq:coarse_WDD_LB}) can be derived as follows:
\begin{equation}
    \begin{split}
        \delta_m (\ket{\widetilde{\psi}_M}) &\geq L_1 (\ket{\widetilde{\psi}_M})\\
    &= \frac{1}{M} \left\{\min_{\Lambda \in \mathcal{M}_{1}} H_{\Lambda;\ket{\widetilde{\psi}_M}}(y_1) + \sum_{n=1}^{M-1} \min_{\Lambda \in \mathcal{N}_{n:1}} H_{\Lambda;\ket{\widetilde{\psi}_M}}(y_{n+1}|Z_{n}) \right \}\\
    &\geq  \frac{1}{M} \left\{ \min_{\Lambda \in \mathcal{M}_{1}} H_{\Lambda;\ket{\Xi_0}}(y_1) + \sum_{n=1}^{M-1} \min_{\Lambda \in \mathcal{N}_{n:1}} H_{\Lambda;\ket{\Xi_n}} (y_{n+1}|Z_{n}) \right\} - \Tilde{\mathcal{O}} (e^{-\frac{m}{2\xi}})\\
    &\geq  \frac{S_{\rm EE}}{2} - \Tilde{\mathcal{O}} (e^{-\frac{m}{2\xi}}). \\
    \end{split}
\end{equation}
Here, the inequality in the third line follows from Lemma~\ref{Slem:cont_ent},~\ref{Slem:cont_cond}, and~\ref{Slem:coarse_approx}.
In the derivation of the fourth line, we utilize the following inequalities:
\begin{alignat}{1}
    &\min_{\Lambda \in \mathcal{M}_{1}} H_{\Lambda;\ket{\Xi_0}}(y_1) \geq S_{\rm EE}, \label{SMeq:coarse_LB_n1}\\
    &\min_{\Lambda \in \mathcal{N}_{n:1}} H_{\Lambda;\ket{\Xi_n}} (y_{n+1}|Z_{n}) \geq \frac{S_{\rm EE}}{2} \quad\quad (1 \leq n \leq M-2), \label{SMeq:coarse_LB_n2L-1}\\ 
    &\min_{\Lambda \in \mathcal{N}_{M-1:1}} H_{\Lambda;\ket{\Xi_{M-1}}} (y_{M}|Z_{M-1}) \geq 0.  \label{SMeq:coarse_LB_nL}
\end{alignat}
Since Eqs.~(\ref{SMeq:coarse_LB_n1}) and (\ref{SMeq:coarse_LB_nL}) follow from simple calculations, their derivations are omitted here. 
Equation~(\ref{SMeq:coarse_LB_n2L-1}) can be derived as follows:
\begin{equation}
\label{SMeq:LB_Stau}
    \begin{split}
        \min_{\Lambda \in \mathcal{N}_{n:1}} H_{\Lambda;\ket{\Xi_n}} (y_{n+1}|Z_{n}) &= \min_{\mu \in \mathcal{E}_{n}} \sum_{Z_{n}} P_\mu[Z_{n}] S(\rho_{n+1}^{\mu,Z_{n}}) \\
        &\geq \min_{\mu \in \mathcal{E}_{n}, Z_n}  S(\rho_{n+1}^{\mu, Z_{n}}) \\
        &\geq S(\tau) = \frac{S_{\rm EE}}{2},
    \end{split}
\end{equation}
where $\mathcal{E}_{n}$ is the set of arbitrary POVMs on parties $\{b_i\}_{i=1}^n$. By denoting the POVM operators for $\mu \in \mathcal{E}_{n}$ as $\{E_{Z_{n}}\}$, $P_\mu[Z_{n}]$ and $\rho_{n+1}^{\mu,Z_{n}}$ are defined as $P_\mu[Z_{n}] \rho_{n+1}^{\mu,Z_{n}} \equiv \tr_{\overline{n+1}}[E_{Z_{n}}\ket{\Xi_n}\bra{\Xi_n}]$.
The inequality in the third line follows from the fact that the conditional density operator $\rho_{n+1}^{\mu,Z_{n}}$ is represented in the form of $\rho_{n+1}^{\mu,Z_{n}} \propto \Tilde{\tau}^{\mu,Z_n} \otimes \tau$ with a certain positive semidefinite operator $\Tilde{\tau}^{\mu,Z_n}$, as shown in Fig.~\ref{SMfig:coarse_LB_approx}.

\begin{figure*}[]
    \centering
    \includegraphics[width=1\textwidth]{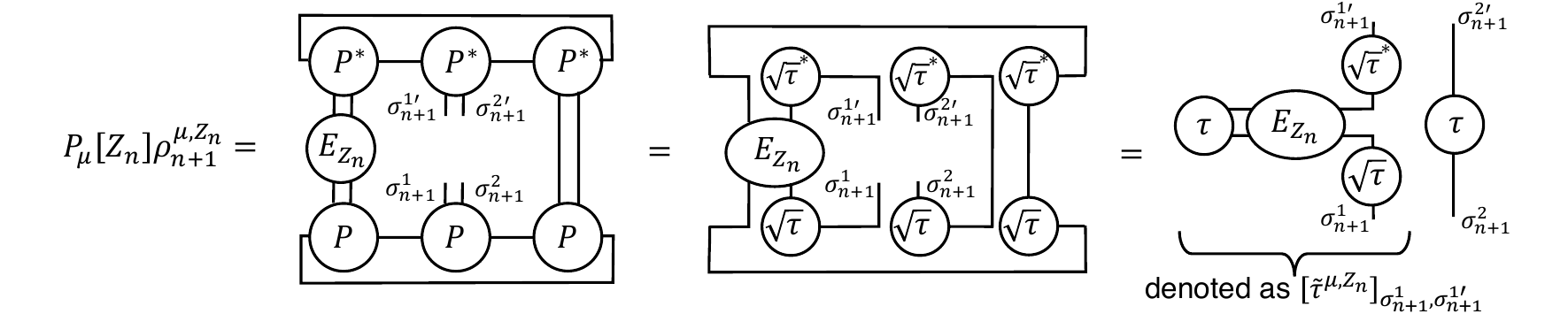}
    \caption{Diagrams for $P_\mu[Z_{n}] \rho_{n+1}^{\mu,Z_{n}} \equiv \tr_{\overline{n+1}}[E_{Z_{n}}\ket{\Xi_n}\bra{\Xi_n}]$. 
    Here, $\ket{\Xi_n}$ is the state defined in Fig.~\ref{SMfig:coarse_approx}(d), but the isometries $V^{[n]}_B$ and $V^{[M-n-1]}_B$ acting on parties $\{b_i\}_{i=1}^{n}$ and $\{b_i\}_{i=n+2}^{M}$, respectively, are abbreviated for simplicity. The rightmost diagram illustrates that the conditional density operator is represented as $\rho_{n+1}^{\mu,Z_{n}} \propto \Tilde{\tau}^{\mu,Z_n} \otimes \tau$ with a positive semidefinite operator $\Tilde{\tau}^{\mu,Z_n}$, which impies $S(\rho_{n+1}^{\mu,Z_{n}}) \geq S(\tau)$.}
    \label{SMfig:coarse_LB_approx}
\end{figure*}

We next derive Eq.~(\ref{SMeq:coarse_WDD_UB}) as follows:
\begin{equation}
\label{SMeq:coarse_WDD_UB_deri}
\begin{split}
    \delta_m (\ket{\widetilde{\psi}_M}) &= \frac{1}{M}\min_{\Lambda \in \mathcal{M}_M} \sum_{n=0}^{M-1} H_{\Lambda; \ket{\widetilde{\psi}_M}} (y_{n+1}|Y_{n}) \\
    &\leq \frac{1}{M} \sum_{n=0}^{M-1} H_{\Lambda^*; \ket{\widetilde{\psi}_M}} (y_{n+1}|Y_{n}) \\
    &\leq \frac{1}{M} \left\{ H_{\Lambda^*; \ket{\widetilde{\psi}_M}} (y_{1}) + \sum_{n=1}^{M-2} H_{\Lambda^*; \ket{\widetilde{\psi}_M}} (y_{n+1}|y_{n}) + H_{\Lambda^*; \ket{\widetilde{\psi}_M}} (y_{M}|y_{M-1},y_1) \right\} \\
    &\leq \frac{1}{M} \left\{ H_{\Lambda^*; \ket{\Xi_0}} (y_{1}) + \sum_{n=1}^{M-2} H_{\Lambda^*; \ket{\Phi_n}} (y_{n+1}|y_{n}) + H_{\Lambda^*; \ket{\Phi_{M-1}}} (y_{M}|y_{M-1},y_1)  \right\} + \Tilde{\mathcal{O}} (e^{-\frac{m}{2\xi}}) \\
    &= \frac{S_{\rm EE}}{2} + \Tilde{\mathcal{O}} (e^{-\frac{m}{2\xi}}),
\end{split}
\end{equation}
where the measurement protocol $\Lambda^* \in \mathcal{M}_M$ is defined as the projective measurements on the diagonal basis of $\tau$ for all parties, as illustrated in Figs.~\ref{SMfig:coarse_UB}(a) and (b).
The inequality in the fourth line is derived from Lemma~\ref{Slem:coarse_approx}, in essentially the same discussion as Lemma~\ref{Slem:cont_ent} and~\ref{Slem:cont_cond}.
The fifth line is derived from the following equations, which can be shown with simple calculations:
\begin{alignat}{1}
    &H_{\Lambda^*; \ket{\Xi_0}} (y_{1}) = S_{\rm EE}, \label{SMeq:coarse_UB_n1} \\
    &H_{\Lambda^*; \ket{\Phi_n}} (y_{n+1}|y_{n}) = \frac{S_{\rm EE}}{2}\quad \quad   (1 \leq n \leq M-2),\label{SMeq:coarse_UB_n2L-1}\\ 
    &H_{\Lambda^*; \ket{\Phi_{M-1}}} (y_{M}|y_{M-1},y_1)=0, \label{SMeq:coarse_UB_nL}
\end{alignat} 
For instance, Eq.~(\ref{SMeq:coarse_UB_n2L-1}) is the consequence of the following equality, which is derived through the tensor transformation illustrated in Fig.~\ref{SMfig:coarse_UB}(c):
\begin{equation}
\label{SMeq:P_Lambda}
    P_{\Lambda^*; \ket{\Phi_n}}[y_{n}^1,y_{n}^2,y_{n+1}^1,y_{n+1}^2] = q_{y_{n}^1} q_{y_{n}^2} \delta_{y_{n}^2,y_{n+1}^1} q_{y_{n+1}^2},
\end{equation}
where $(y_{n}^1,y_{n}^2) \equiv y_n$ and $(y_{n+1}^1,y_{n+1}^2) \equiv y_{n+1}$ represent the measurement outcomes from the parties $b_n$ and $b_{n+1}$, respectively, and $\{q_i\}_i$ denotes the diagonal probability distribution for the density operator $\tau$.

\begin{figure*}[]
    \centering
    \includegraphics[width=1\textwidth]{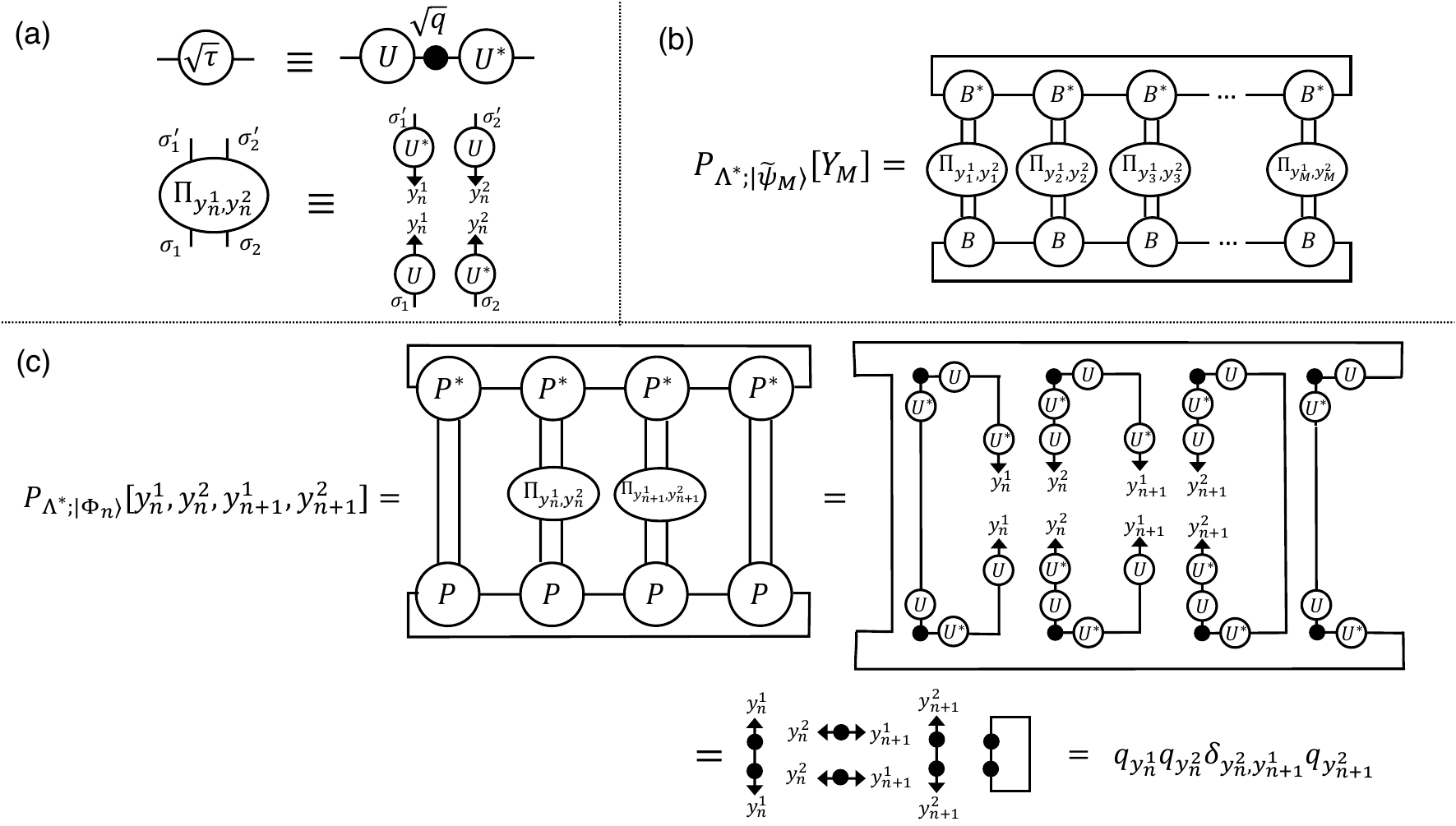}
    \caption{Ansatz measurement protocol $\Lambda^* \in \mathcal{M}_M$.
    (a) Diagram for the rank-one projector on each party. The square root of the fixed point tensor is diagonalized as $\sqrt{\tau}= U \sqrt{q} U^\dag$, where $U$ is $D_B$-dimensional unitary matrix and $\sqrt{q}$ is $D_B$-dimensional diagonal matrix. The measurement outcomes from $n$-th party is defined as $y_n\equiv (y_n^1,y_n^2)$ and the corresponding rank-one projector is denoted as $\Pi_{y_n^1,y_n^2}$. The triangles in this diagram represent that the indices of the tensor legs are fixed to the denoted labels. Therefore, the projector $\Pi_{y_n^1,y_n^2}$ is defined as $[\Pi_{y_n^1,y_n^2}]_{\sigma_1,\sigma_2,\sigma_1^\prime,\sigma_2^\prime} \equiv U_{\sigma_1,y_n^1}U^*_{\sigma_2,y_n^2} U^*_{\sigma_1^\prime,y_n^1}U_{\sigma_2^\prime,y_n^1}$.
    (b) Diagram for the probability $P_{\Lambda^*;\ket{\widetilde{\psi}_M}} [Y_M]$. The ansatz measurement protocol $\Lambda^*\in \mathcal{M}_M$ is to perform projective measurement with the projectors $\{\Pi_{y_n^1,y_n^2}\}_{y_n^1,y_n^2}$ on all parties $\{b_n\}_{n=1}^M$.
    (c) Diagrams for the probability $P_{\Lambda^*;\ket{\Phi_n}} [y_{n}^1,y_{n}^2,y_{n+1}^1,y_{n+1}^2]$. As the consequence of the tensor transformation, the probability is calculated as $P_{\Lambda^*; \ket{\Phi_n}}[y_{n}^1,y_{n}^2,y_{n+1}^1,y_{n+1}^2] = q_{y_{n}^1} q_{y_{n}^2} \delta_{y_{n}^2,y_{n+1}^1} q_{y_{n+1}^2}$.}
    \label{SMfig:coarse_UB}
\end{figure*}

\end{proof}

\section{Work deficit density in the MPS family under coarse-graining} \label{Ss:num_calc}

In this section, we calculate the work deficit density in the MPS family defined as Eqs.~(\ref{SMeq:MPS_fam_gnega}) and (\ref{SMeq:MPS_fam_gposi}), in a coarse-grained picture. We express the number of sites per single party as $m$, the total number of sites as $N$, and the number of parties as $M$, where the equality $N=mM$ is satisfied.
We consider the thermodynamic limit, where $N\to \infty$ and $M \to \infty$ is taken while keeping $m = \frac{N}{M}$ constant.
In the following, we evaluate the behavior of $\delta_m$ in two parameter ranges $\xi_g \gg m$ and $\xi_g \ll m$, where $\xi_g$ is the correlation length of the MPS family calculated as
\begin{equation*}
    \xi_g = \left( \ln \frac{1+|g|}{1-|g|} \right)^{-1}.
\end{equation*}

In the parameter region $m\gg \xi_g$, the work deficit density $\delta_m$ converges to half of the entanglement entropy $\frac{S_{\rm EE}}{2}$, which follows from Theorem~\ref{SThm4:coarse-graining}.
Since the fixed point tensor $\tau$ is calculated as Eq.~(\ref{Seq:MPSfam_RDM}), the entanglement entropy $S_{\rm EE} \equiv 2 S(\tau)$ is quantified as follows:
\begin{equation}
    S_{\rm EE} = \left\{
    \begin{alignedat}{2}
        &2\ln 2  \quad  &(g < 0) \\
        &2 h\left( \frac{1}{2}-\frac{\sqrt{g}}{1+g} \right) \quad &(g > 0)
    \end{alignedat}
    \right. ,\label{Seq:MPSfam_EE}
\end{equation}
where $h\left(p\right) \equiv -p \ln p - (1-p) \ln (1-p)$ is the binary entropy function.

On the other hand, in the parameter region $m\ll \xi_g$, $\delta_m$ is not reduced to the entanglement entropy, and therefore we need to evaluate it in a different manner.
Specifically, we here show that $\delta_m$ is upper-bounded as
\begin{equation}
\label{Seq:WD_coarse_UB}
    \delta_m \leq \Tilde{\mathcal{O}} (m|g|).
\end{equation}
Since $m |g| \ll 1$ is satisfied if $m\ll \xi_g$, Eq.~(\ref{Seq:WD_coarse_UB}) implies that the work deficit density approaches zero around $g=0$.
The derivation of Eq.~(\ref{Seq:WD_coarse_UB}) is as follows:
\begin{align*}
    \delta_m &\leq U_{\Lambda^{\otimes M}} \equiv \frac{1}{M}H_{\Lambda^{\otimes M}}(Y_M) \\
    &\leq \frac{1}{M} \left\{ H_{\Lambda} (y_1) + \sum_{n=1}^{M-1} H_{\Lambda^{\otimes 2}}(y_{n+1}|y_n) \right\} \\
    &\leq \frac{2\ln 2}{M} + H_{\Lambda^{\otimes 2}}(y_{2}|y_1) \\
    &\leq \Tilde{\mathcal{O}} (m|g|).
\end{align*}
Here, $\Lambda^{\otimes M}$ is a translationally invariant protocol to perform identical measurement $\Lambda$ on $M$ parties, whose specific measurement basis is defined later. The third line follows from the translation invariance of the state and the measurement protocol.
The inequality in the final line is derived by upper-bounding both terms in the third line. 
Since we can immediately show that the first term $\frac{2\ln 2}{M}$ is negligible in the thermodynamic limit $M \to \infty$, what remains to be shown is the inequality 
\begin{equation}
\label{SMeq:UB_ansatz}
H_{\Lambda^{\otimes 2}}(y_{2}|y_1) \leq \Tilde{\mathcal{O}} (m|g|),    
\end{equation}
which is derived in the rest of this section.

\begin{figure*}[]
    \centering
    \includegraphics[width=1\textwidth]{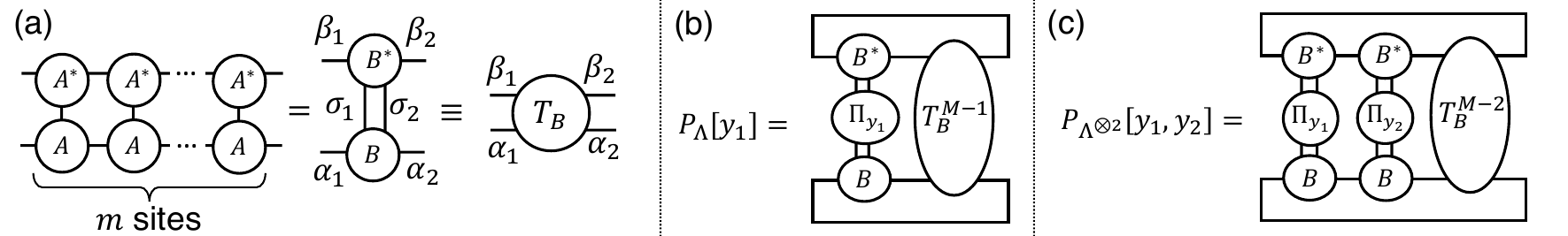}
    \caption{Calculation in the MPS family (Eqs.~(\ref{SMeq:MPS_fam_gnega}) and (\ref{SMeq:MPS_fam_gposi})) in a coarse-grained picture. 
    (a) Diagram for the $m$-site tensor $B$ and its transfer matrix $T_B$. 
    (b) Diagram for the probability $P_{\Lambda}[y_1]$. Here, $\Pi_{y_1}$ represents a rank-one projector on the party $b_1$. 
    (c) Diagram for the probability $P_{\Lambda}[y_1, y_{2}]$. Here, $\Pi_{y_1}$ and $\Pi_{y_{2}}$ represent rank-one projectors on parties $b_1$ and $b_2$, respectively.}
    \label{SMfig:coarse_MPS_fam_direct}
\end{figure*}

To derive Eq.~(\ref{SMeq:UB_ansatz}), we first calculate the $m$-site tensor $B$ and its transfer matrix $T_B$, defined as Fig.~\ref{SMfig:coarse_MPS_fam_direct}(a).
The transfer matrix is calculated as
\begin{equation}
    [T_B]_{(\alpha_1,\alpha_2);(\beta_1,\beta_2)} = \left\{
    \begin{alignedat}{2}
    &\left(
    \begin{array}{cccc}
    \frac{1+\lambda_2^m}{2} & \frac{-\sqrt{|g|}\lambda_2^m}{1+g} &0 & 0\\
    \frac{-\sqrt{|g|}\lambda_2^m}{1+g}& \frac{1-\lambda_2^m}{2} &0 &0\\
    0 & 0 & \frac{1-\lambda_2^m}{2} &\frac{\sqrt{|g|}\lambda_2^m}{1+g}\\
    0 & 0 &\frac{\sqrt{|g|}\lambda_2^m}{1+g} & \frac{1+\lambda_2^m}{2}
    \end{array}
    \right)  \quad  &(g < 0)\\
        &\left(
    \begin{array}{cccc}
    \frac{1+\lambda^m_2}{2} & \frac{\sqrt{g}}{1+g} &0 & 0\\
    \frac{\sqrt{g}}{1+g} & \frac{1-\lambda_2^m}{2} &0 &0\\
    0 & 0 & \frac{1-\lambda_2^m}{2} &\frac{\sqrt{g}}{1+g} \\
    0 & 0 &\frac{\sqrt{g}}{1+g} & \frac{1+\lambda_2^m}{2}
    \end{array}
    \right)  \quad  &(g > 0) 
    \end{alignedat}
    \right. 
\end{equation}
where $(\alpha_1,\alpha_2)$ is the row index and $(\beta_1,\beta_2)$ is the column index.
Here, $\lambda_2$ denotes the subleading eigenvalue of the transfer matrix (i.e., $\lambda_2 = \frac{1+g}{1-g}$ for $g<0$ and $\lambda_2 = \frac{1-g}{1+g}$ for $g>0$).
By taking the standard square root of $T_B$, the $m$-site tensor $B$ is calculated as follows:
\begin{equation}
    [B]_{(\alpha_1,\alpha_2);(\sigma_1,\sigma_2)} =
        \left(
    \begin{array}{cccc}
    a & b^* &0 & 0\\
    b & c & 0 & 0\\
    0 & 0 & d & e^* \\
    0 & 0 & e & f
    \end{array}
    \right),
\end{equation}
where $a$, $c$, $d$, and $f$ are real numbers and $b$ and $e$ are complex numbers, whose sizes are evaluated as follows both in $g<0$ and $g>0$:
\begin{equation}
\label{SMeq:coarse_mat_sizes}
    \begin{split}
        &1-a = \mathcal{O}(m|g|), \quad |b| = \mathcal{O}(\sqrt{|g|}), \quad c = \mathcal{O}(\sqrt{m|g|}), \\
    &1-f = \mathcal{O}(m|g|), \quad |e| = \mathcal{O}(\sqrt{|g|}), \quad d = \mathcal{O}(\sqrt{m|g|}).
    \end{split}
\end{equation}
In the derivation of Eq.~(\ref{SMeq:coarse_mat_sizes}), we utilize the relationship $B B^\dag = T_B$ and $1 - \lambda_2^m = \mathcal{O}(m |g|)$.

We next determine the ansatz measurement protocol $\Lambda$ and derive the inequality (\ref{SMeq:UB_ansatz}).
The measurement basis for $\Lambda$ on the index $(\sigma_1,\sigma_2)$ is defined as follows:
\begin{equation}
    \Pi_{y=1} = \left(
    \begin{array}{cccc}
    1 & 0 & 0 & 0 \\
    0 & 0 & 0 & 0 \\
    0 & 0 & 0 & 0 \\
    0 & 0 & 0 & 0 
    \end{array}
    \right),  \ 
    \Pi_{y=2} = \left(
    \begin{array}{cccc}
    0 & 0 & 0 & 0 \\
    0 & 1 & 0 & 0 \\
    0 & 0 & 0 & 0 \\
    0 & 0 & 0 & 0 
    \end{array}
    \right), \ 
    \Pi_{y=3} = \left(
    \begin{array}{cccc}
    0 & 0 & 0 & 0 \\
    0 & 0 & 0 & 0 \\
    0 & 0 & 1 & 0 \\
    0 & 0 & 0 & 0
    \end{array}
    \right),  \ 
    \Pi_{y=4} = \left(
    \begin{array}{cccc}
    0 & 0 & 0 & 0 \\
    0 & 0 & 0 & 0 \\
    0 & 0 & 0 & 0 \\
    0 & 0 & 0 & 1
    \end{array}
    \right). \ 
\end{equation}
For such a measurement protocol $\Lambda$, we can derive the following inequality:
\begin{equation}
\label{SMeq:coarse_MPSfam_ansatz}
    \begin{split}
        H_{\Lambda^{\otimes 2}}(y_{2}|y_1)  &=\sum_{i=1}^4 P_\Lambda[y_{1}=i] H_{\Lambda^{\otimes 2}}(y_{2}|y_1=i)\\
        &= P_\Lambda[y_{1}=1] H_{\Lambda^{\otimes 2}}(y_{2}|y_1 =1) + P_\Lambda[y_{1}=4] H_{\Lambda^{\otimes 2}}(y_{2}|y_1=4) + \mathcal{O}(m|g|) \\
        &\leq \Tilde{\mathcal{O}}(m|g|).
    \end{split}
\end{equation}
The second line follows from the following equations, which are derived by taking the tensor contraction in Fig.~\ref{SMfig:coarse_MPS_fam_direct}(b):
\begin{equation}
    \begin{split}
    P_\Lambda [y_{1}=2] &=\frac{1+\lambda_{2}^{N-m}}{2(1+\lambda_{2}^{N})}|b|^2 + \frac{1-\lambda_{2}^{N-m}}{2(1+\lambda_{2}^{N})} c^2= \mathcal{O}(m|g|),  \\
    P_\Lambda [y_{1}=3] &=\frac{1+\lambda_{2}^{N-m}}{2(1+\lambda_{2}^{N})}|e|^2 + \frac{1-\lambda_{2}^{N-m}}{2(1+\lambda_{2}^{N})} d^2= \mathcal{O}(m|g|).
    \end{split}
\end{equation}
The third line of Eq.~(\ref{SMeq:coarse_MPSfam_ansatz}) follows from the inequalities $H_{\Lambda^{\otimes 2}}(y_{2}|y_{1}=1)\leq \Tilde{\mathcal{O}}(m|g|)$ and $H_{\Lambda^{\otimes 2}}(y_{2}|y_{1}=4) \leq \Tilde{\mathcal{O}}(m|g|)$, which are directly derived from the Alicki-Fannes inequality \cite{alicki2004continuity} and the following equations:
\begin{equation}
\label{SMeq:coarse_MPSfam_prob_y1y2}
\begin{split}
    P_{\Lambda^{\otimes 2}} [y_{2}=1|y_{1}=1] &= \frac{(1+\lambda_{2}^{N-2m}) a^2+ (1-\lambda_{2}^{N-2m}) |b|^2}{(1+\lambda_{2}^{N-m}) a^2+ (1-\lambda_{2}^{N-m})|b|^2} a^2 = 1 - \mathcal{O}(m|g|), \\
    P_{\Lambda^{\otimes 2}} [y_{2}=4|y_{1}=4] &= \frac{(1+\lambda_{2}^{N-2m}) f^2+ (1-\lambda_{2}^{N-2m}) |e|^2}{(1+\lambda_{2}^{N-m}) f^2+ (1-\lambda_{2}^{N-m})|e|^2} f^2 = 1 - \mathcal{O}(m|g|).
\end{split}
\end{equation}
These equations are derived through the calculation shown in Fig.~\ref{SMfig:coarse_MPS_fam_direct}(c).

\bibliography{biblio.bib}